\documentclass[11pt,letterpaper]{amsart}

\usepackage{tikz}
\usetikzlibrary{topaths,calc}

\usepackage{mathrsfs, amssymb,amsmath,url,amsthm,color,comment}
\usepackage{hyperref}
\usepackage{thm-restate}
\usepackage{thmtools}
\usepackage{todonotes}
\usepackage{algorithm}
\usepackage{algpseudocode}
\usepackage{hyperref,multirow}
\hypersetup{
    colorlinks = true,
    linkcolor = blue
    }
\usepackage[nameinlink,capitalise]{cleveref}

\usepackage{graphicx}

\usepackage[letterpaper,left=1in,top=1in,right=1in,bottom=1in]{geometry}

\theoremstyle{plain}

    \declaretheorem[name=Theorem]{theorem}
    
  \declaretheorem[name=Corollary,numberlike=theorem]{corollary}
  \declaretheorem[name=Conjecture,numberlike=theorem]{conjecture}
  \declaretheorem[name=Observation,numberlike=theorem]{observation}
  
  \declaretheorem[name=Lemma,numberlike=theorem]{lemma}

\theoremstyle{definition}
\declaretheorem[name=Definition,numberlike=theorem]{definition}
     \declaretheorem[name=Example,numberlike=theorem]{example}

 \newcommand{\ignore}[1]{}

\renewcommand\thmcontinues[1]{continued}

\newcommand{\actson}{\curvearrowright}

\DeclareMathOperator{\csp}{CSP}
\DeclareMathOperator{\Csp}{CSP}
\DeclareMathOperator{\CSP}{CSP}
\DeclareMathOperator{\Pol}{Pol}
\DeclareMathOperator{\Aut}{Aut}
\DeclareMathOperator{\End}{End}

\DeclareMathOperator{\proj}{proj}
\DeclareMathOperator{\V}{\mathcal{V}}
\DeclareMathOperator{\constraints}{\mathcal{C}}
\DeclareMathOperator{\instance}{\mathcal{I}}

\DeclareMathOperator{\id}{id}

\renewcommand\vec{\overrightarrow}

\newcommand\Kall{\ensuremath{\mathcal K^\ell_{\textnormal{all}}}}
\newcommand{\inj}{\textnormal{inj}}
\newcommand\injtuples[1][\V]{[#1]^\ell}
\newcommand\fin{\textrm{fin}}
\newcommand\fininstance[1][\instance]{#1_{\fin}}
\newcommand\injinstance[1][\instance]{#1^{(\inj)}}
\newcommand\injinstances{\textnormal{Inj}}

\newcommand\restrinstance[2][\instance]{#1|_{#2}}

\newcommand\rel[1]{\mathbb{#1}}

\newcommand{\tuple}[1]{\mathbf{#1}}

\newcommand{\NP}{\mathrm{NP}}

\renewcommand{\P}{\mathrm{P}}

\newcommand{\sH}{\mathbb H}
\newcommand{\sA}{\mathbb A}
\newcommand{\sB}{\mathbb B}
\newcommand{\sC}{\mathbb C}

\newcommand\sX{\mathbb X}

\newcommand{\CA}{\Pol(\sA)}
\newcommand{\CAH}{\mathscr{C}_{\mathbb A}^{\sH, \inj}}
\newcommand{\CAA}{\mathscr{C}_{\mathbb A}^{\mathbb A, \inj}}

\newcommand{\en}{\{E,N\}}

\newcommand{\cC}{\mathscr C}
\newcommand{\cM}{\mathscr M}
\newcommand{\cD}{\mathscr D}

\newcommand{\fC}{\mathscr C}
\newcommand{\fD}{\mathscr D}

\newcommand{\fG}{\mathscr G}

\newcommand{\cL}{\mathcal L}
\newcommand{\cF}{\mathcal F}

\newcommand{\cJ}{\mathcal J}
\newcommand{\gG}{\mathscr G}

\newcommand{\Projs}{\mathscr P}

\author[A.~Mottet]
{Antoine Mottet}
	\address{Hamburg University of Technology, Research Group on Theoretical Computer Science, Germany}
	\email{antoine.mottet@tuhh.de}
	\urladdr{https://amottet.github.io/}

\author[T.~Nagy]
{Tom\'a\v s Nagy}
	\address{Theoretical Computer Science Department, Jagiellonian University, Poland}
	\email{tomas.nagy@email.com}
    \urladdr{https://nagyto.github.io/}

\author[M.~Pinsker]
{Michael Pinsker}
	\address{Institut f\"{u}r Diskrete Mathematik und Geometrie, FG Algebra, TU Wien}
	\email{marula@gmx.at}
    \urladdr{http://dmg.tuwien.ac.at/pinsker/}

\thanks{This research was partially funded by the Deutsche Forschungsgemeinschaft (DFG, German Research
Foundation) – project number 534904934.
This research was funded in whole or in part by the Austrian Science Fund (FWF) [P 32337, I 5948]. This research was funded in whole or in part by National Science Centre, Poland 2021/03/Y/ST6/00171. For the purpose of Open Access, the authors have applied a CC BY public copyright licence to any Author Accepted Manuscript (AAM) version arising from this submission. This research is also funded by the European Union (ERC, POCOCOP, 101071674). Views and opinions expressed are however those of the author(s) only and do not necessarily reflect those of the  European Union or the European Research Council Executive Agency. Neither the European Union nor the granting authority can be held responsible for them.}

\begin{document}

\title[An order out of nowhere: a new algorithm for infinite-domain {CSP}s]{
An order out of nowhere: \\a new algorithm for infinite-domain {CSP}s
}

\begin{abstract}
    We consider the problem of satisfiability of sets of constraints in a given set of finite uniform hypergraphs.
    While the problem under consideration is similar in nature to the problem of satisfiability of constraints in graphs, the classical complexity reduction to finite-domain CSPs that was used in the proof of the complexity dichotomy for such problems cannot be used as a black box in our case.
    We therefore introduce an algorithmic technique inspired by classical notions from the theory of finite-domain CSPs, and prove its correctness based on symmetries that depend on a linear order that is external to the structures under consideration.
    Our second main result is a P/NP-complete complexity dichotomy for such problems over many sets of uniform hypergraphs.
    The proof is based on the translation of the problem into the framework of constraint satisfaction problems (CSPs) over infinite uniform hypergraphs.
    Our result confirms in particular the Bodirsky-Pinsker conjecture for CSPs of first-order reducts of %
    {many} homogeneous hypergraphs {including the random hypergraphs and hypergraphs omitting a generalised clique}. %
    This forms a vast generalisation of previous work by Bodirsky-Pinsker (STOC'11) and Bodirsky-Martin-Pinsker-Pongr\'acz (ICALP'16) on graph satisfiability.
\end{abstract}

\maketitle

\section{Introduction}

In~\cite{Schaefer-Graphs}, Bodirsky and the third author introduced the computational problem Graph-SAT
as a generalisation of systematic restrictions of the Boolean satisfiability problem studied by Schaefer in~\cite{Schaefer}. A \emph{graph formula} is a formula formed from the atomic formulas $E(x,y)$ and $x=y$ using negation, conjunction and disjunction, where $E$ is interpreted as the edge relation of a simple undirected graph.
Given a finite set $\Psi$ of graph formulas, the \emph{graph satisfiability problem} Graph-SAT($\Psi$) gets as an input a finite set $\V$ of variables and a graph formula $\Phi=\phi_1\wedge\dots\wedge \phi_n$, where every $\phi_i$ is obtained from a formula $\psi\in\Psi$ by substituting the variables of $\psi$ by variables from $\V$; the goal is to decide the existence of a graph satisfying $\Phi$. 
Any instance of a Boolean satisfiability problem can indeed  easily be reduced to a problem of this form, roughly by  replacing Boolean variables by pairs of variables which are to be assigned vertices in a graph, and by translating the potential Boolean values 0 and 1 into the non-existence or existence of an edge between these two variables. The main result of~\cite{Schaefer-Graphs} states that this computational problem is either solvable in polynomial time or is NP-complete. This can be put in contrast with the theorem of Ladner~\cite{Ladner} stating that if P $\neq$ NP, then there exist computational problems that are neither solvable in polynomial-time nor NP-complete.
Similar \emph{dichotomy theorems} have been established for related problems concerning the satisfaction of constraints by linear orders~\cite{temporalCSP}, partially ordered sets~\cite{PosetCSP}, tournaments~\cite{SmoothApproximations}, or phylogenetic trees~\cite{PhylogenyCSP}.
It is conjectured that such dichotomies defying Ladner's theorem are common; we refer to~\Cref{sect:conj} and~\Cref{conj:Bod-Pin} for a precise statement.

In order to develop our understanding of such natural generalisations of the classical Boolean satisfiability problem, we consider in this article the complexity of Graph-SAT where graph formulas are replaced by \emph{$\ell$-hypergraph formulas} for some fixed $\ell\geq 2$.
More precisely, we consider formulas where $E$ is an $\ell$-ary symbol denoting the edge relation of an $\ell$-uniform undirected hypergraph; in the following, since all our hypergraphs are uniform and undirected, we simply write \emph{$\ell$-hypergraph}. 
The problem $\ell$-Hypergraph-SAT is then defined in the same way as the problem Graph-SAT above. 
We also study the complexity of the natural variant of the $\ell$-Hypergraph-SAT problem, investigated in~\cite{HomogeneousGraphs} for the special case of graphs, where we ask for the existence of a satisfying hypergraph that belongs to a prescribed set $\mathcal K$ of finite $\ell$-hypergraphs.
This corresponds to imposing structural restrictions on the possible satisfying hypergraph solutions.
For example, it is natural to ask for the existence of a solution in the class $\mathcal K^\ell_r$ of all finite $\ell$-hypergraphs omitting a generalised clique on $r$ vertices.
We use the notation $\ell$-Hypergraph-SAT($\Psi,\mathcal{K}$) to denote this problem.

Surprisingly, it turns out that $\ell$-hypergraph problems behave very differently from the corresponding graph problems, requiring in particular genuinely novel algorithmic methods to handle them.
A natural attempt to solve hypergraph satisfiability in polynomial time is to use a generic reduction to constraint satisfaction problems (CSPs) whose domain consists of the hypergraphs with at most $\ell$ elements~\cite{ReductionFinite}.
While this reduction always works in the (graph) case of $\ell=2$~\cite{SmoothApproximations}, it can happen for $\ell>2$ that the resulting finite-domain CSP is an NP-complete problem, although  the original hypergraph satisfiability problem is solvable in polynomial time.
Our main result, \Cref{thm:main-algo}, is an algorithm running in polynomial time and solving the $\ell$-Hypergraph-SAT($\Psi,\mathcal K$) problem under some general algebraic assumptions.

The next example illustrates that in this setting our algorithm is strictly more powerful than the reduction of~\cite{ReductionFinite}.

\begin{example}[label=ex:order]
    Let $\ell=3$, let $\psi$ be a formula with 4 free variables that holds precisely for the hypergraphs in~\Cref{fig:picture-example}, and let $\Psi$ be the set consisting of $\psi$. 
    This is an example where the reduction to the finite from~\cite{ReductionFinite} cannot be applied to prove the tractability of $\ell$-Hypergraph-SAT($\Psi$).
   However, the problem is solvable in polynomial time, as it can be solved by the algorithm introduced in \Cref{sect:zhuk}.
\begin{center}
\begin{figure}[!htbp]
\begin{tikzpicture}
\begin{scope}
	\node[label=right:4] (a1) at (0,0) {};
	\node[label=left:3] (a2) [left=of a1] {};
	\node[label=right:2] (a3) [above=of a1] {};
	\node[label=left:1] (a4) [above=of a2] {};

    \filldraw[red!40] (a1) circle (3pt);
    \filldraw[red!40] (a2) circle (3pt);
    \filldraw[blue!40] (a3) circle (3pt);
    \filldraw[blue!40] (a4) circle (3pt);
\end{scope}

\begin{scope}[yshift=-3cm]
	\node (a1) at (0,0) {};
	\node (a2) [left=of a1] {};
	\node (a3) [above=of a1] {};
	\node (a4) [above=of a2] {};

    \filldraw[red!40] (a1) circle (3pt);
    \filldraw[blue!40] (a2) circle (3pt);
    \filldraw[red!40] (a3) circle (3pt);
    \filldraw[blue!40] (a4) circle (3pt);
\end{scope}

\begin{scope}[xshift=6cm]
	\node (a1) at (0,0) {};
	\node (a2) [left=of a1] {};
	\node (a3) [above=of a1] {};
	\node (a4) [above=of a2] {};
	
	\filldraw[fill=blue!30, fill opacity=0.3] ($(a1)+(0.3,-0.3)$)
		to[out=-135,in=-45] ($(a2)+(-0.3,-0.3)$)
		to[out=135,in=-180] ($(a3)+(0,0.3)$)
		to[out=0,in=45] ($(a1)+(0.3,-0.3)$);
		
	\filldraw[fill=red!30, fill opacity=0.3] ($(a2)+(-0.3,-0.3)$)
		to[out=-45,in=-135] ($(a1)+(0.3,-0.3)$)
		to[out=45,in=0] ($(a4)+(0,0.3)$)
		to[out=180,in=135] ($(a2)+(-0.3,-0.3)$);

  \draw (a1) circle (3pt);
    \draw (a2) circle (3pt);
    \draw (a3) circle (3pt);
    \draw (a4) circle (3pt);
\end{scope}

\begin{scope}[xshift=3cm]
		\node (a1) at (0,0) {};
	\node (a2) [left=of a1] {};
	\node (a3) [above=of a1] {};
	\node (a4) [above=of a2] {};
	
	\filldraw[fill=blue!30, fill opacity=0.3] ($(a1)+(0.3,-0.1)$)
		to[out=45,in=-45] ($(a3)+(0.3,0.3)$)
		to[out=135,in=45] ($(a4)+(-0.3,0.3)$)
		to[out=-135,in=-135] ($(a1)+(0.3,-0.1)$);
		
	\filldraw[fill=red!30, fill opacity=0.3] ($(a2)+(-0.3,-0.1)$)
		to[out=-45,in=-45] ($(a3)+(0.3,0.3)$)
		to[out=135,in=45] ($(a4)+(-0.3,0.3)$)
		to[out=-135,in=135] ($(a2)+(-0.3,-0.1)$);
  \draw (a1) circle (3pt);
    \draw (a2) circle (3pt);
    \draw (a3) circle (3pt);
    \draw (a4) circle (3pt);
\end{scope}

\begin{scope}[yshift=-3cm,xshift=3cm]
		\node (a1) at (0,0) {};
	\node (a2) [left=of a1] {};
	\node (a3) [above=of a1] {};
	\node (a4) [above=of a2] {};
	
	\filldraw[fill=blue!30, fill opacity=0.3] ($(a2)+(-0.3,-0.3)$)
		to[out=-45,in=-135] ($(a1)+(0.3,-0.3)$)
		to[out=45,in=0] ($(a4)+(0,0.3)$)
		to[out=180,in=135] ($(a2)+(-0.3,-0.3)$);
		
	\filldraw[fill=red!30, fill opacity=0.3] ($(a2)+(-0.3,-0.1)$)
		to[out=-45,in=-45] ($(a3)+(0.3,0.3)$)
		to[out=135,in=45] ($(a4)+(-0.3,0.3)$)
		to[out=-135,in=135] ($(a2)+(-0.3,-0.1)$);
  \draw (a1) circle (3pt);
    \draw (a2) circle (3pt);
    \draw (a3) circle (3pt);
    \draw (a4) circle (3pt);
\end{scope}

\begin{scope}[yshift=-3cm,xshift=6cm]
		\node (a1) at (0,0) {};
	\node (a2) [left=of a1] {};
	\node (a3) [above=of a1] {};
	\node (a4) [above=of a2] {};
	
	\filldraw[fill=blue!30, fill opacity=0.3] ($(a1)+(0.3,-0.3)$)
		to[out=-135,in=-45] ($(a2)+(-0.3,-0.3)$)
		to[out=135,in=-180] ($(a3)+(0,0.3)$)
		to[out=0,in=45] ($(a1)+(0.3,-0.3)$);
  \filldraw[fill=red!30, fill opacity=0.3] ($(a1)+(0.3,-0.1)$)
		to[out=45,in=-45] ($(a3)+(0.3,0.3)$)
		to[out=135,in=45] ($(a4)+(-0.3,0.3)$)
		to[out=-135,in=-135] ($(a1)+(0.3,-0.1)$);
  \draw (a1) circle (3pt);
    \draw (a2) circle (3pt);
    \draw (a3) circle (3pt);
    \draw (a4) circle (3pt);
\end{scope}
\end{tikzpicture}
\caption{Hypergraphs on at most $4$ elements satisfying $\psi$ in~\Cref{ex:order}. The vertices are labeled to represent each of the four free variables of $\psi$ (only shown on one of the hypergraphs for readability). The two 3-hypergraphs on the left have 2 vertices, and the color coding denotes vertices that are equal. The other four 3-hypergraphs have four vertices each and precisely two hyperedges.}
\label{fig:picture-example}
\end{figure}
\end{center}
\end{example}

Building on~\Cref{thm:main-algo}, our second contribution is a full complexity dichotomy for the problems $\ell$-Hypergraph-SAT($\Psi,\mathcal K$) where $\mathcal K$ is the class $\Kall$ of all finite $\ell$-hypergraphs, 
or $\mathcal K^\ell_r$.

\begin{theorem}\label{thm:main-1}
    Let $\ell\geq 3$, let $\mathcal K$ be either the class $\Kall$ of all finite $\ell$-hypergraphs or the class $\mathcal K^\ell_r$ for some $r>\ell$, and let $\Psi$ be a set of $\ell$-hypergraph formulas.
    Then $\ell$-Hypergraph-SAT$(\Psi,\mathcal K)$ is either in P, or it is NP-complete.
    Moreover, given $\Psi$, one can algorithmically decide which of the cases holds.
\end{theorem}

In fact, our polynomial-time algorithm in~\Cref{thm:main-algo} solves the hypergraph satisfiability problem for all classes $\mathcal K$ of hypergraphs satisfying certain assumptions that we introduce in~\Cref{sect:zhuk}.
Likewise, our results imply a dichotomy result as in~\Cref{thm:main-1} for every class $\mathcal K$ satisfying certain structural assumptions that we introduce in~\Cref{sect:overview}, see~\Cref{thm:hypergraph-dichotomy}.

\subsection{Connection to Constraint Satisfaction Problems}

The \emph{constraint satisfaction problem} with template $\rel A=(A;R_1,\dots,R_n)$ is the computational problem $\CSP(\rel A)$ of deciding, given an instance with variables $\V$ and constraints $\phi(x_{i_1},\dots,x_{i_r})$ with $\phi\in\{R_1,\dots,R_n\}$ and $x_{i_1},\dots,x_{i_r}\in\V$, whether there exists an assignment $f\colon\V\to A$ that satisfies all the constraints.

Note how the problem $\ell$-Hypergraph-SAT$(\Psi)$ is similar in nature to a constraint satisfaction problem, where the difference lies in the fact that we are not asking for a labelling of variables to elements of a structure $\rel A$, but rather for a consistent labelling of $\ell$-tuples of variables to a finite set describing all the possible $\ell$-hypergraphs on at most $\ell$ elements. For example, in the case of $\ell=2$, this set contains 3 elements (for the graph on a single vertex, and the two undirected graphs on 2 vertices), while for $\ell=3$ this set contains 6 elements (there is one labeled 3-hypergraph on a single element, three on 2 elements, and two on 3 elements).

It was already noticed in~\cite{Schaefer-Graphs} that it is possible to design a structure $\rel A$ (which is necessarily infinite) such that Graph-SAT$(\Psi)$ is equivalent to $\CSP(\rel A)$, and this observation also carries out to the hypergraph setting as follows.

Fix $\ell\geq 3$ and a class $\mathcal K$ of finite $\ell$-hypergraphs.
Let us assume that $\mathcal K$ is an \emph{amalgamation class}: an isomorphism-closed class that is closed under induced sub-hypergraphs and with the property that for any two hypergraphs $\rel H_1,\rel H_2\in\mathcal K$  having a common hypergraph $\rel H$ as intersection, there exists $\rel H'\in\mathcal K$ and embeddings of $\rel H_1,\rel H_2$ into $\rel H'$ which agree on $\rel H$.
A classical result of Fra\"iss\'e~\cite{Fraisse} yields that there exists an infinite limit hypergraph $\rel H_{\mathcal K}$, called the \emph{Fra\"iss\'e limit of $\mathcal K$}, with the property that the finite induced sub-hypergraphs of $\rel H_{\mathcal K}$ are precisely the hypergraphs in $\mathcal K$.
Moreover, this limit can be taken to be \emph{homogeneous}, i.e., 
highly symmetric in a certain precise sense -- see~\Cref{sect:Prelims} for  precise definitions of these concepts.
If $\Psi$ is a set of $\ell$-hypergraph formulas, then it defines in $\rel H_{\mathcal K}$ a set of relations that one can view as a CSP template $\rel A_{\mathcal K,\Psi}$.
It follows that the problem $\ell$-Hypergraph-SAT($\mathcal K,\Psi$) is precisely the same as $\CSP(\sA_{\mathcal K,\Psi})$.
The assumption that $\mathcal K$ is an amalgamation class is rather mild and is for example fulfilled by the classes of interest for~\Cref{thm:main-1}, namely by the class $\Kall$ of all finite $\ell$-hypergraphs, or for any $\ell<r$ by the class $\mathcal K_r^\ell$.
We refer to~\cite{3uniformhypergraphs} for a discussion of amalgamation classes of 3-hypergraphs; importantly, while in the case of $\ell=2$ all such classes are known~\cite{LachlanWoodrow}, it seems very difficult to obtain a similar classification in general since there are uncountably-many such classes already for $\ell=3$. 
The latter fact obliges us to build on and refine abstract methods rather than relying on the comfort of a classification in our  general dichotomy result (\Cref{thm:hypergraph-dichotomy}), which contrasts with the approach for graphs in~\cite{HomogeneousGraphs}.

Using the reformulation of $\ell$-hypergraph problems as constraint satisfaction problems, we show that the border between tractability and NP-hardness in \Cref{thm:main-1} can be described algebraically by  properties of the \emph{polymorphisms} of the structures $\sA_{\mathcal K,\Psi}$, i.e., by the functions preserving all relations of $\sA_{\mathcal K,\Psi}$.
This implies, in particular, the above-mentioned  decidability of this border. Roughly speaking, the tractable case corresponds to the CSP template enjoying some non-trivial algebraic invariants in the form of polymorphisms, whereas the hard case is characterised precisely by the absence of such invariants.
For the precise definitions of all the mentioned concepts, see~\Cref{sect:Prelims,subsect:overview-UA}.

\begin{theorem}\label{thm:main-1+2}
Let $\ell\geq 3$, let $\mathcal K$ be either the class $\Kall$ of all finite $\ell$-hypergraphs or the class $\mathcal K^\ell_r$ for some $r>\ell$, and let $\Psi$ be a set of $\ell$-hypergraph formulas. Then precisely one of the following applies.

\begin{enumerate}
     \item\label{itm:tractable-case} The clone of polymorphisms of $\sA_{\mathcal K,\Psi}$ has no uniformly continuous minion homomorphism to the clone of projections $\Projs$, and $\CSP(\rel A_{\mathcal K,\Psi})$ is in $\P$.
     \item The clone of polymorphisms of $\sA_{\mathcal K,\Psi}$ has a uniformly continuous minion homomorphism to the clone of projections $\Projs$, and $\CSP(\rel A_{\mathcal K,\Psi})$ is $\NP$-complete.
\end{enumerate}
\end{theorem}

The algebraic assumptions in the second item of the theorem correspond to the clone of polymorphisms being trivial in a certain sense (i.e., containing only polymorphisms that imitate the behaviour of projections when restricted to a certain set).
For the precise definitions, see~\Cref{subsect:overview-UA} or~\cite{wonderland}.

\subsection{Related work on constraint satisfaction problems}\label{sect:conj}

In the framework of CSPs, it is natural to consider not only classes of finite $\ell$-hypergraphs but also classes of different finite structures in a fixed relational signature.
If such a class $\mathcal K$ is an amalgamation class, then there exists a countably infinite homogeneous structure $\sB_{\mathcal K}$ whose finite substructures are precisely the structures in $\mathcal K$.
However, $\Csp(\sB_{\mathcal K})$ is not guaranteed to be contained in the complexity class NP since the class $\mathcal{K}$ does not have to be algorithmically enumerable (as mentioned above, there are uncountably many amalgamation classes of 3-hypergraphs; hence there exists such a $\mathcal K$ such that $\CSP(\sB_{\mathcal K})$ is undecidable).
A natural way of achieving the algorithmical enumerability of $\mathcal K$ is to require that there exists a natural number $b_{\mathcal K}$ such that a structure is contained in $\mathcal K$ if, and only if, all its substructures of size at most $b_{\mathcal K}$ are in $\mathcal K$.
In this case, we say that $\mathcal K$ (or its Fra\"iss\'e limit $\sB_{\mathcal K}$) is \emph{finitely bounded}. 
For every set $\Psi$ of formulas in the language of the structures at hand, one then gets as in the previous section a structure $\sA_{\mathcal K,\Psi}$ whose domain is the same as $\sB_{\mathcal K}$ and whose relations are definable in first-order logic from the relations of $\sB_{\mathcal K}$ -- we say that $\sA_{\mathcal K,\Psi}$ is a \emph{first-order reduct} of $\sB_{\mathcal K}$.

Thus, for every set $\Psi$ of formulas and every finitely bounded amalgamation class $\mathcal K$, the generalised satisfiability problem parametrised by $\Psi$ and $\mathcal K$ is the CSP of a first-order reduct $\sA_{\mathcal K,\Psi}$ of a finitely bounded homogeneous structure.
It is known that the complexity of the CSP over any such template depends solely on the polymorphisms of $\sA_{\mathcal K,\Psi}$~\cite{Topo-Birk}.
This motivates the following conjecture generalising the dichotomy for Graph-SAT which was formulated by Bodirsky and {the third author} %
in 2011 (see~\cite{BPP-projective-homomorphisms}). The modern formulation of the conjecture based on recent progress \cite{BKO+17,BKO+19,wonderland} is the following:
\begin{conjecture}\label{conj:Bod-Pin}
Let $\sA$ be a CSP template which is a first-order reduct of
a finitely bounded homogeneous structure. Then one of the following applies.
\begin{enumerate}
    \item The clone of polymorphisms of $\sA$ has no uniformly continuous minion homomorphism to the clone of projections $\Projs$, and $\CSP(\sA)$ is in $\P$. 
    \item The clone of polymorphisms of $\sA$ has a uniformly continuous minion homomorphism to the clone of projections $\Projs$, and $\CSP(\sA)$ is $\NP$-complete.
\end{enumerate}
\end{conjecture}

It follows that \Cref{thm:main-1+2} is a special case of \Cref{conj:Bod-Pin}. It is known that if the clone of polymorphisms of any CSP template within the range of \Cref{conj:Bod-Pin} has a uniformly continuous minion homomorphism to $\Projs$, then the CSP of such template is NP-hard~\cite{wonderland}.
Already before \Cref{conj:Bod-Pin} was introduced, a similar conjecture was formulated by Feder and Vardi~\cite{FederVardi} for CSPs over templates with finite domains and it was  confirmed independently by Bulatov and Zhuk~\cite{Bulatov:2017,Zhuk:2017,Zhuk:2020} recently. \Cref{conj:Bod-Pin} itself has been confirmed for many subclasses: for example for $\Csp$s of all structures first-order definable in finitely bounded homogeneous graphs \cite{Schaefer-Graphs,HomogeneousGraphs}, in $(\mathbb Q;<)$ \cite{temporalCSP}, in any unary structure \cite{ReductsUnary}, in the random poset \cite{PosetCSP}, in the random tournament \cite{SmoothApproximations}, or in the homogeneous branching C-relation \cite{PhylogenyCSP}, in $\omega$-categorical monadically stable structures \cite{BodorDiss}, as well as for all $\Csp$s in the class MMSNP \cite{MMSNP}, and for $\Csp$s of representations of some relational algebras \cite{NetworkCSP}.

\subsection{Novelty of the methods and significance of the results}

We prove that under the algebraic assumption in item (\ref{itm:tractable-case}) of~\Cref{thm:main-1+2}, $\sA_{\mathcal K,\Psi}$ admits non-trivial symmetries that can be seen as operations acting on the set of \emph{linearly ordered} $\ell$-hypergraphs with at most $\ell$ elements.
{Considering symmetries of the associated CSP templates which act on a set of finite structures was a crucial part of the complexity classification of the Graph-SAT problems as well as of all the other above-mentioned subclasses of templates for which \Cref{conj:Bod-Pin} was confirmed. The case of hypergraphs is, as observed in~\cite{infinitesheep}, to date the only example of structures which do not themselves induce linear orders (more precisely, in model-theoretic terms, hypergraphs form a class having the \emph{non-strict order property}, see~\cite{Simon}), and where}
the introduction of a linear order ``out of nowhere'' is unavoidable, in the sense that the symmetries of $\sA_{\mathcal K,\Psi}$ acting on unordered $\ell$-hypergraphs can be trivial even if $\CSP(\sA_{\mathcal K,\Psi})$ is solvable in polynomial time.
{For all the other classes of structures not encoding linear orders, all tractable CSP templates posses non-trivial symmetries acting on all finite structures in the respective classes.}
As a consequence, the aforementioned ``reduction to the finite'' introduced in~\cite{ReductionFinite}, which is enough to prove the tractability part of most of the complexity dichotomies mentioned in the previous section, cannot be used in the hypergraph satisfiability setting.
In order to prove the tractability part of~\Cref{thm:main-1+2}, we thus introduce new algorithmic techniques inspired by results in the theory of constraint satisfaction problems with \emph{finite} domains, in particular by absorption theory~\cite{cyclic} and Zhuk's theory~\cite{Zhuk:2017,Zhuk:2020}.
More precisely, let $\instance$ be an instance of $\Csp(\sA_{\mathcal K,\Psi})$.
Our algorithm transforms $\instance$ into an equi-satisfiable instance $\instance'$ that is sufficiently locally consistent, such that the solution set of a certain relaxation of $\instance'$ does not imply any restrictions on the solution set of the whole instance, and that satisfies an additional condition resembling Zhuk's notion of irreducibility~\cite{Zhuk:2017,Zhuk:2020}. 
We then prove that any 
non-trivial instance satisfying those properties has an injective solution. 
This step is to be compared with the case of \emph{absorbing reductions} in Zhuk's algorithm. 
The existence of an injective solution can then be checked by the aforementioned reduction to the finite from~\cite{ReductionFinite,ReductsUnary}. In this way, we resolve
the trouble with hitherto standard methods pointed out in~\cite{infinitesheep}. A positive resolution of the general~\Cref{conj:Bod-Pin} will likely have to proceed in a similar spirit, albeit at a yet higher level of sophistication. In the case of graphs, the above described algorithm is not necessary since every instance can be immediately reduced to a finite-domain CSP by the black box reduction.

We use the recently developed theory of smooth approximations~\cite{SmoothApproximations} to prove the dichotomy, i.e., that $\sA_{\mathcal K,\Psi}$ satisfies one of the two items of~\Cref{thm:main-1+2}, for all $\Psi$.
The classification of the complexity of graph-satisfiability problems from~\cite{Schaefer-Graphs} used a demanding case distinction over the possible automorphisms groups of the structures $\sA_{\mathcal K,\Psi}$ (where $\Psi$ is a set of graph formulas, and $\mathcal K$ is the class of all finite simple undirected graphs) -- it was known previously that there are exactly 5 such groups~\cite{ThomasGraphs}.
Our result relies neither on such a classification of the automorphism groups of the structures under consideration, nor on the classification of the hypergraphs of which they are first-order reducts;
as mentioned above, no such classification is available.
While Thomas~\cite{ThomasClassification} obtained a classification of the mentioned automorphism groups for every fixed $\ell$ and for $\mathcal K$ consisting of all finite $\ell$-hypergraphs, this number grows with $\ell$ and makes an exhaustive case distinction impossible.
To overcome the absence of such classifications, we rely on the scalability of the theory of smooth approximations, i.e., on the fact that the main results of the theory can be used without knowing the base structures under consideration.
This was claimed to be one of the main contributions of this theory; \Cref{thm:main-1+2} and its generalisation~\Cref{thm:hypergraph-dichotomy} are the first complexity classification using smooth approximations that truly exemplifies this promise. 

\subsection{Bonus track: local consistency}

Our structural analysis of $\ell$-hypergraph problems allows us to obtain, in \Cref{sect:bwidth}, as an easy consequence a
description of the hypergraph satisfiability problems $\ell$-Hypergraph-SAT($\Psi,\mathcal K$) that are solvable by local consistency methods, assuming that $\Psi$ contains the atomic formula $E$. Similar classifications, and general results on the amount of local consistency needed in those cases, had previously {been} obtained for various other problems (including Graph-SAT problems) which can be modeled as CSPs of first-order reducts of finitely bounded homogeneous structures~\cite{SmoothApproxCollapses, SmoothApproximations}.

\begin{theorem}\label{thm:hypergraphs_bwidth_intro}
    Let $r>\ell\geq 3$, let $\mathcal K$ be either the class $\Kall$ of all finite $\ell$-hypergraphs or the class $\mathcal K^\ell_r$, and let $\Psi$ be a set of $\ell$-hypergraph formulas containing $E(x_1,\dots,x_\ell)$. Then precisely one of the following applies.

\begin{enumerate}
    \item The clone $\CA$ has no uniformly continuous minion homomorphism to the clone of affine maps over a finite module, and $\Csp(\sA)$ has relational width $(2\ell,\max(3\ell,{r}))$.
    \item The clone $\CA$ has a uniformly continuous minion homomorphism to the clone of affine maps over a finite module.
\end{enumerate}
\end{theorem}

\subsection{Future work}

This work is concerned with the complexity of the decision version of constraint satisfaction problems whose study is motivated by~\Cref{conj:Bod-Pin}.
A natural variant of such problems is the optimisation version, where one is interested in finding a solution to an instance of the CSP that minimises the number of unsatisfied constraints. The complexity of such problems (called MinCSPs) has mostly been investigated for finite templates, but recently also in the case of infinite templates falling within the scope of~\Cref{conj:Bod-Pin} from the point of view of exact optimisation and approximation~{\cite{MakarychevPhylogeny,GuruswamiPermutationCSP,GuruswamiOrderingCSP,TemporalVCSPs,PointAlg,LinEq}}, as well as from the point of view of parametrised complexity~\cite{OsipovWahlstroem}.

Our complexity classification for the decision CSP (\Cref{thm:main-1+2} and its generalisation \Cref{thm:hypergraph-dichotomy}) can be seen as a foundation for a systematic structural study of optimisation problems over hypergraphs.

\subsection{Organisation of the present article}

After introducing a few notions needed for the formulation of the main algorithm in~\Cref{sect:Prelims}, we introduce the algorithm and prove its correctness in~\Cref{sect:zhuk}. In~\Cref{sect:overview}, we finally give an overview of the proof of~\Cref{thm:main-1+2} and introduce all the remaining notions. In~\Cref{sect:binarypol}, we prove a few technical results which are needed in~\Cref{sect:NP-hard}, where we prove that if the assumptions under which the presented algorithm does not work are not satisfied, then the CSP under consideration is NP-hard. Finally, in~\Cref{sect:bwidth}, we prove the above-mentioned result about hypergraph satisfiability problems that are solvable by local consistency methods.

\section{Preliminaries}\label{sect:Prelims}

For any $k\geq 1$, we write $[k]$ to denote the set $\{1,\ldots,k\}$.
A tuple is called \emph{injective} if its entries are pairwise distinct. 
In the entire article, we consider only relational structures in a finite signature.

A \emph{primitive-positive} (pp-)formula is a first-order formula built only from atomic formulas, existential quantification, and conjunction. A relation $R\subseteq A^n$ is \emph{pp-definable} in a relational structure $\sA$ if there exists a pp-formula $\phi(x_1,\dots,x_n)$ such that the tuples in $R$ are precisely the tuples satisfying $\phi$.

Let $\ell\geq 2$. A structure $\sH=(H;E)$ is an \emph{$\ell$-hypergraph}  if the relation $E$ is of arity $\ell$, contains only injective tuples (called \emph{hyperedges}), and is \emph{fully symmetric}, i.e., every tuple obtained by permuting the components of a hyperedge is a hyperedge as well. Given any $\ell$-hypergraph $\sH=(H;E)$, we write $N$ for the set of all injective $\ell$-tuples in $H$ that are not hyperedges, and we call this set the \emph{non-hyperedge} relation.

\subsection{CSPs and Relational Width}

A \emph{CSP instance} over a set $A$ is a pair $\mathcal \instance=(\V,\mathcal C)$, where $\V$ is a non-empty finite set of variables, and $\mathcal C$ is a set of \emph{constraints}; each constraint $C\in \mathcal C$ is a subset of $A^U$ for some non-empty $U\subseteq \V$ ($U$ is called the \emph{scope} of $C$).
For a relational structure $\sA$, we say that $\instance$ is an \emph{instance of $\Csp(\sA)$} if for every $C\in\mathcal{C}$ with scope $U$, there exists an enumeration $u_1,\ldots,u_k$ of the elements of $U$ and 
a $k$-ary relation $R$ of $\sA$ such that for all $f\colon U\to A$ we have 
$f\in C\Leftrightarrow (f(u_1),\dots,f(u_k))\in R$. 
A mapping $s\colon \V\to A$ is a~\emph{solution} of the instance $\instance$ if we have $s|_U\in C$ for every $C\in\mathcal C$ with scope $U$.
Given a constraint $C\subseteq A^U$ and a tuple $\tuple v\in U^k$ for some $k\geq 1$, the \emph{projection of $C$ onto $\tuple v$} is defined by $\proj_{\tuple v}(C):=\{f(\tuple v)\colon f\in C\}$.
Let $U\subseteq \V$. We define the \emph{restriction} of $\instance$ to $U$ to be an instance $\instance|_U=(U,\constraints|_U)$ where the set of constraints $\constraints|_U$ contains for every $C\in\constraints$ the constraint $C|_{U}=\{g|_U\mid g\in C\}$.

We denote by $\Csp_{\injinstances}(\sA)$ the restriction of $\Csp(\sA)$ to those instances of $\Csp(\sA)$ where for every constraint $C$ and for every pair of distinct variables $u,v$ in its scope, $\proj_{(u,v)}(C)\subseteq \{(a,b)\in A^2\mid a\neq b\}$.

\begin{definition}
Let $1\leq m\leq n$. We say that an instance $\instance=(\V,\constraints)$ is \emph{$(m,n)$-minimal} if both of the following hold:
\begin{itemize}
\item every non-empty subset of at most $n$ variables in $\V$ is contained in the scope of some constraint in $\instance$;
\item for every at most $m$-element tuple of variables $\tuple v$ and any two constraints $C_1, C_2 \in \constraints$ whose scopes contain all variables of $\tuple v$, the projections of $C_1$ and $C_2$ onto $\tuple v$ coincide.
\end{itemize}
\end{definition}

For $m\geq 1$, we say that an instance is \emph{$m$-minimal} if it is $(m,m)$-minimal.
We say that an instance $\instance$ of the CSP is \emph{non-trivial} if it does not contain any empty constraint.
Otherwise, $\instance$ is \emph{trivial}.

For all $1\leq m\leq n$ and for every instance $\instance$ of a $\Csp(\sA)$ for some finite-domain structure $\sA$, an $(m,n)$-minimal instance with the same solution set as $\instance$ can be computed from $\instance$ in polynomial time.
The same holds for any $\omega$-categorical structure $\sA$ (see \Cref{subsect:mt} for the definition of $\omega$-categoricity, and see e.g., Section~2.3 in~\cite{SymmetriesEnough} for a description of the $(m,n)$-minimality algorithm in this setting). The resulting instance $\instance'$ is called the \emph{$(m,n)$-minimal instance equivalent to $\instance$} and the algorithm that computes this instance is called the~\emph{$(m,n)$-minimality algorithm}. Note that the instance $\instance'$ is not necessarily an instance of $\Csp(\sA)$. However, $\instance'$ is an instance of $\Csp(\sA')$ where $\sA'$ is the expansion of $\sA$ by all at most $n$-ary relations pp-definable in $\sA$. Moreover, $\Csp(\sA')$ has the same complexity as $\Csp(\sA)$.

If $\instance$ is $m$-minimal and $\tuple v$ is a tuple of variables of length at most $m$, then by definition there exists a constraint of $\instance$ whose scope contains all variables in $\tuple v$, and all the constraints who do have the same projection on $\tuple v$.
We write $\proj_{\tuple v}(\instance)$ for this projection, and call it the \emph{projection of\/ $\instance$ onto $\tuple v$}.

\begin{definition}
Let $1\leq m\leq n$, and let $\sA$ be a relational structure.
We say that $\Csp(\sA)$ has relational width $(m,n)$ if every non-trivial $(m,n)$-minimal instance equivalent to an instance of $\Csp(\sA)$ has a solution. $\Csp(\sA)$ has \emph{bounded width} if it has relational width $(m,n)$ for some natural numbers $m \leq n$.
\end{definition}

\subsection{Basic model-theoretic definitions}\label{subsect:mt}

Let $\sB$ and $\sC$ be relational structures in the same signature.
A \emph{homomorphism} from $\sB$ to $\sC$ is a mapping $f\colon B\rightarrow C$ with the property that for every relational symbol $R$ from the signature of $\sB$ and for every $\tuple b\in R^\sB$, it holds that $f(\tuple b)\in R^\sC$.
An embedding of $\sB$ into $\sC$ is an injective homomorphism $f\colon \sB\to\sC$ such that $f^{-1}$ is a homomorphism from the structure induced by the image of $f$ in $\sC$ to $\sB$, and an \emph{isomorphism} from $\sB$ to $\sC$ is a bijective embedding of $\sB$ into $\sC$.  
An endomorphism of $\sB$ is a homomorphism from $\sB$ to $\sB$, an \emph{automorphism} of $\sB$ is an isomorphism from $\sB$ to $\sB$. 
We denote the set of endomorphisms of $\sB$ by $\End(\sB)$ and the set of its automorphisms by $\Aut(\sB)$.

Let $\ell\geq 2$, and let $\mathcal{K}$ be an isomorphism-closed class of finite $\ell$-hypergraphs.
We say that $\mathcal K$ is an \emph{amalgamation class} if the following two conditions are satisfied: It is closed under induced substructures and for any $\ell$-hypergraphs $\sH,\sH_1,\sH_2\in \mathcal K$ and for any embeddings $f_i$ of $\sH$ into $\sH_i$ ($i\in\{1,2\}$), there exists an $\ell$-hypergraph $\sH'$ and embeddings $g_i$ of $\sH_i$ into $\sH'$ ($i\in\{1,2\}$) such that $g_1\circ f_1=g_2\circ f_2$.
We write $\vec{\mathcal K}$ for the class which contains,
for every $\ell$-hypergraph $\sH$ from $\mathcal K$, all ordered $\ell$-hypergraphs obtained by linearly ordering $\sH$.

A relational structure $\sB$ is \emph{homogeneous} if every isomorphism between finite induced substructures of $\sB$ extends to an automorphism of $\sB$.
The class of finite substructures of a homogeneous structure $\sB$ is an amalgamation class; and conversely, for every amalgamation class $\mathcal K$ there exists a homogeneous structure $\sB_{\mathcal K}$ whose finite induced substructures are exactly the structures in $\mathcal K$ (see e.g.~\cite{HodgesLong} for this as well as the other claims in this section). The structure $\sB_{\mathcal K}$ is called the Fra\"iss\'e limit of $\mathcal K$.
The \emph{universal homogeneous $\ell$-hypergraph} is the Fra\"iss\'e limit of $\Kall$.

A \emph{first-order reduct} of a structure $\sB$ is a structure $\sA$ on the same domain whose relations are definable over $\sB$ by first-order formulas without parameters.
Recall that for any amalgamation class $\mathcal K$ and for any set $\Psi$ of $\ell$-hypergraph formulas, $\sA_{\mathcal K,\Psi}$ denotes the first-order reduct of the Fra\"iss\'e limit $\sH_{\mathcal K}$ of $\mathcal K$ whose relations are defined by the formulas in $\Psi$. 
We remark that if $\sB$ is the Fra\"iss\'e limit of a finitely bounded class, then every first-order formula is equivalent to one without quantifiers.

A countable relational structure is \emph{$\omega$-categorical} if its automorphism group has finitely many orbits in its componentwise action on $n$-tuples of elements for all $n \geq 1$.
This is equivalent to saying that there are only finitely many relations of any fixed arity $n\geq 1$ that are first-order definable from $\sA$. 
Every first-order reduct of a finitely bounded homogeneous structure is $\omega$-categorical. 

\subsection{Polymorphisms}
A \emph{polymorphism} of a relational structure $\sA$ is a function from $A^n$ to $A$ for some $n\geq 1$ which \emph{preserves} all relations of $\sA$, i.e., for every such relation $R$ of arity $m$ and for all tuples $(a^1_1, \ldots,a^1_m), \ldots, (a^n_1, \ldots, a^n_m)\in R$, it holds that $(f(a^1_1, \ldots, a^n_1),\ldots, f(a^1_m, \ldots, a^n_m)) \in R$.
We also say that a polymorphism of $\sA$ preserves a constraint $C\subseteq A^U$ if for all $g_1,\dots,g_n\in C$, it holds that $f\circ(g_1,\dots,g_n)\in C$.
The set of all polymorphisms of a structure $\sA$, denoted by $\CA$, is a \emph{function clone}, i.e., a set of finitary operations on a fixed set which contains all projections and which is closed under arbitrary compositions. Every relation that is pp-definable in a relational structure $\sA$ is preserved by all polymorphisms of $\sA$.

Let $S\subseteq R\subseteq A^n$ be relations pp-definable in a structure $\sA$.
We say that \emph{$S$ is a binary absorbing subuniverse of $R$ in $\sA$} if there exists a binary operation $f\in\Pol(\sA)$ such that for every $\tuple s\in S, \tuple r\in R$, we have that $f(\tuple s,\tuple r), f(\tuple r,\tuple s)\in S$. In this case, we write $S\trianglelefteq_{\sA} R$, and we say that $f$ \emph{witnesses} the binary absorption.

Let $\sA$ be a relational structure, and let $\gG=\Aut(\sA)$ be the group of its automorphisms.
For $n\geq 1$, a $k$-ary operation $f$ defined on the domain of $\sA$ is \emph{$n$-canonical} with respect to $\sA$ if for all $\tuple a_1,\dots,\tuple a_k\in A^n$ and all $\alpha_1,\dots,\alpha_k\in\gG$, there exists $\beta\in\gG$ such that $f(\tuple a_1,\dots,\tuple a_k)=\beta\circ f(\alpha_1(\tuple a_1),\dots,\alpha_k(\tuple a_k))$.
A function $f$ that is $n$-canonical with respect to $\sA$ for all $n \geq 1$ is called \emph{canonical} with respect to $\sA$. 
In particular, $f$ induces an operation on the set $A^n/\gG$ of orbits of $n$-tuples under $\gG$ for every $n\geq 1$.
In our setting, we are interested in operations that are canonical with respect to a homogeneous $\ell$-hypergraph $\sH$ or to a homogeneous linearly ordered $\ell$-hypergraph $(\sH,<)$.
In this case, an operation canonical with respect to $\sH$ can simply be seen as an operation on labeled $\ell$-hypergraphs with at most $n$ elements, while an operation canonical with respect to $(\sH,<)$ can be seen as an operation on labeled $\ell$-hypergraphs with at most $n$ elements which carry a weak linear order. 

\section{Polynomial-Time Algorithms From Symmetries}\label{sect:zhuk}

In this section, we fix $\ell\geq 3$ and a finitely bounded class $\mathcal K$ of $\ell$-hypergraphs
such that $\vec{\mathcal K}$ is an amalgamation class.
We write $(\sH,<)$ for the Fra\"iss\'e limit of $\vec{\mathcal K}$, $I_n$ for the set of injective $n$-tuples of elements from $H$ for any $n\geq 1$, $I$ for $I_\ell$, and $b_{\sH}$
for an integer witnessing that $\mathcal K$ is finitely bounded.
We also fix a first-order reduct $\sA$ of $\sH$. 
We say that $\sA$ \emph{admits an injective linear symmetry} if it has a ternary injective polymorphism $m$ which is canonical with respect to $(\sH,<)$, and which has the property that for any $\tuple a,\tuple b\in I$, the orbits under $\Aut(\sH)$ of $m(\tuple a,\tuple a,\tuple b),m(\tuple a,\tuple b,\tuple a),m(\tuple b,\tuple a,\tuple a)$ and $\tuple b$ agree. Note that in this case, $m$ induces an operation on the set $\en$ of orbits of injective $\ell$-tuples under $\Aut(\sH)$. 
We say that $m$ \emph{acts as a minority operation on $\en$} since the second condition on $m$ can be equivalently written as $m(X,X,Y)=m(X,Y,X)=m(Y,X,X)=Y$ for all $X,Y\in\en$.

We prove the following.

\begin{restatable}{theorem}{mainalgo}\label{thm:main-algo}
	Let $\ell\geq 3$, let $\mathcal K$ be a finitely bounded class of $\ell$-hypergraphs such that $\vec{\mathcal K}$ is an amalgamation class.
    Let $\sA$ be a first-order reduct of the Fra\"iss\'e limit $\sH$ of $\mathcal K$.
	Suppose that $I\trianglelefteq_{\sA} H^\ell$,
	and that $\sA$ admits an injective linear symmetry or is such that $\CSP_{\injinstances}(\sA)$ has bounded width.
	Then $\CSP(\sA)$ is solvable in polynomial time.
\end{restatable}

\subsection{\texorpdfstring{$\sA$}{[math]} admits an injective linear symmetry}\label{subsect:algorithm}

Let $\sA$ be a first-order reduct of $\sH$ admitting an injective linear symmetry. 
Set $p_1(x,y):=m(x,y,y)$. It follows that $p_1$ is canonical with respect to $(\sH,<)$ and that it acts as the first projection on $\en$, i.e., it satisfies for any $\tuple a,\tuple b\in I$ that the orbit of $p_1(\tuple a,\tuple b)$ under $\Aut(\sH)$ is equal to the orbit of $\tuple a$.
Moreover, by composing $p_1$ with a suitable endomorphism of $\sH$, we can assume that $p_1(y,x)$ acts lexicographically on the order, i.e., $p_1(x,y)<p_1(x',y')$ if $y<y'$ or $y=y'$ and $x<x'$ (for more details, see \Cref{sect:binarypol}).

In the remainder of this section, we present an algorithm solving
$\Csp(\sA)$ in polynomial time, 
given that $\sA$ has among its polymorphisms operations $p_1$ and $m$ with the properties derived above.
Before giving the technical details, we give here an overview of the methods we employ.
Let $\instance$ be an instance of $\Csp(\sA)$.
Our algorithm transforms $\instance$ into an equi-satisfiable instance $\instance'$ that is sufficiently minimal, such that the solution set of a certain relaxation  of $\instance'$ is subdirect on all projections to an $\ell$-tuple $\tuple v$ of pairwise distinct variables (i.e., for every tuple $\tuple a$ in this projection, this relaxation of $\instance'$ has a solution where the variables from $\tuple v$ are assigned values from $\tuple a$), and that additionally satisfies a condition which we call \emph{inj-irreducibility}, inspired by Zhuk's notion of irreducibility~\cite{Zhuk:2017,Zhuk:2020}. 
We then prove that any 
non-trivial instance satisfying those properties has an injective solution.
This step is to be compared with the case of \emph{absorbing reductions} in Zhuk's algorithm, and in particular with Theorem 5.5 in~\cite{Zhuk:2020}, in which it is proved that any sufficiently minimal and irreducible instance that has a solution also has a solution where an arbitrary variable is constrained to belong to an absorbing subuniverse. Since in our setting $I$ is an absorbing subuniverse of $H^\ell$ in $\sA$, this fully establishes a parallel between the present work and~\cite{Zhuk:2020}. The algorithm that we  introduce works with infinite sets which are however always unions of orbits of $\ell$-tuples under $\Aut(\sH)$. 
$\Aut(\sH)$ is oligomorphic, i.e., it has only finitely many orbits in its action on $H^k$ for every $k\geq 1$; in particular, there are only finitely many orbits of $\ell$-tuples under $\Aut(\sH)$, whence we can represent every union of such orbits by listing all orbits included in this union.

\subsubsection{Finitisation of instances}
Let $\sA$ be a first-order reduct of $\sH$. 
Let $\instance=(\V,\constraints)$ be an instance of $\CSP(\sA)$. In this section, we always assume that the variable set $\V$ is equipped with an arbitrary linear order; this assumption is however inessential and only  used to formulate the statements and proofs in a more concise way.
We denote by $\injtuples$ the set of injective increasing $\ell$-tuples of variables from $\V$.
Given any instance $\instance$ of $\CSP(\sA)$, consider the following CSP instance $\fininstance$ over the set $\mathcal O$ of orbits of $\ell$-tuples under $\Aut(\sH)$, called the \emph{finitisation of $\instance$}:
\begin{itemize}
    \item The variable set of $\fininstance$ is the set $\injtuples$.
    \item For every constraint $C\subseteq A^U$ in $\instance$, $\fininstance$ contains the constraint $C'$ containing the maps $g\colon \injtuples[U]\to\mathcal O$ such that there exists $f\in C$ satisfying $f(\tuple v) \in g(\tuple v)$ for every $\tuple v\in\injtuples[U]$.
\end{itemize}

This instance corresponds to the instance $\instance_{\Aut(\sH),\ell}$ from~\cite[Definition 3.1]{SymmetriesEnough},
with the difference that there the 
$\ell$-element subsets of $\V$ were used as variables, and the domain consisted of orbits of maps. However, the translation between the two definitions is straightforward. Note that if a mapping $f\colon \V\rightarrow A$ is a solution of $\instance$, then the mapping $h\colon [\V]^\ell\rightarrow\mathcal O$, where $h(\tuple v)$ is the orbit of $f(\tuple v)$ under $\Aut(\sH)$ for every $\tuple v\in[\V]^\ell$ is a solution of $\fininstance$.

Let $\cJ=(S,\constraints)$ be an instance over the set $\mathcal O$ of orbits of $\ell$-tuples under $\Aut(\sH)$, e.g., $\cJ=\fininstance$ for some $\instance$. The~\emph{injectivisation} of $\cJ$, denoted by $\cJ^{(\inj)}$, is the instance obtained by removing from all constraints all maps taking some value outside the two injective orbits $E$ and $N$.

Let $\instance=(\V,\constraints)$ be an instance of $\Csp(\sA)$; the \emph{injective finitisation of $\instance$} is the instance $\injinstance[({\fininstance})]$. 
Let $S\subseteq\injtuples$. The \emph{injective finitisation of $\instance$ on $S$} is the restriction of the injective finitisation of $\instance$ to $S$.
For any constraint $C\in\constraints$, the corresponding constraint in the injective finitisation of $\instance$ is called the injective finitisation of $C$.
Note that if $\sA$ admits an injective linear symmetry, then for any instance $\instance=(\V,\constraints)$ of $\Csp(\sA)$ and for any $S\subseteq\injtuples$, the injective finitisation of $\instance$ on $S$ is solvable in polynomial time. This follows from Lemma 3.4 in~\cite{SymmetriesEnough} and from the dichotomy theorem for finite-domain $\Csp$s~\cite{Zhuk:2017,Zhuk:2020,Bulatov:2017}.

Let $\sA$ be a first-order reduct of $\sH$ preserved by $m$ and by $p_1$. We can assume that $\sA$ has among its relations all unions of orbits of $\ell$-tuples under $\Aut(\sH)$ that are preserved by $p_1$ and by the ternary injection $m$. Otherwise, we expand $\sA$ by these finitely many relations and we prove that the $\Csp$ of this expanded structure is solvable in polynomial time. Note that in particular, every orbit of $\ell$-tuples under $\Aut(\sH)$ is a relation of $\sA$.
Moreover, we suppose that $\sA$ has the property that for every instance $\instance$ of $\Csp(\sA)$, the $(2\ell,\max(3\ell,b_{\sH}))$-minimal instance equivalent to $\instance$ is again an instance of $\Csp(\sA)$.
This can be achieved without loss of generality since it is enough to expand $\sA$ by finitely many pp-definable relations, which are also preserved by $m$ and $p_1$. Note that if $\instance$ is a $(2\ell,\max(3\ell,b_{\sH}))$-minimal instance of $\Csp(\sA)$, then its injective finitisation $\fininstance$ is $(2,3)$-minimal by~\cite[Lemma 3.2]{SymmetriesEnough}; in particular, $\fininstance$ is \emph{cycle consistent}, i.e., it satisfies one of the basic consistency notions used in Zhuk's algorithm~\cite{Zhuk:2017}. Moreover, if $\fininstance$ is $(2,3)$-minimal, then for any solution $h\colon [\V]^\ell\rightarrow\mathcal O$ of $\fininstance$, any mapping $f\colon \V\rightarrow A$ with $f(\tuple v)\in h(\tuple v)$ for every $\tuple v\in[\V]^\ell$ is a solution of $\instance$ by~\cite[Lemma 3.3]{SymmetriesEnough}.

Let $\instance=(\V,\constraints)$ be an instance of $\CSP(\sA)$, and let $C\in \constraints$.
Since $\sA$ is preserved by $m$, there exists a set of linear equations over $\mathbb Z_2$ associated with the injective finitisation of $C$.
By abuse of notation, we write every linear equation as $\sum\limits_{\tuple v\in S}X_{\tuple v}=P$, where $P\in\en$ and $S\subseteq[\V]^\ell$ is a set of injective $\ell$-tuples of variables from the scope of $C$.
In these linear equations, we identify $E$ with $1$ and $N$ with $0$, so that e.g. $E+E=N$ and $N+E=E$.
Using this notation, the canonical behaviour of the function $m$ on $\en$ can be written as $m(X,Y,Z)=X+Y+Z$ which justifies the notion of $\sA$ admitting linear symmetries.

We can assume that no equation $\sum\limits_{\tuple v\in S}X_{\tuple v}=P$ associated with the injective finitisation of any constraint $C\in \constraints$ splits into two equations $\sum\limits_{\tuple v\in S_1}X_{\tuple v}=P_1$ and $\sum\limits_{\tuple v\in S_2}X_{\tuple v}=P_2$, where $S_1\cap S_2=\emptyset$, $S_1\cup S_2=S$, $P_1+P_2=P$, and such that every 
mapping contained in the injective finitisation of $C$ satisfies both these equations. If this is not the case, let us consider only equations satisfying this assumption -- it is clear that this new set of equations is satisfiable if, and only if, the original set of equations is satisfiable. We call every equation satisfying this assumption \emph{unsplittable}.

For an instance $\instance=(\V,\constraints)$ of $\CSP(\sA)$, we define an instance $\instance_{\textrm{eq}}=(\V,\constraints_{\textrm{eq}})$ of the equality-$\CSP$ (i.e., CSP over structures first-order definable over $(H;=)$) over the same base set $H$ corresponding to the closure of the constraints under the full symmetric group on $H$. Formally, for every constraint $C\in\constraints$, the corresponding constraint $C_{\textrm{eq}}\in \constraints_{\textrm{eq}}$ contains all functions $\alpha h$ for all $h\in C$ and $\alpha\in \textrm{Sym}(H)$.
Since $\sA$ is preserved by a binary injection, the constraints of $\instance_{\textrm{eq}}$ are preserved by the same or indeed any binary injection and hence, its $\Csp$ has relational width $(2,3)$ by the classification of equality $\CSP$s \cite{ecsps}.

Let $\instance$ be an $\ell$-minimal instance of $\CSP(\sA)$, let $\tuple v\in\injtuples$, and let $R\subseteq\proj_{\tuple v}(\instance)$ be an $\ell$-ary relation from the signature of $\sA$. 
Let $\instance^{\tuple v\in R}$ be the instance obtained from $\instance$ by replacing every constraint $C$ containing all variables from $\tuple v$ by $\{g\in C \mid g(\tuple v)\in R\}$.

We call an $\ell$-minimal instance of $\CSP(\sA)$ \emph{$\textrm{eq}$-subdirect} if for every $\tuple v\in\injtuples$ and for every non-injective orbit $O\subseteq\proj_{\tuple v}(\instance)$ under $\Aut(\sH)$, the instance $(\instance^{\tuple v\in O})_{\textrm{eq}}$ has a solution.
Note that by $\ell$-minimality and since all constraints of the instance are preserved by a binary injection, the instance $(\instance^{\tuple v\in O})_{\textrm{eq}}$ has a solution for every injective orbit $O\subseteq\proj_{\tuple v}(\instance)$ under $\Aut(\sH)$. Indeed, any injective mapping from $\V$ to $H$ is a solution of $(\instance^{\tuple v\in O})_{\textrm{eq}}$.

\begin{example}[label=ex:eq-sd]
    Let $\sH$ be the universal homogeneous $\ell$-hypergraph. Let $\tuple u=(u_1,\ldots,u_\ell), \tuple v=(v_1,\ldots,v_\ell)$ be disjoint $\ell$-tuples of variables, and let $\V$ be the set of all variables contained in these tuples. We define a CSP instance $\instance=(\V,\constraints)$ over the set $H$ as follows. Let $\tuple u'=(u_2,\ldots,u_\ell,u_1)$. We set $\constraints$ to contain two constraints $C,C'$ such that $C$ contains all mappings $f\colon \V\rightarrow H$ such that $f(\tuple u)$ and $f(\tuple v)$ belong to the same orbit under $\Aut(\sH)$, and $C'$ contains all mappings $f\colon \V\rightarrow H$ such that $f(\tuple u')$ and $f(\tuple v)$ belong to the same orbit. Note that the constraints $C,C'$ are preserved by any function which is canonical with respect to $\sH$. It is easy to see that $\instance$ is non-trivial and $\ell$-minimal, but it is not eq-subdirect.
    Indeed, for any non-injective and non-constant mapping $g$ from the set of variables of $\tuple u$ to $H$, it holds that $g(\tuple u)$ and $g(\tuple u')$ are contained in different orbits under $\Aut(\sH)$.
\end{example}

It is clear that we can obtain an $\textrm{eq}$-subdirect instance out of an $\ell$-minimal instance in polynomial time by the  algorithm in~\Cref{alg:eq-subdirect}. The algorithm successively shrinks for every $\tuple v\in[\V]^\ell$ the projection $\proj_{\tuple v}(\instance)$ to the union of those orbits $O\subseteq\proj_{\tuple v}(\instance)$ under $\Aut(\sH)$ for which the instance $(\instance^{\tuple v\in O})_{\textrm{eq}}$ has a solution; it stops when no more orbits can be removed from $\proj_{\tuple v}(\instance)$ for any $\tuple v\in[\V]^\ell$.

\begin{figure}
\begin{algorithmic}[1]
\State \textbf{INPUT:} $\ell$-minimal instance $\instance=(\V,\constraints)$ of $\CSP(\sA)$;
\State \textbf{OUTPUT:} $\textrm{eq}$-subdirect instance $\instance'$;
\Repeat
\State changed:=false;
\For {$\tuple v \in \injtuples$}
    \If{not changed}
        \State{$P:=\proj_{\tuple v}(\instance)\cap I$};
        \For {$O\subseteq\proj_{\tuple v}(\instance)$ non-injective orbit}
            \If{$(\instance^{\tuple v\in O})_{\textrm{eq}}$ has a solution}
                \State $P:=P\cup O$;
            \EndIf
        \EndFor
        \For{$C\in\constraints$ containing all variables of $\tuple v$ in its scope}
            \State replace $C$ by $\{g\in C \mid g(\tuple v)\in P\}$;
        \EndFor
        \If{$\instance$ is not $\ell$-minimal}
            \State make $\instance$ $\ell$-minimal;
            \State changed:=true;
    \EndIf
    \EndIf
\EndFor
\Until{not changed}
\\ \Return $\instance$
\end{algorithmic}
\caption{Procedure \textsc{Eq-Subdirect}}
\label{alg:eq-subdirect}
\end{figure}

\begin{restatable}{lemma}{eqsubdirect}
    The instance $\instance'$ outputted by the algorithm in~\Cref{alg:eq-subdirect} is $\ell$-minimal, eq-subdirect, and it is an instance of $\CSP(\sA)$. Moreover, it has the same solution set as $\instance$.
\end{restatable}

\begin{proof}
It is easy to see that $\instance'$ is $\ell$-minimal since in every run of the repeat... until not changed-loop (lines 3-22), we check $\ell$-minimality of the produced instance. We also do not remove any solutions of $\instance'$ since we remove only orbits $O\subseteq\proj_{\tuple v}(\instance)$ such that $(\instance^{\tuple v\in O})_{\textrm{eq}}$ does not have a solution, which in particular implies that $\instance$ does not have a solution $f\colon \V\rightarrow H$ with $f(\tuple v)\in O$.

To prove that $\instance'$ is an instance of $\CSP(\sA)$, it is sufficient to show that for every $\tuple v \in \injtuples$ the last $P$ on line 10 is preserved by the binary injection $p_1$ as well as by the ternary injection $m$. To this end, let $\tuple v\in\injtuples$ be injective, let $P$ be as in the algorithm, let $O_1,O_2,O_3$ be orbits of $\ell$-tuples from $P$, and let $g_i, i\in\{1,2,3\}$ be solutions to $(\instance^{\tuple v\in O_i})_{\textrm{eq}}$, $i\in\{1,2,3\}$.

Let $C\in\constraints$ be arbitrary, and let $U$ be its scope. By the definition of $\instance_{\textrm{eq}}$, there exist $h_1,h_2,h_3\in C$ and $\alpha_1,\alpha_2,\alpha_3\in \textrm{Sym}(H)$ such that $g_i|_U=\alpha_i h_i$ for every $i\in\{1,2,3\}$. Since every binary injection is canonical in its action on $\{=,\neq\}$, it follows that $p_1(g_1,g_2)|_U=p_1(\alpha_1 h_1,\alpha_2 h_2)=\beta_1 p_1(h_1,h_2)\in C$ and $m(g_1,g_2,g_3)|_U=m(\alpha_1 h_1,\alpha_2 h_2,\alpha_3 h_3)=\beta_2 m(h_1,h_2,h_3)\in C$ for some $\beta_1,\beta_2\in \textrm{Sym}(H)$. Hence, $p_1(g_1,g_2)|_U,m(g_1,g_2,g_3)|_U\in C$ and since $C$ was chosen arbitrarily, both $p_1(g_1,g_2)$ and $m(g_1,g_2,g_3)$ are solutions to $(\instance^{\tuple v\in O_i})_{\textrm{eq}}$.

Let $\tuple v\in \injtuples$. If $C$ contains all variables from $\tuple v$ in its scope, then either $p_1(g_1,g_2)(\tuple v)$ is injective and $p_1(g_1,g_2)(\tuple v)\in P$, or $p_1(g_1,g_2)(\tuple v)$ and $p_1(h_1,h_2)(\tuple v)$ are non-injective and contained in the same orbit under $\Aut(\sH)$ and it follows that $p_1(h_1,h_2)(\tuple v)\in P$. Similar argument holds for $m(h_1,h_2,h_3)$ as well. Therefore, $P$ is preserved by $p_1$ and $m$ as desired.
\end{proof}

Note that for any $1\leq m\leq n$ and any instance $\instance$ of $\Csp(\sA)$, we can compute an instance that is both eq-subdirect and $(m,n)$-minimal and that has the same solution set as $\instance$ in polynomial time.
Indeed, it is enough to repeat the above-mentioned algorithm and the $(m,n)$-minimality algorithm until no orbits under $\Aut(\sH)$ are removed from any constraint.

\begin{example}[continues=ex:eq-sd]
    If we run the algorithm in~\Cref{alg:eq-subdirect} on the instance from \Cref{ex:eq-sd}, we remove from $\proj_{\tuple u}(\instance)$ and $\proj_{\tuple v}(\instance)$ all orbits under $\Aut(\sH)$ containing a non-constant, non-injective tuple.
\end{example}

\subsubsection{Inj-irreducibility}

For any $\tuple a \in H^{\ell}$, we write $O(\tuple a)$ for the orbit of $\tuple a$ under $\Aut(\sH)$ and $O_<(\tuple a)$ for the orbit of $\tuple a$ under $\Aut(\sH,<)$. Recall that being canonical with respect to $(\sH,<)$, the function $p_1$ acts naturally on orbits under $\Aut(\sH,<)$; we can therefore abuse the notation and write $p_1(O,P)$ for orbits $O,P$ under $\Aut(\sH,<)$. We will say that a non-injective orbit $O$ of $\ell$-tuples under $\Aut(\sH)$ is:

\begin{itemize}
    \item \emph{deterministic} if for every $\tuple a\in O$, there exists $\alpha\in\Aut(\sH)$ such that $p_1(O_<(\alpha(\tuple a)),O_<(\tuple e))=p_1(O_<(\alpha(\tuple a)),O_<(\tuple n))$, where $\tuple e,\tuple n$ are arbitrary strictly increasing $\ell$-tuples of elements of $\sH$ with $\tuple e\in E$, $\tuple n\in N$,
        
    \item \emph{non-deterministic} otherwise.
\end{itemize}

For a tuple $\tuple a$ contained in a deterministic orbit $O$, we call any $\alpha\in\Aut(\sH)$ with the property that $p_1(O_<(\alpha(\tuple a)),E)=p_1(O_<(\alpha(\tuple a)),N)$ for the strictly increasing ordering in the second coordinate \emph{deterministic for $\tuple a$}. Note that for any $\beta\in\Aut(\sH,<)$, $\beta\alpha\in\Aut(\sH)$ is deterministic for $\tuple a$ as well since $p_1$ is canonical with respect to $(\sH,<)$.

Let $\cJ=(\V,\constraints)$ be a CSP instance over a set $B$.
A sequence $v_1, C_1, v_2,\ldots, C_k, v_{k+1}$, where $k\geq 1$, $v_i\in \V$ for every $i\in[k+1]$, $C_i\in\constraints$ for every $i\in[k]$, and $v_i,v_{i+1}$ are contained in the scope of $C_i$ for every $i\in[k]$, is called a~\emph{path} in $\cJ$.
We say that two elements $a,b\in B$ are \emph{connected} by a path $v_1, C_1, v_2,\ldots, C_k, v_{k+1}$ if there exists a tuple $(c_1,\ldots,c_{k+1})\in B^{k+1}$ such that $c_1=a, c_{k+1}=b$, and such that $(c_i,c_{i+1})\in\proj_{(v_i,v_{i+1})}(C_i)$ for every $i\in[k]$. Suppose that $\cJ$ is $1$-minimal. 
The \emph{linkedness congruence} on $\proj_v(\cJ)$ is the equivalence relation $\lambda$ on $\proj_v(\cJ)$ defined by $(a,b)\in\lambda$ if there exists a path $v_1, C_1, v_2,\ldots, C_k, v_{k+1}$ from $a$ to $b$ in $\cJ$ such that $v_1=v_{k+1}=v$. 
Note that for a finite relational structure $\sB$, for a $(2,3)$-minimal instance $\cJ=(\V,\constraints)$ of $\Csp(\sB)$, and for any $v\in\V$, the linkedness congruence $\lambda$ on $\proj_v(\cJ)$ is a relation pp-definable in $\sB$. Indeed, it is easy to see that the binary relation containing precisely the pairs $(a,b)\in B^2$ that are connected by a particular path in $\instance$ is pp-definable in $\sB$. 
If we concatenate all paths that connect two elements $(a,b)\in\lambda$, the resulting path connects every pair $(a,b)\in\lambda$ since by the $(2,3)$-minimality of $\cJ$, every path from $v$ to $v$ connects $c$ to $c$ for every $c\in\proj_v(\cJ)$. It follows that $\lambda$ is pp-definable.

\begin{definition}\label{def:inj-irreducible}
Let $\sA$ be a first-order reduct of $\sH$, and let $\instance=(\V,\constraints)$ be a non-trivial  $\ell$-minimal instance of $\Csp(\sA)$.
We call $\instance$ \emph{inj-irreducible} 
if for every set $S\subseteq\injtuples$, one of the following holds for the instance $\mathcal J=\restrinstance[\fininstance]{S}$:
\begin{itemize}
\item $\injinstance[\mathcal J]$ has a solution,
\item for some $\tuple v\in S$, $\proj_{\tuple v}(\mathcal J)$ contains the two injective orbits and the linkedness congruence on $\proj_{\tuple v}(\mathcal J)$ does not connect them,
\item for some $\tuple v\in S$, the linkedness congruence on $\proj_{\tuple v}(\mathcal J)$ links an injective orbit to a non-deterministic orbit.
\end{itemize}
\end{definition}

\begin{example}[label=ex:inj-irred]
We illustrate the concept of inj-irreducibility on the following instance.
    Let $\ell=3$, and let $\sH$ be the universal homogeneous $\ell$-hypergraph. Let $a\neq b\in H$ be arbitrary; we call the orbits $O(a,a,b), O(a,b,a), O(b,a,a)$ under $\Aut(\sH)$ \emph{half-injective}. 
    Let us define an instance $\instance=(\V,\constraints)$ over the set $H$ as follows. Above, we identified $E$ with $1$ and $N$ with $0$ in the linear equations associated to injective finitisations of constraints. In this example, we identify also all half-injective orbits under $\Aut(\sH)$ with $1$. Hence, we can write, e.g., $E+O(a,a,b)=0$.
    
    Let $\tuple v_1,\tuple v_2,\tuple v_3$ be increasing triples of pairwise disjoint variables, and set $\V$ to be the union of all variables of these tuples. We define $\constraints$ to be a set consisting of two constraints, $C_0$ and $C_1$, defined as follows. For $i\in\{0,1\}$, we set $C_i$ to contain all mappings $f\in H^{\V}$ such that both of the following hold:
    \begin{itemize}
        \item either $O(f(\tuple v_1))+O(f(\tuple v_2))+O(f(\tuple v_3))=i$, or $O(f(\tuple v_1))=O(f(\tuple v_2))=O(f(\tuple v_3))=\{(a,a,a)\mid a\in H\}$,
        \item $f(x)\neq f(y)$ for all $x,y\in \V$ belonging to different triples from $\{\tuple v_1,\tuple v_2,\tuple v_3\}$.
    \end{itemize}

    We show that the constraints $C_0,C_1$ are preserved by a binary injection $p_1$ and a ternary injection $m$ that are both canonical with respect to $\sH$.
    To this end, we define the canonical behaviours of $p_1$ and $m$ on the orbits of triples under $\Aut(\sH)$ as follows. We set $p_1$ to act as the first projection on the numbers associated to the respective orbits, and to satisfy $p_1(O(a,a,a),P)=p_1(P,O(a,a,a))=P$  for an arbitrary $a\in H$ and for an arbitrary orbit $P$. Note that these assumptions together with the requirement that $p_1$ is an injection uniquely determine the behaviour of $p_1$ -- e.g., $p_1(P,N)=E$ for an arbitrary half-injective orbit $P$.
    We set $m$ to act as an idempotent minority on the numbers associated to the orbits, and to act as $p_1(x,p_1(y,z))$ on the orbits where the action is not determined by the previous condition. It is easy to verify that the constraints $C_0$ and $C_1$ are preserved by both $p_1$ and $m$, and hence $\instance$ is an instance of $\Csp(\sA)$ for some first-order reduct $\sA$ of $\sH$ which falls into the scope of this section.

    It immediately follows that all the half-injective orbits under $\Aut(\sH)$ are deterministic since $p_1(O_<(\tuple c),E)=p_1(O_<(\tuple c),N)=E$ for any $\tuple c\in H^3$ contained in a half-injective orbit; the orbit $P$ of the constant tuples is non-deterministic since $p_1(P,E)=E$, and $p_1(P,N)=N$.
    It is also easy to see that $\instance$ is non-trivial and $(6,9)$-minimal.
    Moreover, $\instance$ is not inj-irreducible. Indeed, setting $S:=\{\tuple v_1,\tuple v_2,\tuple v_3\}$, the linkedness congruence on $\proj_{\tuple v_1}(\fininstance|_S)$ connects precisely all injective and half-injective orbits, and the injective finitisation of $\instance$ on $S$ does not have a solution.
\end{example}

Note that if $\sA$ is the structure from \Cref{ex:inj-irred}, its CSP can be solved by the reduction to the finite from~\cite{ReductionFinite}.
For simplicity, we choose to illustrate the concepts that we have just introduced on this example rather than on an example where the canonical behaviour of the functions $p_1$ and $m$ depends on the linear order.
However, \Cref{ex:order} provides us with a structure admitting linear symmetries where the canonical behaviour of any polymorphism satisfying the assumptions on $m$ or on $p_1$ depends on the additional linear order (see \Cref{subsect:inj-irr} for the proof that \Cref{ex:order} satisfies the assumptions of this section).

\begin{lemma}\label{lemma:split}
Let $C$ be a constraint of an instance of $\Csp(\sA)$ which contains an injective mapping, and let $S$ 
be a set of variables appearing together in an unsplittable linear equation associated with the injective finitisation of $C$.
Then for every $g\in C$, either $g(\tuple{v})$ is in an injective or deterministic orbit for all $\tuple v\in S$, or $g(\tuple{v})$ is in a non-deterministic orbit for all $\tuple v\in S$.
\end{lemma}

\begin{proof}
Suppose not, and let $g\in C$ be a counterexample. Let $\emptyset\neq S'\subsetneq S$ be the set of all $\tuple v\in S$ such that $g(\tuple v)$ is in an injective or deterministic orbit.
Let $\sum\limits_{\tuple v\in S}X_{\tuple v}=P$ be an unsplittable linear equation associated with the injective finitisation of $C$ and containing all the variables from $S$.

Let $\{\tuple{v}_1,\ldots,\tuple{v}_n\}$ be the set of all $\tuple v_i\in S'$ such that $g(\tuple v_i)$ is in a deterministic orbit. For every $i\in[n]$, let $\alpha_i\in\Aut(\sH)$ be deterministic for $g(\tuple{v}_i)$. Let $p_1'(x,y):=p_1(\alpha_1 x, p_1(\alpha_2 x, p_1(\ldots,p_1(\alpha_n x,y)\dots)))$.
{By the definition of $S$, it follows that}
for all $\tuple v\in {S'}$, we have that $p_1'(O_<(g(\tuple v)),E)$ and $p_1'(O_<(g(\tuple{v})),N)$ are subsets of the same injective orbit $O_{\tuple v}$ under $\Aut(\sH)$ for the increasing order in the second coordinate.
For all $\tuple v\in S\backslash S'$, let $P_{\tuple v}$ be $N$ if $p_1'(O_<(g(\tuple v)),E)\subseteq E$ for the increasing order in the second coordinate, and let $P_{\tuple v}$ be $E$ otherwise. {It follows that}
{\begin{equation}
    \sum\limits_{\tuple{v}\in S\backslash S'}O(h(\tuple{v}) = \sum\limits_{\tuple{v}\in S\backslash S'}O(g'(\tuple{v}))
    +\sum\limits_{\tuple{v}\in S\backslash S'}P_{\tuple{v}}.\label{eq:Pv}
\end{equation}}

Since $C$ contains an injective mapping, 
there exists a monotone injective $g'\in C$. 
Moreover, since $S'\subsetneq S$ and the equation under consideration is unsplittable, {we can choose $g'$ so that $\sum\limits_{\tuple v\in S'}O(g'(\tuple v))$ is equal to a fixed injective orbit. We pick $g'$ so that} %
\begin{equation}
    \sum\limits_{\tuple v\in S'}O(g'(\tuple v))=\sum\limits_{\tuple v\in S'}O_{\tuple v}+\sum\limits_{\tuple v\in S\backslash S'}P_{\tuple v}+E.\label{eq:Sprime}
\end{equation}
Let us consider $h=p_1'(g,g')\in C$. $h$ is clearly injective and we obtain:
\begin{align*}
    \sum\limits_{\tuple{v}\in S}O(h(\tuple{v})) 
    &=\sum\limits_{\tuple{v}\in S'}O(h(\tuple{v}))+\sum\limits_{\tuple{v}\in S\backslash S'}O(h(\tuple{v}))
    {\stackrel{(\ref{eq:Pv})}{=}}\sum\limits_{\tuple{v}\in S'}O_{\tuple{v}}+\sum\limits_{\tuple{v}\in S\backslash S'}O(g'(\tuple{v}))
    +\sum\limits_{\tuple{v}\in S\backslash S'}P_{\tuple{v}}\\
    &{\stackrel{(\ref{eq:Sprime})}{=}}\sum\limits_{\tuple{v}\in S'}O(g'(\tuple{v}))+E
    +\sum\limits_{\tuple{v}\in S\backslash S'}O(g'(\tuple{v}))
    =\sum\limits_{\tuple{v}\in S}O(g'(\tuple{v}))+E\\
    &=P+E
\end{align*}
Hence, the mapping $h'\colon S\to \en$ defined by $h'(\tuple v):=O(h(\tuple v))$ is a solution of the injective finitisation of $C$ on $S$ but it does not satisfy the unsplittable linear equation, a contradiction.
\end{proof}

\begin{theorem}\label{thm:inj-irreducibility}
    Let $\sA$ be a first-order reduct of $\sH$ that admits {an injective linear symmetry.}
    Let $\instance$ be a $(2\ell,\max(3\ell,b_{\sH}))$-minimal,
    inj-irreducible instance of $\CSP(\sA)$ with variables $\V$  such that for every distinct $u,v\in \V$, $\proj_{(u,v)}(\instance)\cap I_2\neq \emptyset$. Then $\instance$ has an injective solution.
\end{theorem}

\begin{proof}
Note that if $\instance$ has fewer than $\ell$ variables, it has an injective solution by the assumption on binary projections of $\instance$ and since all constraints of $\instance$ are preserved by the binary injection $p_1$. 
Let us therefore suppose that $\instance$ has at least $\ell$ variables. Let us assume for the sake of contradiction that $\instance$ does not have an injective solution.
Let $\mathcal J$ be $\fininstance$, and let $\mathcal C$ be the set of its constraints.
By assumption, $\injinstance[\mathcal J]$ does not have a solution.
Note that $\injinstance[\mathcal J]$ corresponds to a system of linear equations over $\mathbb Z_2$, which is therefore unsatisfiable. In case this system can be written as a diagonal block matrix, there exists a set $S\subseteq\injtuples$ of variables such that the system of equations associated with the injectivisation of $\cL:=\restrinstance[\mathcal J]{S}=(S,\mathcal C')$ corresponds to a minimal unsatisfiable block. By definition, this means that $\injinstance[\cL]$ is unsatisfiable. 
The instance $\cL$ has the property that for every non-trivial partition of $S$ into parts $S_1,S_2$, there exists an unsplittable equation associated with the injectivisation of a constraint $C\in\mathcal C'$ which contains variables from both $S_1$ and $S_2$.

Since $\instance$ is inj-irreducible, there exists $\tuple v\in S$ such that the two injective orbits are elements of $\proj_{\tuple v}(\cL)$ and are not linked, or some injective orbit in $\proj_{\tuple v}(\cL)$ is linked to a non-deterministic orbit in $\proj_{\tuple v}(\cL)$.

In the first case, we note that for \emph{all} $\tuple w\in S$ such that $\proj_{\tuple w}(\cL)$ contains the two injective orbits, the two injective orbits are not linked. 
Indeed, suppose that there exists $\tuple w\in S$ such that $E,N\in\proj_{\tuple w}(\cL)$ are linked, i.e., there exists a path $\tuple v_1=\tuple w,C_1,\ldots, C_k,\tuple v_{k+1}=\tuple w$ in $\cL$ connecting $E$ and $N$. Since $\instance$ is $(2\ell, \max(3\ell,b_{\sH}))$-minimal, Lemma 3.2 in~\cite{SymmetriesEnough} yields that $\cJ$ and hence also $\cL$ is $(2,3)$-minimal. In particular, there exists $C\in\constraints'$ containing in its scope both $\tuple v$ and $\tuple w$. Let $O_1,O_2\in \en$ be disjoint such that there exist $g_1,g_2\in C$ with $g_1(\tuple v)= O_1, g_1(\tuple w)= E, g_2(\tuple v)= O_2, g_2(\tuple w)= N$. It follows that the path $\tuple v, C,\tuple w=\tuple v_1,C_1,\ldots,C_k,\tuple v_{k+1}=\tuple w,C,\tuple v$ connects $O_1$ with $O_2$ in $\proj_{\tuple v}(\cJ)$, a contradiction.
Let $g\colon S\to\en$ be defined as follows.
For a fixed $\tuple v\in S$, let $g(\tuple v)$ be an arbitrary element of $\proj_{\tuple v}(\injinstance[\cL])$.
Next, for $\tuple w\in S$, define $g(\tuple w)$ to be the unique injective orbit $O$ such that there exists a constraint $C\in \constraints'$ containing both $\tuple v$ and $\tuple w$ in its scope and such that there exists $g'\in C$ with $g'(\tuple v)=g(\tuple v)$ and $g'(\tuple w)=O$. 
This $g$ is a solution of $\injinstance[\cL]$, a contradiction.

Thus, it must be that a non-deterministic orbit in $\proj_{\tuple v}(\cL)$ is linked to an injective orbit in $\proj_{\tuple v}(\cL)$. 
Hence, there exists a path in $\cL$ from $\tuple v$ to $\tuple v$ and connecting an injective orbit to a non-deterministic one.
Moreover, up to composing this path with additional constraints, one can assume that this path goes through all the variables in $S$. This follows by the $(2,3)$-minimality of $\cJ$.
Define a partition of $S$ where $\tuple w\in S_1$ if the first time that $\tuple w$ appears in the path, the element associated with $\tuple w$ is in an injective orbit, and $\tuple w\in S_2$ otherwise. Since the system of unsplittable equations associated with $\injinstance[\cL]$ cannot be decomposed as a diagonal block matrix, some constraint $C\in\mathcal C'$ gives an equation in that system containing $\tuple u_1\in S_1$ and $\tuple u_2\in S_2$.
Thus, there exists $g\in C$ with $g(\tuple u_1)$ injective, and $g(\tuple u_2)$ non-deterministic.
This contradicts~\Cref{lemma:split}.
\end{proof}

\subsubsection{Establishing inj-irreducibility}\label{subsect:inj-irr}

We introduce a polynomial-time algorithm which
produces, given an instance $\instance$ of $\CSP(\rel A)$, an instance $\instance'$ of $\CSP(\rel A)$ that is either inj-irreducible or trivial and that has a solution if, and only if, $\instance$ has a solution.
It uses the fact that the injective finitisation of an instance $\instance$ of $\Csp(\sA)$ on $S$ is solvable in polynomial time for any set $S\subseteq \injtuples$. This follows from the fact that the constraints of the injective finitisation of $\instance$ are preserved by a ternary minority by Lemma 3.4 from~\cite{SymmetriesEnough}.

Let us first give a brief description of the algorithm. It gradually ensures that the instance is $(2\ell,\max(3\ell,b_{\sH}))$-minimal, eq-subdirect, and so that for no distinct variables $u,v\in\V$, it holds that $\proj_{(u,v)}(\instance)=\{(a,a)\mid a\in H\}$. If the instance satisfies these assumptions, we consider for every $\tuple u\in[\V]^{\ell}$ every partition $\{E^1_{\tuple u},\ldots,E^s_{\tuple u}\}$ on $\proj_{\tuple u}(\instance)$ with pp-definable classes satisfying that $E^1_{\tuple u}$ contains no non-deterministic orbit.
We find the maximal subset $S\subseteq[\V]^{\ell}$, such that for every $\tuple w\in S$, the set $\{E^1_{\tuple w},\ldots,E^s_{\tuple w}\}$ defined by
$E^i_{\tuple w}:=\{\tuple a\in H^{\ell}\mid \exists \tuple b\in H^{\ell}\colon (\tuple b,\tuple a)\in \proj_{(\tuple u,\tuple w)}(\instance)\}$ for every $i\in[s]$ forms a partition on $\proj_{\tuple w}(\instance)$ with the property that $E^1_{\tuple w}$ contains no non-deterministic orbit. For every such partition, the algorithm checks if the injective finitisation of $\instance$ on $S$ has a solution; if not, we constrain every $\tuple w\in S$ to not take any value from $E^1_{\tuple w}$. It is \emph{a priori} unclear that adding these constraints on the one hand yields an instance of $\csp(\rel A)$, and on the other hand that we do not transform a satisfiable instance into an unsatisfiable one in this way. This is the heart of our construction and is proved in~\Cref{thm:inj-irreducibility} after a series of lemmas.

\begin{figure}
\begin{algorithmic}[1]
\State \textbf{INPUT:} instance $\instance=(\V,\constraints)$ of $\CSP(\sA)$;
\State \textbf{OUTPUT:} instance $\instance'$ that is either inj-irreducible or trivial and that has a solution if, and only if, $\instance$ has a solution;
\Repeat
\State changed := false;
\State $\mathcal J$ := $(2\ell,\max(3\ell,b_{\sH}))$-minimal $\textrm{eq}$-subdirect instance with the same solution set as $\instance$;
\State $\instance:=\mathcal J$;
\State $\instance:=$ \textsc{IdentifyAllEqual}($\instance$);
\For{$\tuple u\in \injtuples$}
\For{$\{E^1_{\tuple u},\ldots,E^s_{\tuple u}\}$ partition on $\proj_{\tuple u}(\instance)$ with pp-definable classes such that $\proj_{\tuple u}(\instance)\cap I\subseteq E^1_{\tuple u}$ and $E^1_{\tuple u}$ contains no non-deterministic orbit}
    \State $S,\{E_{\tuple w}^i\mid i\in [s], \tuple w\in S\}:=$ \textsc{ExtendPartition}($\instance,\{\tuple u\},\{E_{\tuple u}^i\mid i\in [s]\}$);
    \State solve the injective finitisation of $\instance$ on $S$;
    \If{it does not have a solution and not changed}
        \State changed := true;
        \For{$\tuple v\in S$}
        \For{$C\in\constraints$ containing all variables of $\tuple v$ in its scope}
            \State           replace $C$ by $\{f\in C\mid f(\tuple v)\notin E^1_{\tuple v}\}$;
        \EndFor
        \EndFor
    \EndIf
\EndFor
\EndFor
\Until{not changed};
\\ \Return $\instance$
\end{algorithmic}
\caption{Procedure \textsc{InjIrreducibility}}
\label{alg:inj-irreducibility}
\end{figure}

The algorithm uses the subroutines \textsc{IdentifyAllEqual} and \textsc{ExtendPartition}. The subroutine \textsc{IdentifyAllEqual} takes as an input a $2$-minimal instance $\instance=(\V,\constraints)$, and returns an instance where all variables $u,v\in\V$ with $\proj_{(u,v)}(\instance)=\{(a,a)\mid a\in H\}$ are identified. It is clear that this can be implemented in polynomial time. Note moreover that for a $(2\ell,\max(3\ell, b_{\sH}))$-minimal instance $\instance$, the resulting instance $\textsc{IdentifyAllEqual}(\instance)$ is again $(2\ell,\max(3\ell, b_{\sH}))$-minimal.
The subroutine \textsc{ExtendPartition} is given in~\Cref{alg:extend-partition}.
We note that if $(S,\{E^i_{\tuple w}\})$ is a set of partitions returned by \textsc{ExtendPartition}, then for every $\tuple w\in S$, the linkedness congruence on $\proj_{\tuple w}(\fininstance)$ defined by the instance $\restrinstance[\fininstance]{S}$ is a refinement of the partition $\{E^i_{\tuple w}\}$.

\begin{figure}
\begin{algorithmic}[1]
\State \textbf{INPUT:} a tuple $(\instance,S,\{E_{\tuple w}^i\mid i\in [s], \tuple w\in S\})$ where
\begin{itemize}
    \item $\instance=(\V,\constraints)$ is a $((2\ell,\max(3\ell,b_{\sH})))$-minimal instance of $\Csp(\sA)$,
    \item $S\subseteq\injtuples$,
    \item $\{E^1_{\tuple w},\ldots,E^s_{\tuple w}\}$ is a partition on $\proj_{\tuple w}(\instance)$ with pp-definable classes for every $\tuple w\in S$ such that $E^1_{\tuple w}$ contains no non-deterministic orbit and such that $E^i_{\tuple w}:=\{\tuple a\in H^{\ell}\mid \exists \tuple b\in H^{\ell}\colon (\tuple b,\tuple a)\in \proj_{(\tuple v,\tuple w)}(\instance)\}$ for every $\tuple v,\tuple w\in S$ and every $i\in[s]$;
\end{itemize}  
\State \textbf{OUTPUT:} $S, \{E^i_{\tuple w} : i\in[s], \tuple w\in S\}$ as above such that no tuple $\tuple w$ can be added to $S$ where a partition on $\proj_{\tuple w}(\instance)$ as above exists;
\Repeat
    \State added := false;
    \For{$\tuple v\in S, \tuple w\in \injtuples$ with $\tuple w\notin S$}
    \State{$D:=\proj_{(\tuple v,\tuple w)}(\instance)$};
    \For{$t=1,\ldots,s$}
        \State $E^t_{\tuple w}:=\{\tuple a\in H^{\ell}\mid \exists \tuple b\in E^t_{\tuple v}\colon (\tuple b,\tuple a)\in D\}$;
    \EndFor
    \If{$E^1_{\tuple w},\ldots,E^s_{\tuple w}$ are disjoint and $E^1_{\tuple w}$ contains no non-deterministic orbit}
            \State $S:=S\cup \{\tuple w\}$;
            \State {added := true};
    \EndIf
    \EndFor
\Until{not added}\\
\Return{$S, \{E^i_{\tuple w} : i\in[s], \tuple w\in S\}$}
\end{algorithmic}
\caption{Procedure \textsc{ExtendPartition}}
\label{alg:extend-partition}
\end{figure}

First of all, note that the instance $\instance'$ outputted by the algorithm in~\Cref{alg:inj-irreducibility} is $(2\ell,\max(3\ell,b_{\sH}))$-minimal, and for every $u\neq v\in\V$, either $\instance'_{(u,v)}$ is empty or $\instance'_{(u,v)}\cap I_2\neq\emptyset$. We show in~\Cref{lemma:ppdef} that $\instance'$ is an instance of $\CSP(\sA)$. Then it follows that if $\instance'$ is non-trivial, then every constraint $C'$ of $\instance'$ contains an injective tuple since $C'$ is preserved by the binary injection $p_1$.

Let $S\subseteq \injtuples$ be the set outputted by \textsc{ExtendPartition}($\instance,\{\tuple u\},\{E_{\tuple u}^i\mid i\in [s]\}$) for some $\tuple u\in\injtuples$ as in the algorithm for an instance $\instance$ with $\instance=$ \textsc{IdentifyAllEqual}($\instance$) -- note that the output of this subroutine is used in the algorithm only for instances satisfying this assumption.
Note that for every $\tuple w\in S$, $E^1_{\tuple w}$ contains all injective orbits under $\Aut(\sH)$ contained in $\proj_{\tuple w}(\instance)$.
This can be easily shown by induction on the size of $S$. For $S$ containing just the tuple $\tuple u$, there is nothing to prove. Suppose that it holds for $S$, let $\tuple w\in [\V]^{\ell}$ be a tuple added to $S$, and let $\tuple v\in S$ be such that $E^1_{\tuple w}=\{\tuple a\in H^{\ell}\mid \exists\tuple b\in H^{\ell}\colon (\tuple b,\tuple a)\in\proj_{(\tuple v,\tuple w)}(\instance)\}$.
Let $O\subseteq \proj_{\tuple w}(\instance)$ be an injective orbit under $\Aut(\sH)$, and let $C\in\constraints$ be a constraint containing all variables from the tuples $\tuple v$ and $\tuple w$ in its scope.
Let $g_1\in C$ be such that $g_1(\tuple w)\in O$, and let $g_2\in C$ be injective (such $g_2$ exists since $\instance=$ \textsc{IdentifyAllEqual}($\instance$)). 
Then $p_1(g_1,g_2)\in C$ is injective and hence, $p_1(g_1(\tuple v),g_2(\tuple v))\in E^1_{\tuple v}$ by the induction assumption. Moreover, $p_1(g_1(\tuple w),g_2(\tuple w))\in O$ and $p_1(g_1(\tuple w),g_2(\tuple w))\in E^1_{\tuple w}$ by the definition of $E^1_{\tuple w}$.

It is easy to see that the algorithm runs in polynomial time: the algorithm of checking $\textrm{eq}$-subdirectness as well as the $(2\ell,\max(3\ell,b_{\sH}))$-minimality algorithm both run in polynomial time. Moreover, the number of runs of every for- or repeat...until-loop in the main algorithm as well as in the subroutine \textsc{ExtendPartition} is bounded by $c^c (d+{\ell}^{\max(3\ell,b_{\sH})})\binom{|\V|}{\ell}^2$, where $c$ is the number of orbits of $\ell$-tuples under $\Aut(\sH)$ and $d$ is the number of constraints of the original instance $\instance$.

Before proving that $\instance'$ is an inj-irreducible instance of $\CSP(\sA)$ and that it has a solution if, and only if, $\instance$ has a solution, we show a few auxiliary statements.

\begin{lemma}\label{lemma:NoSolutionRemoved}
Let $\instance=(\V,\constraints)$ be an $\ell$-minimal non-trivial instance of $\CSP(\sA)$ such that $\proj_{(u,v)}(\instance)\cap I_2\neq\emptyset$ for every $u\neq v\in\V$.
Let $S\subseteq\injtuples$ be such that $\proj_{\tuple{v}}(\instance)$ contains some injective orbit and no non-deterministic orbit for every $\tuple{v}\in S$. Suppose that $\instance$ has a solution.
Then the injective finitisation of\/ $\instance$ on $S$ has a solution as well.
\end{lemma}
\begin{proof}
Let $g\colon \V\to H$ be a solution of $\instance$ such that $g(\tuple{v})$ is not injective for some $\tuple{v}\in S$. Let $\{\tuple{v}_1,\ldots,\tuple{v}_n\}$ be the set of all $\tuple v_i\in S$ such that $g(\tuple v_i)$ is in a non-injective orbit under $\Aut(\sH)$. For every $i\in[n]$, let $\alpha_i\in\Aut(\sH)$ be deterministic for $g(\tuple{v}_i)$.
Let \[p_1'(x,y):=p_1(\alpha_1 x, p_1(\alpha_2 x, p_1(\ldots,p_1(\alpha_n x,y)))).\]

It is easy to see that for all $\tuple v\in S$, we have that $p_1'(O_<(g(\tuple v)),E)$ and $p_1'(O_<(g(\tuple{v})),N)$ are subsets of the same injective orbit $O_{\tuple v}$ under $\Aut(\sH)$ for the increasing order in the second coordinate.

We claim that $f\colon S\to \en$ defined by $f(\tuple v)=O_{\tuple v}$ is a solution of the injective finitisation of $\instance$ on $S$.
To this end, let $C\in\constraints$, and let $g'\in C$ be monotone and injective. It follows that $p_1'(g,g')\in C$, and moreover $O(p_1'(g(\tuple v),g'(\tuple v))=O_{\tuple v}$ for all $\tuple v\in S$ such that all variables from $\tuple v$ are in the scope of $C$, and the lemma follows.
\end{proof}

Let $\sA$ be a first-order reduct of $\sH$, let $\instance=(\V,\constraints)$ be an instance of $\CSP(\sA)$, and let $W\subseteq \V$. Let $g\colon W\to A$, and let $h\colon \V\to A$, we say that $g$ is \emph{ordered by $h$} if for all $u,v\in W$, we have $g(u)<g(v)$ if, and only if, $h(u)<h(v)$.

\begin{lemma}\label{lemma:ppdef}
Let $\instance=(\V,\constraints)$ be a $(2\ell,\max(3\ell,b_{\sH}))$-minimal, non-trivial, $\textrm{eq}$-subdirect instance of $\CSP(\sA)$ such that $\proj_{(u,v)}(\instance)\cap I_2\neq\emptyset$ for every $u\neq v\in\V$. Let $S\subseteq\injtuples$, and let $\{E^{i}_{\tuple v}\mid \tuple v\in S, i\in [s]\}$ be sets of classes of the partitions from the algorithm. Suppose that there exists $\tuple v'\in S$ such that the relation $\bigcup\limits_{i\in\{2,\ldots, s\}}E^i_{\tuple v'}$ is not preserved by $p_1$ or by $m$. Then the injective finitisation of $\instance$ on $S$ has a solution.
\end{lemma}
\begin{proof}
Suppose that the relation $R:=\bigcup\limits_{i\in\{2,\ldots, s\}}E^i_{\tuple v'}$ is not preserved by $p_1$. Let $\tuple a_1, \tuple a_2\in R$ be such that $p_1(\tuple a_1,\tuple a_2)\in E^1_{\tuple v'}$.

For $j\in[2]$, let $O^j$ be the orbit of $\tuple a_j$ under $\Aut(\sH)$. Let $h^j\colon \V\to H$ be a solution of $(\instance^{\tuple v'\in O^j})_{\textrm{eq}}$ such that $h^j(\tuple v')=\tuple a_j$.
Let $h:=p_1(h^1,h^2)$.
It follows that $h(\tuple v')\in E^1_{\tuple v'}$ and, by the way how the partitions are obtained, $h(\tuple v)\in E^1_{\tuple v}$ for every $\tuple v\in S$. In particular, $h(\tuple v)$ lies in an injective or deterministic orbit under $\Aut(\sH)$ for every $\tuple v\in S$.

Let $\tuple v_1,\ldots,\tuple v_n$ be all tuples in $S$ such that $h(\tuple v_i)$ is not injective. For every $i\in [n]$, let $\beta_i\in \Aut(\sH)$ be such that $\beta_i$ is deterministic for $h(\tuple v_i)$.
Let now $p''_1(x,y):=p_1(\beta_1 x,p_1(\beta_2 x,p_1(\ldots,p_1(\beta_n x,y))\dots))$.
Observe that for all $\tuple v\in S$, we have that $p''_1(O_<(h(\tuple v)),E)$ and $p''_1(O_<(h(\tuple v)),N)$ are subsets of the same orbit $O_{\tuple v}\in\en$ under $\Aut(\sH)$.
Let $f\colon S\to\en$ be defined as $f(\tuple v):=O_{\tuple v}$.
We claim that $f$ is a solution of the injective finitisation of $\instance$ on $S$.

To this end, let $C\in \constraints$. By the choice of $h^1,h^2$, there exist $g^1,g^2\in C$ ordered by $h^1$ and $h^2$, respectively. It follows that $g:=p_1(g^1,g^2)$ is ordered by $h$ and moreover, $g\in C$. For every $i\in[n]$, let $\alpha_i\in\Aut(\sH)$ be such that $\alpha_i g$ is ordered by $\beta_i h$. Let $p'_1(x,y):=p_1(\alpha_1 x,p_1(\alpha_2 x,p_1(\ldots,p_1(\alpha_n x,y))\dots))$, and let $g'\in C$ be monotone and injective. It follows that for every $\tuple v\in S$ such that all variables from $\tuple v$ are contained in the scope of $C$ and for every $i\in[n]$, $O_<(\alpha_i g(\tuple v))=O_<(\beta_i h(\tuple v))$. In particular, $p_1'(g(\tuple v),g'(\tuple v))\in f(\tuple v)$ as desired.

The case when $R$ is not preserved by $m$ is similar.
\end{proof}

Using the auxiliary statements above, we are able to prove the correctness of the main algorithm.

\begin{theorem}\label{thm:algorithm_correct}
The instance $\instance'$ produced by the algorithm in~\Cref{alg:inj-irreducibility} is an instance of $\CSP(\sA)$ and it has a solution if, and only if, the original instance has a solution.
Moreover, $\instance'$ is either trivial or inj-irreducible.
\end{theorem}

\begin{proof}
It is easy to see that $\instance'$ is an instance of $\CSP(\sA)$ by the algorithm -- the only thing that needs to be proven is that if the algorithm constrains an injective tuple $\tuple v\in\V^{\ell}$ by $\proj_{\tuple v}(\instance)\backslash E^1_{\tuple v}$, where $\instance=(\V,\constraints)$ is an $\textrm{eq}$-subdirect, $(2\ell,\max(3\ell,b_{\sH}))$-minimal instance obtained during the run of the algorithm, then $\proj_{\tuple v}(\instance)\backslash E^1_{\tuple v}$ is preserved by $p_1$ and by $m$. But this follows directly from~\Cref{lemma:ppdef}.

We now show that if the original instance has a solution, then the instance $\instance'$ has a solution as well. Suppose that in the repeat...\ until not changed loop (lines 3-22), we get an instance $\instance=(\V,\constraints)$ and we removed a solution of $\instance$ during one run of this loop. Clearly, we didn't remove any solutions by making the instance $\textrm{eq}$-subdirect, $(2\ell,\max(3\ell,b_{\sH}))$-minimal and by identifying variables in the \textsc{IdentifyAllEqual} subroutine. Hence, the only problem could occur when removing the classes $E^1_{\tuple w}$.

Suppose therefore that there is a set $S\subseteq \injtuples$ and a partition $\{E^j_{\tuple v}\mid j\in \{1,\ldots,s\}, \tuple v\in S\}$ on the projections of $\instance$ to the tuples from $S$. Suppose moreover that $\instance$ had a solution $s\colon \V\to H$ such that $s(\tuple w)\in E^1_{\tuple w}$ for some $\tuple w\in S$. Then $s(\tuple v)\in E^1_{\tuple v}$ for every $\tuple v\in S$ by the definition of the partition.

Recall that $E^1_{\tuple v}$ contains all injective orbits under $\Aut(\sH)$ that are contained in $\proj_{\tuple v}(\instance)$ for all tuples $\tuple v\in S$ by the reasoning below the algorithm. Let $\instance''$ be an $\ell$-minimal instance equivalent to the instance obtained from $\instance$ by adding for every $\tuple v=(v_1,\ldots,v_{\ell})\in S$ the constraint $\{g\colon \{v_1,\ldots,v_{\ell}\} \rightarrow H\mid g(\tuple v)\in E^1_{\tuple v}\}$.
Since $E^1_{\tuple v}$ is pp-definable in $\sA$, it is by assumption one of the relations of $\sA$, and therefore $\instance''$ is an instance of $\CSP(\sA)$.
Since $s$ is a solution of $\instance$, and hence also of $\instance''$, $\instance''$ is non-trivial.
Since the classes $\{E^1_{\tuple v}\mid \tuple v\in S\}$ were removed, the injective finitisation of $\instance$ on $S$ does not have a solution, and it follows that the injective finitisation of $\instance''$ on $S$ does not have a solution either. Hence, $s$ is not a solution of $\instance''$ by~\Cref{lemma:NoSolutionRemoved}, a contradiction.

We finally prove that if $\instance'$ is non-trivial, it is inj-irreducible.
Note that by the algorithm, $\instance'$ is $(2\ell,\max(3\ell, b_{\sH}))$-minimal and if it is non-trivial, then any projection of $\instance'$ to a pair of distinct variables has a non-empty intersection with $I_2$.
In particular, for every injective tuple $\tuple v$ of variables of $\instance'$, $\proj_{\tuple v}(\instance')$ contains an injective orbit under $\Aut(\sH)$.
Let $S$ be a subset of variables of $\fininstance[\instance']$. 

Suppose that for some variable $\tuple w$ in $S$, the linkedness congruence on $\proj_{\tuple w}(\restrinstance[\fininstance']{S})$ links the injective orbits within this set, and separates the injective orbits from the non-deterministic ones.

Each block $B_{\tuple w}^i$ of the linkedness congruence on $\proj_{\tuple w}(\restrinstance[\fininstance']{S})$ defines a subset of $\proj_{\tuple w}(\instance')$ by taking $E_{\tuple w}^i:=\bigcup B_{\tuple w}^i$. 
By assumption, this partition of $\proj_{\tuple w}(\instance')$ extends to a partition $\{E^1_{\tuple v},\dots,E^s_{\tuple v}\}$ on $\proj_{\tuple v}(\instance')$ for $\tuple v\in S$.
Each $E_{\tuple v}^i$ is pp-definable: for an arbitrary orbit $O$ in $B_{\tuple w}^i$, $E_{\tuple v}^i$ consists of the tuples that are reachable from some $\tuple a\in O$ by a suitable path of constraints.
Since $\sA$ contains all orbits of $\ell$-tuples under $\Aut(\sH)$ by our assumptions, $O$ is a relation of $\sA$.
Thus, the algorithm has checked that the injective finitisation of $\instance'$ on $S$ has a solution, showing the inj-irreducibility of $\instance'$.
\end{proof}

\begin{example}[continues=ex:inj-irred]
    Let $\instance$ be the instance from \Cref{ex:inj-irred}. Observe that for every $S\subseteq[\V]^3$ which does not contain all of $\tuple{v_1},\tuple{v_2},\tuple{v_3}$, the injective finitisation of $\instance$ on $S$ has a solution. It is easy to see that when $\tuple u\in[\V]^3$ is different from $\tuple{v_1},\tuple{v_2},\tuple{v_3}$, no partition on $\proj_{\tuple u}(\instance)$ with at least two classes can be extended to the projections of $\instance$ to all of the tuples $\tuple{v_1},\tuple{v_2},\tuple{v_3}$, hence the algorithm does not remove any orbits under $\Aut(\sH)$ from $\proj_{\tuple u}(\instance)$.

    If $\tuple u\in\{\tuple{v_1},\tuple{v_2},\tuple{v_3}\}$, then starting with any non-constant tuple $\tuple a\in\proj_{\tuple u}(\instance)$, we get that for every $\tuple u\neq\tuple w\in\{\tuple{v_1},\tuple{v_2},\tuple{v_3}\}$, and for every non-constant $\tuple b\in H^\ell$, it holds that $(\tuple a,\tuple b)\in\proj_{(\tuple u,\tuple w)}(\instance)$. It follows that the only partition on $\proj_{\tuple u}(\instance)$ which can be extended to $\proj_{\tuple w}(\instance)$ for some $\tuple u\neq\tuple w\in\{\tuple{v_1},\tuple{v_2},\tuple{v_3}\}$ is defined by $E^1_{\tuple u}:=\{\tuple a\in H^\ell\mid \tuple a\neq (b,b,b)\text{ for any }b\in H\}$, and $E^2_{\tuple u}:=\{(b,b,b)\mid b\in H\}$. It is easy to observe that the classes of this partition are preserved by the functions $p_1$ and by $m$ defined in the first part of this example, and that $E^1_{\tuple u}$ contains no non-deterministic orbit under $\Aut(\sH)$. Moreover, this partition extends to the projections of $\instance$ to all the tuples from $S:=\{\tuple{v_1},\tuple{v_2},\tuple{v_3}\}$ such that $E^1_{\tuple u}=E^1_{\tuple w}$ for every $\tuple w\in S$. Hence, the algorithm from \Cref{alg:inj-irreducibility} calls the subroutine \textsc{ExtendPartition} for $(\instance,\{\tuple u\},\{E^1_{\tuple u}, E^2_{\tuple u}\})$, and this subroutine returns $S=\{\tuple{v_1}, \tuple{v_2}, \tuple{v_3}\}$, and the partitions $E^1_{\tuple{v_1}}=E^1_{\tuple{v_2}}=E^1_{\tuple{v_3}}$, $E^2_{\tuple{v_1}}=E^2_{\tuple{v_2}}=E^2_{\tuple{v_3}}$.

    We observed  already in the first part of this example that the injective finitisation of $\instance$ on $S$ does not have a solution. Hence, the algorithm from \Cref{alg:inj-irreducibility} removes $E^1_{\tuple w}$ from $\proj_{\tuple w}(\instance)$ for every $\tuple w\in S$ and replaces $\instance$ with the instance obtained by this removal. Now, $\proj_{\tuple{v_1}}(\instance)=\proj_{\tuple{v_2}}(\instance)=\proj_{\tuple{v_3}}(\instance)=\{(a,a,a)\mid a\in H\}$, and $\instance$ is still eq-subdirect, $(6,9)$-minimal, and non-trivial. In the next run of the repeat... until not changed loop, all variables from $\tuple{v_1}$ are identified, and similarly for $\tuple{v_2}$ and $\tuple{v_3}$. In this way, we obtain the instance $\instance=(\V,\constraints)$ with $|\V|=3$, where for every $i\in\{0,1\}$, $C_i=\{f\colon \V\rightarrow H\mid f\text{ is injective}\}$. It follows that any injective mapping $f\colon \V\rightarrow H$ is a solution of $\instance$, and in particular, the injective finitisation of $\instance$ on $S$ has a solution, where $S$ is a set containing the unique tuple in $[\V]^3$. Hence, no other class of some partition is removed, and the algorithm outputs the instance $\instance$ which has an injective solution.
\end{example}

We now show that the structure defined by the formula from \Cref{ex:order} has polymorphisms satisfying our assumptions on $p_1$ and on $m$.

\begin{example}[continues=ex:order]\label{ex:order-continued}
    Let $R\subseteq H^4$ be the relation defined by $\psi$, and let $\sA:=(H;R)$.
    We define the canonical behaviour of a binary injection $p_1$ with respect to $(\sH,<)$ as follows.
    We require that $p_1$ acts as the first projection on $\en$, if $O$ is a non-injective orbit of triples under $\Aut(\sH)$, and $P$ is an injective orbit, then we require that $p_1(O,P)=p_1(P,O)=P$.
    Finally, for two non-injective orbits $O_1,O_2$ of triples under $\Aut(\sH,<)$ such that $p_1(O_1,O_2)$ should be injective, we require that $p_1(O_1,O_2)=E$ if the minimum of any triple in $O_1$ appears only once in this triple, and $p_1(O_1,O_2)=N$ otherwise.
    Now, we take a ternary injection $m'$ canonical with respect to $(\sH,<)$ which behaves like a minority on $\en$, and we define $m:=m'(p_1(x,p_1(y,z)),p_1(y,p_1(z,x)),p_1(z,p_1(x,y)))$.
    It is easy to see that $p_1$ and $m$ preserve $R$, hence $\sA$ satisfies the assumptions from \Cref{thm:main-algo}.

    Let $\sim$ denote the $6$-ary relation containing the tuples $(\tuple a,\tuple b)$ where $\tuple a,\tuple b$ are triples that are in the same orbit under $\Aut(\sH)$.
    Note that the relation $T$ defined by \[(x_1,x_2,x_3,x_4)\in T :\Longleftrightarrow R(x_1,x_2,x_3,x_4)\land (x_1,x_3,x_2)\sim (x_4,x_2,x_3)\]
    is preserved by all polymorphisms of $\sA$ that are canonical with respect to $\sH$ and that it
    contains precisely those tuples of the form $(a,a,b,b)$ and $(a,b,a,b)$ for arbitrary $a\neq b$.
    It can be seen (e.g., from~\cite{ecsps}) that $\Pol(H;T)$ only contains \emph{essentially unary} operations, of the form $(x_1,\dots,x_n)\mapsto \alpha(x_i)$ for arbitrary permutations $\alpha$ of $H$, and therefore the polymorphisms of $\sA$ that are canonical with respect to $\sH$ are also essentially unary.
    It follows that the finite-domain CSP used in the reduction from~\cite{ReductionFinite} is NP-complete.
\end{example}

\subsection{\texorpdfstring{$\Csp_{\injinstances}(\sA)$}{[math]} has bounded width}\label{sect:bwidth-reduction}

In this section, we consider the case when $\Csp_{\injinstances}(\sA)$ is solvable by local consistency -- we show that if $I$ is a binary absorbing subuniverse of $\sH^\ell$, then also $\Csp(\sA)$ is solvable by local consistency.

\begin{lemma}\label{lemma:bwidth_injinst}
Let $\sA$ be an $\omega$-categorical structure, let $1\leq k\leq m\leq n$, let $S,R$ with $S\subseteq R\subseteq A^k$ be relations pp-definable in $\sA$, and suppose that $S\trianglelefteq_{\sA} R$.
Let $\instance=(\V,\constraints)$ be a non-trivial, $(m,n)$-minimal instance of $\Csp(\sA)$. Let $\instance'$ be an instance obtained from $\instance$ by adding for every $\tuple v=(v_1,\ldots, v_k)\in \V^k$ with $\proj_{\tuple v}(\mathcal I)\subseteq R$, $S\cap\proj_{\tuple v}(\mathcal I)\neq\emptyset$ and such that $S\cap\proj_{\tuple v}(\mathcal I)\trianglelefteq_{\sA} R$ a constraint $\{c\colon \{v_1,\ldots,v_k\}\to A\mid c(\tuple v)\in S\}$. Then the $(m,n)$-minimal instance equivalent to $\instance'$ is non-trivial.
\end{lemma}

\begin{proof}
Let $\instance''$ be the $(m,n)$-minimal instance equivalent to $\instance'$. We find a non-trivial, $(m,n)$-minimal instance $\mathcal J$ of $\Csp$ over the set $A$ with the same set of variables as $\mathcal I''$ and such that every constraint of $\mathcal J$ is a subset of a constraint of $\mathcal I''$. Then it follows that $\instance''$ is non-trivial.

Let $f$ be a binary polymorphism of $\sA$ witnessing $S\trianglelefteq_{\sA} R$, let $F:=\{\alpha f(\beta,\gamma)\mid \alpha,\beta,\gamma\in\Aut(\sA)\}$, and let $T:=\{t\in\Pol(\sA)\mid t\text{ is a term in }F\}$. Let $C$ be a constraint of $\mathcal I$ with scope $\{u_1,\ldots,u_q\}$, and let $c_1,\ldots,c_p\in C$ be such that $(c_1(u_1),\ldots,c_1(u_q)),\ldots,(c_p(u_1),\ldots,c_p(u_q))$ are in pairwise different orbits of $q$-tuples under $\Aut(\sA)$ and such that for every $d\in C$ there exists $i\in[p]$ such that $(d(u_1),\ldots,d(u_q))$ is in the same orbit under $\Aut(\sA)$ as $(c_i(u_1),\ldots,c_i(u_q))$.

Let $t\in T$ be of arity $n\geq 1$. For $i\in[n]$, we say that the $i$-th variable of $t$ is \emph{essential} if there exist $a_1,\ldots,a_n,a'_i\in A$ such that $t(a_1,\ldots,a_n)\neq t(a_1,\ldots, a_{i-1}, a'_i,a_{i+1},\ldots,a_n)$. Set for every $C\in\mathcal{C}$, $C':=\{t(c_1,\ldots,c_p,a_1,\ldots,a_r)\mid t\in T, a_1,\dots,a_r\in C, \text{ all variables of }t\text{ essential}\}$. Set now $\mathcal{J}$ to be the instance obtained from $\mathcal{I}$ by replacing every constraint $C$ by $C'$. Clearly, $\mathcal J$ is non-trivial, for every $k$-tuple of variables $\tuple v$ it holds that if $\proj_{\tuple v}(\instance)\subseteq R$ and $S\cap\proj_{\tuple v}(\instance)\neq\emptyset$, then $S\cap\proj_{\tuple v}(\mathcal J)\trianglelefteq_{\sA} R$ and $C'\subseteq C$ for every $C\in \mathcal{C}$. It remains to show that $\mathcal{J}$ is $(m,n)$-minimal.

To this end, let $1\leq m'\leq m$, let $\tuple v\in \V^{m'}$, let $C,D\in\mathcal{C}$ be such that $\proj_{\tuple v}(C)=\proj_{\tuple v}(D)$, and let $c\in C'$. We need to find $d\in D'$ such that $c(\tuple v)=d(\tuple v)$. We can suppose that there exist $d_1,d_2\in D$ such that $d_1(\tuple v),d_2(\tuple v)$ lie in different orbits under $\Aut(\sA)$ since otherwise, $d(\tuple v)$ is in the same orbit of $m'$-tuples under $\Aut(\sA)$ as $c(\tuple v)$ for any $d\in D$ by the $m'$-minimality of $\instance$, and hence, for any $d\in D'$ we can find $\alpha\in\Aut(\sA)$ such that $\alpha(d(\tuple v))=c(\tuple v)$. By the definition of $D'$, it follows that $\alpha d\in D'$.

We have $c=t(c_1,\ldots,c_p,a_1,\ldots,a_r)$ for some $a_1,\ldots,a_r\in C, t\in T$. Suppose that $D'$ is defined using $d_1,\ldots,d_{p'}\in D$, and let $s\in T$ be of arity $p'\geq 2$ (by our assumption from the previous paragraph). Then $\alpha_j\circ s(d_1,\ldots,d_{p'})(\tuple v)=c_j(\tuple v)$ for some $j\in[p]$ and some $\alpha_j\in\Aut(\sA)$. Find $x_i\in D$ with $c_i(\tuple v)=x_i(\tuple v)$ for every $i\in [p]$ and $y_i\in D$ with $a_i(\tuple v)=y_i(\tuple v)$ for every $i\in[r]$. Set now $d:=t(x_1,\ldots,x_{j-1},\alpha_j(s(d_1,\ldots,d_{p'})),x_{j+1},\ldots,x_p,y_1,\ldots,y_r)$. It is easy to see that $d\in D'$.

It follows that $c(\tuple v)=t(c_1,\ldots,c_p,a_1,\ldots,a_r)(\tuple v)=d(\tuple v)$ as desired.
\end{proof}

\Cref{lemma:bwidth_injinst} yields the following corollary.

\begin{corollary}\label{corollary:CAH_eq-non-affine}
Let $\sA$ be a first-order reduct of\/ $\sH$, and let $\ell\leq m\leq n$.
Suppose that $\Csp_{\injinstances}(\sA)$ has relational width $(m,n)$ and that $I\trianglelefteq_{\sA} H^{\ell}$.
Then every non-trivial $(m,n)$-minimal instance equivalent to an instance of $\CSP(\sA)$ has a solution.
Moreover, such a solution $s$ exists where $s(x)\neq s(y)$ for all variables $x,y$ such that $\proj_{(x,y)}(\instance)\cap I_2\neq\emptyset$.
\end{corollary}

\begin{proof}
Let $f\in\CA$ be a function that witnesses that $I\trianglelefteq_{\sA} H^{\ell}$.
It follows that for all $\tuple a,\tuple b\in A^2$ such that $\tuple b$ is injective, $f(\tuple a,\tuple b),f(\tuple b,\tuple a)\in I_2$ and hence, $I_2\trianglelefteq_{\sA} A^2$. Finally,~\Cref{lemma:bwidth_injinst} gives the result.
\end{proof}

It follows from \Cref{corollary:CAH_eq-non-affine} that if $I$ is a binary absorbing subuniverse of $H^\ell$ in $\sA$ and $\Csp_{\injinstances}(\sA)$ has bounded width, then so does $\Csp(\sA)$, which is therefore in particular solvable in polynomial time.

By combining~\Cref{corollary:CAH_eq-non-affine} and \Cref{thm:inj-irreducibility,thm:algorithm_correct}, we obtain the desired result.
\mainalgo*
\begin{proof}
    If $\Csp_{\injinstances}(\sA)$ has bounded width, it has relational width $(m,n)$ for some $1\leq m\leq n$, and hence also relational width $(m',n')$ for any $m'\geq m, n'\geq n$ with $m'\leq n'$. We can therefore assume without loss of generality that $\ell\leq m\leq n$, and \Cref{corollary:CAH_eq-non-affine} yields that $\Csp(\sA)$ has relational width $(m,n)$, hence it is solvable in polynomial time.
    
If $\sA$ admits an injective linear symmetry, given an instance $\instance$ of $\Csp(\sA)$, the algorithm in \Cref{alg:inj-irreducibility} runs in a polynomial time by the discussion below the algorithm, and it outputs an instance $\instance'$ of $\Csp(\sA)$ which is which is either trivial or inj-irreducible and which has a solution if, and only if, $\instance$ has a solution by \Cref{thm:algorithm_correct}. If $\instance'$ is trivial, $\instance$ has no solution. Otherwise, $\instance'$ is easily seen to satisfy the assumptions of \Cref{thm:inj-irreducibility}, hence it has an injective solution. It follows that $\instance$ has a solution.
\end{proof}

\section{Overview of the proof of~\Cref{thm:main-1+2}}\label{sect:overview}

The proof of our dichotomy theorem (\Cref{thm:main-1+2}) relies on proving that every $\sA$ for which the algebraic assumptions implying polynomial-time solvability in~\Cref{thm:main-algo} are not met is such that $\CSP(\sA)$ is NP-complete.
Before giving an outline of the proof, we introduce additional notions from model theory and universal algebra.

\subsection{Model-theoretic notions}
\label{sect:overview-prelims}

A relational structure $\sA$ is a \emph{model-complete core} if for every finite subset $F\subseteq A$ and for every $\alpha\in\Aut(\sA)$, there exists a unary $g\in\Pol(\sA)$ such that $g|_F=\alpha|_F$. By~\cite{cores}, for any $\omega$-categorical structure $\sA$, there exists an up to isomorphism unique $\omega$-categorical model-complete core $\sA'$ and the $\Csp$s of $\sA$ and $\sA'$ are the same computational problem. In any $\omega$-categorical model-complete core $\sA$, all orbits of $n$-tuples with respect to the automorphism group $\Aut(\sA)$ are pp-definable, for all $n \geq 1$.

A class $\mathcal K$ of structures is a \emph{strong amalgamation class} if for all embeddings $e_i\colon\rel H\hookrightarrow\rel H_i$ ($i\in\{1,2\}$) with $\sH,\rel H_1,\rel H_2\in\mathcal K$, there exist $\rel H'\in\mathcal K$ and embeddings $f_i\colon\rel H_i\hookrightarrow\rel H'$ ($i\in\{1,2\}$) such that $f_1\circ e_1=f_2\circ e_2$ and $f_1(H_1)\cap f_2(H_2)=(f_1\circ e_1)(H)$.

A class of structures is \emph{Ramsey} if it satisfies a certain analogue of Ramsey's theorem -- the concrete definition is not needed for our purpose, we only need its consequence from~\cite{BPT-decidability-of-definability} which is stated later in this section.
Every Ramsey class is an amalgamation class~\cite{NesetrilAmalgamation}.
While the class $\Kall$ of all $\ell$-hypergraphs, or the class $\mathcal K^\ell_r$ of $\ell$-hypergraphs omitting a generalised clique of size $r$, are not themselves Ramsey classes, the classes $\vec{\Kall}$ and $\vec{\mathcal K^\ell_r}$ are -- this follows, e.g., from the Ne\v set\v ril-R\"odl theorem \cite{NesetrilRoedl}.
By abuse of language, we also call the Fra\"iss\'e limit of a Ramsey class a \emph{Ramsey structure}.

\begin{definition}[``Primitivity"]
Let $\sA$ be a relational structure, and let $n\geq 1$. We say that $\sA$ is \emph{$n$-``primitive"} if for every orbit $O\subseteq A^n$ under $\Aut(\sA)$, every $\Aut(\sA)$-invariant equivalence relation on $O$ containing some pair $(\tuple a,\tuple b)$ of disjoint tuples is full.
\end{definition}

\begin{example}\label{ex:primitivity}
Let $\mathcal K$ be $\Kall$ or $\mathcal K^\ell_r$ for $r>\ell$, and let $\sH$ be the Fra\"iss\'e limit of $\mathcal K$.
Then $\sH$ is $n$-``primitive" for any $n\geq 1$. Indeed, let $n\geq 1$, let $O$ be an orbit of $n$-tuples under $\Aut(\sH)$, and let $\sim$ be an equivalence relation on $O$ containing $(\tuple a,\tuple b)$ such that $\tuple a,\tuple b$ are disjoint. Let $\tuple c,\tuple d\in O$ be arbitrary.
If $\mathcal K=\Kall$, let $r:=3$. 
We define $\sX$ to be an $\ell$-hypergraph over $(r+1)n$ elements $\{x_i^j\mid i\in[n], j\in[r+1]\}$ such that the following holds. The hypergraph induced by $(x_1^j,\ldots,x_{n}^j,x_1^{j+1},\ldots,x_{n}^{j+1})$ is isomorphic to the structure induced by $(\tuple a,\tuple b)$ in $\sH$ for every $j\in[r]$, and the hypergraph induced by $(x_1^1,\ldots,x_n^1,x_1^{r+1},\ldots,x_n^{r+1})$ is isomorphic to the structure induced by $(\tuple c,\tuple d)$ in $\sH$.
Moreover, we require that $\sX$ does not contain any hyperedge that is not enforced by the conditions from the previous sentence. It follows that if neither the structure induced by $(\tuple a,\tuple b)$ in $\sH$ nor the structure induced by $(\tuple c,\tuple d)$ contain a generalised clique of size $r$, then neither does $\sX$.
Thus $\sX$ is in $\mathcal K$, so that it embeds into $\sH$.
By the homogeneity of $\sH$, we can assume that the embedding maps $(x_1^1,\ldots,x_n^1,x_1^{r+1},\ldots,x_n^{r+1})$ to $(\tuple c,\tuple d)$.
By the transitivity of $\sim$, we get $\tuple c\sim \tuple d$ so that $\sim$ is full.
\end{example}

\begin{center}
\noindent\fbox{%
    \parbox{0.95\columnwidth}{\textbf{In the rest of the article, we fix $\ell\geq 3$ and a strong amalgamation class $\mathcal K$ of finite $\ell$-hypergraphs such that $\vec{\mathcal K}$ is Ramsey and such that
    the Fra\"iss\'e limit $\sH$ of $\mathcal K$ is $n$-``primitive'' for all $n\geq 1$.
}
    }
}
\end{center}

\smallskip

As in \Cref{sect:zhuk}, for $n\geq 1$, we denote by $I_n$ the set of injective $n$-tuples in $\sH$ and we write $I:=I_\ell$.

\subsection{Universal-algebraic notions}\label{subsect:overview-UA}

For a function clone $\cC$, we denote the domain of its functions by $C$; we say that $\cC$ \emph{acts on} $C$.
The clone $\cC$ also naturally acts (componentwise) on $C^n$ for any $n\geq 1$, on any invariant subset $S$ of $C$ (by restriction), and on the classes of any invariant equivalence relation $\sim$ on an invariant subset $S$ of $C$ (by its action on representatives of the classes).
We write $\cC \curvearrowright C^n$, $\cC \curvearrowright S$ and
$\cC \curvearrowright S/{\sim}$ for these actions. Any action $\cC \curvearrowright S/{\sim}$ is called a \emph{subfactor} of $\cC$, and we also call the pair $(S,\sim)$ a subfactor. A subfactor $(S,\sim)$ is \emph{minimal} if the equivalence relation $\sim$ has at least two classes and no proper subset of $S$ that intersects at least two $\sim$-classes is invariant under $\cC$.

If all functions of a function clone $\cC$ are $n$-canonical with respect to $\sA$, then $\cC$ acts on $C^n/{\Aut(\sA)}$ and we write $\cC^n/{\Aut(\sA)}$ for this action; if $\sA$ is $\omega$-categorical then $\cC^n/{\Aut(\sA)}$ is a function clone on a finite set.
We say that a function is \emph{diagonally canonical} if it satisfies the definition of $n$-canonicity in the case $\alpha_1 = \cdots = \alpha_k$ for every $n\geq 1$.

For a set of functions $\mathscr{F}$ over the same fixed set $C$, we write $\overline{\mathscr F}$ for the set of those functions $g$ such that for all finite subsets $F\subseteq C$, there exists a function in $\mathscr{F}$ which agrees with $g$ on $F$. For $k\geq 1$, for $k$-ary functions $f,g$ and for a permutation group $\gG$ such that $f,g$ and $\gG$ act on the same domain, we say that $f$ \emph{locally interpolates} $g$ modulo $\gG$ if $g\in \overline{\{\beta\circ f(\alpha_1, \ldots, \alpha_k)\mid\beta, \alpha_1, \ldots, \alpha_k \in \gG\}}$. 
Similarly, we say that $f$ \emph{diagonally interpolates} $g$ modulo $\gG$ if $f$ locally interpolates $g$ with $\alpha_1 = \cdots = \alpha_k$.
If $\gG$ is the automorphism group of a Ramsey structure $\sA$,
then every function on $A$ locally (diagonally) interpolates a function modulo $\gG$ that is canonical (diagonally canonical) with respect to $\sA$~\cite{BPT-decidability-of-definability, BodPin-CanonicalFunctions}.
We say that a clone $\cD$ locally interpolates a clone $\cC$ modulo a permutation group $\gG$ if for every $g \in \cD$ there exists $f \in \cC$ such that $g$ locally interpolates $f$ modulo $\gG$.

A \emph{weak near-unanimity (WNU) operation of arity $n\geq 2$} is an operation $w$ satisfying the equation $w(x,\ldots,x,y)=\cdots=w(y,x,\ldots,x)$ for all $x,y$ from its domain.
A ternary operation $m$ is a \emph{minority} operation if it satisfies $m(x,x,y)=m(x,y,x)=m(y,x,x)=y$ for all $x,y$ from its domain.
A binary operation $f$ on a two element domain is a~\emph{semilattice operation} if $f(x,y)=f(y,x)=f(x,x)=x$ and $f(y,y)=y$ for some enumeration $\{x,y\}$ of its domain.

A function is \emph{idempotent} if it satisfies $f(x,\ldots,x)=x$ for every $x$ from its domain. A function clone is idempotent if all of its functions are. 
For a function $f$ of arity $n\geq 1$ and for $i\in[n]$, we say that the $i$-th variable of $f$ is \emph{essential} if there exist $a_1,\ldots,a_n,a'_i$ from the domain of $f$ such that $f(a_1,\ldots,a_n)\neq f(a_1,\ldots, a_{i-1}, a'_i,a_{i+1},\ldots,a_n)$.
A function is called \emph{essentially unary} if at most one of its variables is essential, otherwise is the function is \emph{essential}.

An arity-preserving map $\xi\colon \cC \to \cD$ between function clones is called a \emph{minion homomorphism} if it preserves
compositions with projections, i.e., it satisfies $\xi(f \circ (\pi_1, \ldots, \pi_n)) = \xi(f) \circ (\pi_1, \ldots, \pi_n)$ for all $n, m \geq 1$ and all $n$-ary $f \in \cC$ and $m$-ary  projections $\pi_1, \ldots, \pi_n$.
An arity-preserving map $\xi$ is called a \emph{clone homomorphism} if it preserves
projections, i.e., maps every projection in $\cC$ to the corresponding projection in $\cD$, and compositions, i.e., it satisfies $\xi(f \circ (g_1, \ldots, g_n)) = \xi(f) \circ (\xi(g_1), \ldots, \xi(g_n))$ for all $n, m \geq 1$ and all $n$-ary $f \in \cC$ and $m$-ary $g_1, \ldots, g_n \in \cC$.  
We say that a function clone $\cC$ is
\emph{equationally trivial} if it has a clone homomorphism to the clone $\Projs$ of projections on a two-element domain, and \emph{equationally non-trivial} otherwise.
We say that $\cC$ is
\emph{equationally affine} if it has a clone
homomorphism to an \emph{affine clone}, i.e., a clone of affine maps over a finite module. 
It is known that a finite idempotent clone is either equationally affine or it contains WNU operations of all arities $n \geq 3$ (\cite{MarotiMcKenzie}, this stronger version is attributed to E.\ Kiss in~\cite[Theorem 2.8]{Maltsev-Cond}, a different proof can be found in~\cite{StrongSubalgebras}). 

If $\cC$, $\cD$ are function clones and $\cD$ has a finite domain, then a clone (or minion) homomorphism
$\xi\colon \cC \to \cD$ is \emph{uniformly continuous} if for all $n \geq 1$ there exists a finite subset $F$ of $C^n$
such that $\xi(f) = \xi(g)$
for all $n$-ary $f, g \in \cC$ which agree on $F$.

{Given $T\subseteq H^{{k}}$ and a structure $\mathbb B$ with domain $H$ we say that $f\colon H^k\to H$ is \emph{inj-canonical with respect to $\mathbb B$ on $T$} if for all $n\geq 1$, all injective tuples $\tuple a^1,\dots,\tuple a^k\in H^n$ and all $\alpha_1,\dots,\alpha_k\in\Aut(\mathbb B)$ such that 
$(a^1_i,\dots,a^k_i),(\alpha_1a^1_i,\dots,\alpha_ka^k_i)\in T$ for all $i\in\{1,\dots,n\}$, there exists $\beta\in\Aut(\mathbb B)$ such that $f(\tuple a^1,\dots,\tuple a^k)=\beta f(\alpha_1\tuple a^1,\dots,\alpha_k\tuple a^k)$.}
For a first-order reduct $\sA$ of $\sH$, we consider the following two subclones of $\CA$:

\begin{itemize}
    \item $\CAH$ is the clone of those polymorphisms of $\sA$ which preserve the equivalence of orbits of injective tuples under $\Aut(\sH)${, i.e., which act on $I_\ell /{\Aut(\sH)}$}. {Note that $\CAH$ consists, for all $k\geq 1$, of the $k$-ary polymorphisms of $\sA$ that preserve  $I_\ell$ and that are inj-canonical with respect to $\sH$ on $H^k$.}
    \item $\CAA$ is the clone of those polymorphisms of $\sA$ which preserve the equivalence of orbits of injective tuples under $\Aut(\sA)${, i.e., which act on $I_n /{\Aut(\sA)}$ for all $n\geq 1$.
    These are exactly, for all $k\geq 1$,  the $k$-ary polymorphisms of $\sA$ preserving $I_n$ for all $n$ and that are inj-canonical with respect to $\sA$ on $H^k$}.
\end{itemize}

\subsection{Proof of the dichotomy}

We prove the following generalisation of \Cref{thm:main-1+2}; the fact that~\Cref{thm:main-1+2} follows from~\Cref{thm:hypergraph-dichotomy} can be seen by taking $\mathcal K$ to be $\Kall$ or $\mathcal K^\ell_r$ (which are finitely bounded strong amalgamation classes such that $\vec{\Kall}$ and $\vec{\mathcal K^\ell_r}$ are Ramsey) and observing that the respective Fra\"iss\'e limits are $n$-``primitive'' for all $n\geq 1$ by~\Cref{ex:primitivity}. 
\begin{restatable}{theorem}{dichotomy}\label{thm:hypergraph-dichotomy}
Let $\ell\geq 3$, let $\mathcal K$ be a finitely bounded strong amalgamation class of $\ell$-hypergraphs such that $\vec{\mathcal K}$ is Ramsey.
Suppose that the Fra\"iss\'e limit\  $\sH$ of $\mathcal K$ is $n$-``primitive'' for all $n\geq 1$.
Let $\sA$ be a first-order reduct of\/ $\sH$.
 Then precisely one of the following applies.

\begin{enumerate}
    \item The clone of polymorphisms of $\sA$ has no uniformly continuous minion homomorphism to the clone of projections $\Projs$, and $\CSP(\sA)$ is in $\P$.
    \item The clone of polymorphisms of $\sA$ has a uniformly continuous minion homomorphism to the clone of projections $\Projs$, and $\CSP(\sA)$ is $\NP$-complete.
\end{enumerate}
Moreover, given $\sA$, it is possible to algorithmically decide which of the cases holds.
\end{restatable}

We give an outline of the proof; the details are given in~\Cref{sect:binarypol,sect:NP-hard}.

Let $\sA$ be a model-complete core of a first-order reduct $\sA'$ of $\sH$.
By~\cite{wonderland}, it is enough to prove that~\Cref{thm:hypergraph-dichotomy} holds for $\sA$ since there exists a uniformly continuous minion homomorphism from $\Pol(\sA')$ to $\CA$ and from $\CA$ to $\Pol(\sA')$.
By~\Cref{lemma:mc-cores}, $\sA$ is a one-element structure or it is itself a first-order reduct of $\sH$. In the first case, $\Csp(\sA)$ is in $\P$, so we can suppose that the latter case applies. Moreover, it is also enough to prove the last sentence of \Cref{thm:hypergraph-dichotomy} about algorithmic decidability for a first-order reduct of $\sH$ which is a model-complete core since the model-complete core of a first-order reduct of $\sH$ can be computed algorithmically using \Cref{lemma:mc-cores}. 

By applying some folklore results, as well as a compactness argument, we finally obtain in~\Cref{prop:canonical_injection} that $\CA$ contains a binary injective operation with certain properties unless $\CA$ admits a uniformly continuous clone homomorphism to $\Projs$ in which case, $\Csp(\sA)$ is NP-complete by~\cite{Topo-Birk}. 
This binary injection witnesses that $I\trianglelefteq_{\sA} H^\ell$.
We already know from~\Cref{sect:zhuk} that if $\Csp_{\injinstances}(\sA)$ has bounded width, or if $\sA$ admits an injective linear symmetry, then $\CSP(\sA)$ is in P.

Thus, it remains to prove that if neither of these properties hold, then $\CSP(\sA)$ is NP-complete.
We prove that if $\CAH\actson\en$ is equationally trivial, then $\CA$ has a uniformly continuous clone homomorphism to the clone of projections $\Projs$.
This is enough since
it follows from~\cite{Post} that if $\CAH\actson\en$ is equationally non-trivial, then it contains either a binary function acting as a semilattice operation on $\en$ or a ternary function acting as a majority or as a minority on $\en$. Since these properties are stable under local interpolation and since $(\sH,<)$ is Ramsey, we can suppose that these functions are canonical with respect to $(\sH,<)$. If $\CAH$ contains a function acting as a majority on $\en$ or a function acting as a semilattice on $\en$, then $\Csp_{\injinstances}(\sA)$ has bounded width. If $\CAH$ contains only a function $m'$ which acts as a minority on $\en$ and which is canonical with respect to $(\sH,<)$, \Cref{prop:canonical_injection} yields that $\CAH$ contains a binary injection $p_1$ which acts as the first projection on $\en$ and which is canonical with respect to $(\sH,<)$. Setting $m(x,y,z):=m'(p_1(x,p_1(y,z)),p_1(y,p_1(z,x)),p_1(z,p_1(x,y)))$, we get a function witnessing that $\sA$ admits an injective linear symmetry.

Assuming that $\CAH\actson\en$ is equationally trivial, the first step in order to prove that $\CA$ has a uniformly continuous clone homomorphism to $\Projs$ is to establish that $\CAA$ is equationally trivial as well and that {$\CAH\subseteq\CAA$} (\Cref{lemma:CAA_eq-trivial}).
Moreover, Proposition 36 in~\cite{SmoothApproximations} implies that there exists $k\geq \ell$ such that the action of $\CAA$ on orbits of injective $k$-tuples under $\Aut(\sA)$ is equationally trivial. Therefore, there exists a \emph{naked set} $(S,\sim)$ for this action. A naked set of $\CAA\actson I_k/{\Aut(\sA)}$ consists of an invariant subset $S\subseteq I_k/{\Aut(\sA)}$ and an invariant equivalence relation $\sim$ on $S$ such that $\sim$ has at least two equivalence classes and such that $\CAA\actson I_k/{\Aut(\sA)}$ acts on $S/{\sim}$ by projections. By classical results in finite clone theory, the existence of such a naked set is equivalent to the existence of a clone homomorphism from $\CAA\actson I_k/{\Aut(\sA)}$ to $\Projs$, which extends to a uniformly continuous clone homomorphism $\CAA\to\Projs$.

Now, we can employ the theory of smooth approximations to extend this homomorphism further and obtain a uniformly continuous clone homomorphism $\Pol(\sA)\to\Projs$.
We recall below the relevant definitions from the theory of smooth approximations.

\begin{definition}[Smooth approximations]
Let $A$ be a set, and let $\sim$ be an equivalence relation on $S\subseteq A$. We say that an equivalence relation $\eta$ on a set $S'$ with $S\subseteq S'\subseteq A$ \emph{approximates} $\sim$ if the restriction of $\eta$ to $S$ is a refinement of $\sim$. $\eta$ is called an \emph{approximation} of $\sim$.

For a permutation group $\fG$ acting on $A$ and $S\subseteq A$, we say that an equivalence relation $\eta$ is %
\emph{very smooth on $S$ with respect to $\fG$} if orbit-equivalence with respect to $\fG$ is a refinement of $\eta$ on $S$.
\end{definition}

{For example, the equivalence relation $\eta$ on $H^{\ell}$ with two equivalence classes $I_\ell$ and $H^\ell \backslash I_\ell$ is very smooth for $\Aut(\sH)$ as the orbit-equivalence with respect to $\Aut(\sH)$ is a refinement of $\eta$.}

Recall that $\sH$ is a Fra\"iss\'e limit of a strong amalgamation class.
Thus, it has \emph{no algebraicity} (meaning that the pointwise stabiliser in $\Aut(\sH)$ of any finite tuple has no fixpoints), and hence the hypotheses of the loop lemma of smooth approximations \cite[Theorem 10]{SmoothApproximations} are met and we can use the following reformulation of the lemma to our situation.

\begin{theorem}\label{thm:1st_loop_lemma}
Let $k\geq 1$, and suppose that $\CAA\actson I_k/{\Aut(\sA)}$ is equationally trivial. Then there exists a naked set $(S,\sim)$ of 
 {$\CAA\actson I_k$}   
with $\Aut(\sA)$-invariant $\sim$-classes such that one of the following holds:

\begin{itemize}
    \item $\sim$ is approximated by a $\CA$-invariant equivalence relation that is very smooth with respect to $\Aut(\sA)$;
    \item every $\CA$-invariant binary symmetric relation $R\subseteq (I_k)^2$ that contains a pair $(\tuple a,\tuple b)\in S^2$ such that $\tuple a, \tuple b$ are disjoint and such that $\tuple a\not\sim \tuple b$ contains a pseudo-loop modulo $\Aut(\sA)$, i.e., a pair $(\tuple c,\tuple c')$ where $\tuple c,\tuple c'$ belong to the same orbit under $\Aut(\sA)$.
\end{itemize}
\end{theorem}

In \cite[Theorem 10]{SmoothApproximations}, the first item of the statement gives an approximation ${\sim}$ that is not very smooth, but \emph{presmooth}. By \cite[Lemma 8]{SmoothApproximations}, under the assumption that $\Pol(\sA)$ preserves $I_2$ and that $\sA$ is $n$-``primitive'', every presmooth approximation is very smooth; this justifies our reformulation above. 

Suppose that the first case of~\Cref{thm:1st_loop_lemma} applies.
By a minor modification of the smooth approximation toolbox in \Cref{seubsect:case1} ({\Cref{generalized-fundamental} and} \Cref{lem:existssmoothapprox}), this implies that $\CA$ has a uniformly continuous clone homomorphism to $\CAA\actson I_k/{\Aut(\sA)}$, and hence to the clone of projection.

Suppose now that the second case of~\Cref{thm:1st_loop_lemma} applies.
By~\cite[Lemma 13]{SmoothApproximations}, $\CA$ contains a \emph{weakly commutative function}, i.e., a binary operation $f$ with the property that $f(\tuple a, \tuple b) \sim f (\tuple b, \tuple a)$ holds for all $\tuple a, \tuple b \in I_k$ such that $f(\tuple a, \tuple b)$ and $f(\tuple b, \tuple a)$ are in $S$ and disjoint.
It follows from a fairly involved compactness argument in~\Cref{subsect:case2} (\Cref{lemma:weaklycomm-semilattice}) that $\CAH\actson\en$ contains a semilattice operation, and in particular is equationally non-trivial, which is a contradiction.

Finally, it follows that in order to algorithmically decide if $\Csp(\sA)$ is in $\P$, it is enough to check if it has a polymorphism that is canonical with respect to $(\sH,<)$ and that acts as a minority, majority or as a semilattice operation on $\en$. Since there are only finitely many canonical behaviours with respect to $(\sH,<)$ for functions of arity at most $3$, it is enough to go through all of them to decide the tractability of $\Csp(\sA)$.

\section{Model-Complete Cores and Injective Polymorphisms}\label{sect:binarypol}
In this section, we first prove some basic facts about model-complete cores of first-order reducts of $\sH$. 
In the second part, we prove that such first-order reducts that are model-complete cores and that are equationally non-trivial 
have binary injective polymorphisms acting as a projection or as a semilattice operation on $\en$.

\subsection{Model-complete cores}
Let $\gG$ be a permutation group, and let $g\colon G\to G$ be a function. We say that $g$ is \emph{range-rigid} with respect to $\gG$ if all orbits of tuples under $\gG$ that intersect the range of $g$ are invariant under $g$.
The \emph{age} of a relational structure $\sA$ is the set of its finite substructures up to isomorphism.
We use the following theorem to understand the model-complete cores of first-order reducts of $\sA$.

\begin{theorem}[\cite{Cores_Ramsey}]\label{thm:Ramsey_cores}
    Let $\sA$ be a first-order reduct of a homogeneous Ramsey structure $\sB$, and let $\sA'$ be its model-complete core. Then $\sA'$ is a first-order reduct of a homogeneous Ramsey substructure $\sB'$ of\/ $\sB$.
    Moreover, there exists $g\in\End(\sA)$ that is range-rigid with respect to $\Aut(\sB)$ and such that the age of\/ $\sB'$ is equal to the age of the structure induced by the range of $g$ in $\sB$.
\end{theorem}

\begin{restatable}{proposition}{mccores}\label{lemma:mc-cores}
    Let $\sA$ be a first-order reduct of $\sH$. Then the model-complete core of $\sA$ is a one-element structure or a first-order reduct of $\sH$. Moreover, if $\sA$ is a model-complete core that is a first-order reduct of $\sH$ and not of $(H;=)$, then the range of every $f\in\End(\sA)$ intersects every orbit under $\Aut(\sH,<)$.
\end{restatable}

\begin{proof}
    Using~\Cref{thm:Ramsey_cores}, we obtain that the model-complete core $\sA'$ of $\sA$ is a first-order reduct of a homogeneous Ramsey substructure $\sB'$ of $(\sH,<)$. Moreover, there exists $g\in \End(\sA)$ which is range-rigid with respect to $\Aut(\sH,<)$ and such that the age of the structure induced by the range of $g$ in $(\sH,<)$ is equal to the range of $\sB'$.

    If the range of $g$ contains just one element, $\sB'$ is a one-element structure. Otherwise, the range contains at least two elements $a<b$ and by the range-rigidity, it then contains infinitely many elements. In particular, it contains a hyperedge or a non-hyperedge.

    If the range contains only hyperedges, then $g$ sends every $\ell$-tuple in $N$ to an $\ell$-tuple ordered in the same way as the original tuple that lies in the hyperedge relation $E$. It follows that for all injective tuples $\tuple a,\tuple b$ of the same length, there exists an automorphism $\alpha$ of $\sH$ and an embedding $e$ from the range of $g$ into $\sB'$ such that $e\circ g\circ \alpha(\tuple a)=\tuple b$. It follows that $\Aut(\sA')$ is the full symmetric group on the domain of $\sA'$, and hence $\sA'$ is a first-order reduct of $(A',=)$ which is isomorphic to $(H;=)$. If the range of $g$ contains only $\ell$-tuples in $N$, $\sA'$ is a first-order reduct of $(A',=)$ by the same argument where the roles of $E$ and $N$ are switched. Finally, if the range of $g$ contains both a hyperedge as well as a non-hyperedge, it follows from the range-rigidity of $g$ that $\sB'$ is isomorphic to $(\sH,<)$ and $\sA'$ is isomorphic to $\sA$. In particular, $\sA'$ is a first-order reduct of $\sH$.

    Suppose now that $\sA$ is a model-complete core that is a first-order reduct of $\sH$ but not of $(H;=)$, and let $f\in\End(\sA)$. Suppose that the range of $f$ does not intersect every orbit under $\Aut(\sH,<)$. By Lemma 15 and Lemma 11 in~\cite{Cores_Ramsey} applied to $\End(\sA)$, the range-rigid function $g$ does not intersect every orbit under $\Aut(\sH,<)$ either, and we obtain a contradiction with the previous paragraph.
\end{proof}

\subsection{Injective binary polymorphisms}
\begin{lemma}\label{lemma:binary_essential}
Let $\sA$ be a first-order reduct of $\sH$ that is a model-complete core. If $\CA$ does not have a uniformly continuous clone homomorphism to $\Projs$, then it contains a binary essential operation.
\end{lemma}

\begin{proof}
It follows from Corollary 6.9 in~\cite{Topo} that $\CA$ contains a ternary essential operation. Moreover, the binary relation $O:=\{(a,b)\mid a\neq b\in H\}$ is an orbit under $\Aut(\sH)$ that is \emph{free}, i.e., for every $(c,d)\in H^2$, there exists $a\in H$ such that $(a,c),(a,d)\in O$. Now, the lemma follows directly from Proposition 22 in~\cite{SmoothApproximations}. {Another way to obtain this is  the recent result~\cite[Theorem~A]{Binarysymm} stating that every infinite $\omega$-categorical model-complete core without a uniformly continuous clone homomorphism from its polymorphisms  to $\Projs$ satisfies the conclusion of the lemma.}
\end{proof}

{In the following, we say that an $n$-ary function $f$ \emph{acts lexicographically on the order} if it satisfies $f(x_1,\dots,x_n)<f(y_1,\dots,y_n)$ whenever there exists $k\in [n]$ such that $x_j=y_j$ for all $j\in[k-1]$ and $x_k<y_k$.}

\begin{observation}\label{obs:lexicographic}
    {For every {$n$-ary}  %
    injective operation $f$ on $H$, there exists an embedding $e$ of $\sH$ %
    into $\sH$ such that $f':= e\circ f$ acts lexicographically on the order.}
\end{observation}

\begin{proof}
    {Let $Y\subseteq H$ be the range of $f$, and for $f(x_1,\ldots,x_n), f(y_1,\ldots,y_n)\in Y$ set $f(x_1,\ldots,x_n)<'f(y_1,\ldots,y_n)$ if there exists $k\in [n]$ such that $x_j=y_j$ for all $j\in[k-1]$ and $x_k<y_k$.  Extend $<'$ arbitrarily to a linear order on $H$. Then  $(\sH,<')$ embeds into $(\sH,<)$ via an embedding $e$, since $(\sH,<)$ is an expansion of $\sH$ by a linear order that is added freely. Clearly, $f':=e\circ f$ acts lexicographically on the order $<$.}
\end{proof}

We say that a relational structure $\sB$ has \emph{finite duality} if it is finitely bounded and the set $\cF$ from the definition of finite boundedness is closed under homomorphic images.

\begin{restatable}{lemma}{essentialinjection}\label{lemma:essential_injection}
Let $\sA$ be a first-order reduct of $\sH$ that is a model-complete core. Suppose that $\CA$ contains a binary essential operation. Then $\CAH$ contains a binary injection $f$ { canonical with respect to $(\sH,<)$} such that one of the following holds:

\begin{itemize}
    \item $f$ acts like a semilattice operation on $\en$, or
    \item $f$ acts like a projection on $\en$.%
\end{itemize}
\end{restatable}

\begin{proof}
The binary disequality relation $\neq$ is clearly an orbit under $\Aut(\sA)$, and since $\sA$ is a model-complete core, $\sA$ pp-defines $\neq$. $\sH$ is clearly transitive, i.e., $\Aut(\sH)$ has only one orbit in its action on $H$. Moreover, $\Aut(\sH)$ has only one orbit of injective pairs in its action on $H^2$, namely the orbit of $\{(a,b)\mid a\neq b\in H\}$ and the structure $(H;\neq)$ clearly has finite duality. Proposition 25 in~\cite{SmoothApproximations} then implies that $\CA$ contains a binary injection.

Note that if $\sA$ is a first-order reduct of $(H;=)$, then this binary injection can be composed with a unary injection so that the result acts like a semilattice operation on $\en$, so that we can assume below that $\sA$ is not a first-order reduct of $(H;=)$.

Since 
$(\sH,<)$ is a homogeneous Ramsey structure and since any function interpolated by an injective function is injective, $\CA$ contains a binary injection $f$ which is canonical with respect to $(\sH,<)$. Since $\sA$ is a model-complete core and not a first-order reduct of $(H;=)$, we can assume that the action of $f$ on orbits of $\ell$-tuples under $\Aut(\sH,<)$ is idempotent by~\Cref{lemma:mc-cores} -- the range of the endomorphism $e$ defined by $e(x):=f(x,x)$ intersects every orbit under $\Aut(\sH,<)$, and hence, if $f$ does not act idempotently on the orbits of $\ell$-tuples under $\Aut(\sH,<)$, we can compose $f$ finitely many times with itself until it does. Here, by composition of a binary function $g$ with $f$, we mean the function $g(f(x,y),f(x,y))$. 
Moreover, by {\Cref{obs:lexicographic},} %
we can assume that $f$ acts lexicographically on the order $<$.

Let $\alpha$ be a permutation of $\{1,\ldots,\ell\}$.
If $O$ is an orbit of injective $\ell$-tuples under $\Aut(\sH,<)$, then we write $\alpha(O)$ for the orbit obtained by changing the order of the tuples in $O$ according to $\alpha$ (not by permuting the tuples in $O$). That way, $\alpha$ acts naturally on orbits of injective $\ell$-tuples.
We also apply this notation to unions of orbits of injective tuples under $\Aut(\sH,<)$.
Let $J$ be the set of strictly increasing $\ell$-tuples with respect to $<$.
{For a fixed permutation $\alpha$, there is a binary function $f_\alpha$ on $\en$ obtained by considering the action of $f$ on $J\times\alpha(J)$.
Namely, $f_{\alpha}(O,O')$ is the orbit of $f(\tuple s,\tuple t)$ under $\Aut(\sH)$, where $\tuple s\in O\cap J$ and $\tuple t\in O' \cap\alpha(J)$.
This is well-defined since $f$ is canonical with respect to $(\sH,<)$.
We write $\id$ for the identity on $\{1,\ldots,\ell\}$.
We note that $f_{\id}$ is idempotent, by the previous paragraph.
Thus, it is either a projection or a semilattice operation.
We say that $f_\alpha$ is $\lor$ if $f_\alpha(E,N)=f_\alpha(N,E)=E$ and it is $\land$ otherwise.
If $f_\alpha$ is the first projection for all $\alpha$, then we are done as $f$ then acts like the first projection on $\en$; similarly, if $f_\alpha$ is $\lor$ for all $\alpha$, or if $f_\alpha$ is $\land$ for all $\alpha$, then $f$ acts like a semilattice on $\en$.}%

{Suppose that $f_{\id}$ is a projection. Up to replacing $f$ by $g(x,y):=f(y,x)$ if $f_{\id}$ is the second projection, we can assume that $f_{\id}$ is the first projection.
Define $g(x,y):=f(x,f(x,y))$. This is an injective operation that is canonical with respect to $(\sH,<)$.
Moreover, for every permutation $\alpha$ and $\tuple s\in J,\tuple t\in \alpha(J)$, we have $f(\tuple s,\tuple t)\in J$ since $f$ acts lexicographically on the order.
Thus, for all orbits $O,O'\in\en$, $g_\alpha(O,O') = f_{\id}(O,f_\alpha(O,O')) = O$ and therefore all $g_\alpha$ are the first projection, and we are done as $g$ acts like the first projection on $\en$.}

{Suppose now that $f_{\id}$ is a semilattice; without loss of generality, assume it is $\lor$.
We first show that for all $\alpha$, we must have $f_\alpha(N,N)=N$.
Indeed, suppose that there exists $\tuple s_0\in N\cap J$ and $\tuple t_0\in N\cap\alpha(J)$ such that $f(\tuple s_0,\tuple t_0)\in E$.
We define an endomorphism $e$ of $\sA$ with the property that the range of $e$ does not intersect $N$, a contradiction with $\sA$ not being a first-order reduct of $(H;=)$.
By compactness, it suffices to show that for every finite set $F\subseteq J$ there exists an order-preserving endomorphism $e_F$ such that $e_F(\tuple s)\in E$ for all $\tuple s\in F$.
One can take the identity on $H$ for $e_\emptyset$.
Suppose that $e_F$ is already defined and $\tuple s\not\in F$.
If $e_F(\tuple s)\in E$, there is nothing to do and we can define $e_{F\cup\{\tuple s\}}:=e_F$.
Otherwise, let $\beta,\gamma$ be automorphisms of $\sH$ such that {$\beta$ is order-preserving,} $\beta e_{{F}}(\tuple s)=\tuple s_0$ and $\gamma e_{{F}}(\tuple s)=\tuple t_0$.
Then let $e_{F\cup\{\tuple s\}}(x) := f(e_{{F}}(x),f(\beta e_{{F}}(x),\gamma e_{{F}}(x)))$.
We have $e_{F\cup\{\tuple s\}}(\tuple s)= f(e_{{F}}(\tuple s),f(\tuple s_0,\tuple t_0))$. Since $e_{{F}}(\tuple s)\in J, f(\tuple s_0,\tuple t_0)\in E\cap J$ and $f_{\id}$ is $\lor$, we get that $e_{F\cup\{\tuple s\}}(\tuple s)\in E$.
Moreover, for any $\tuple t\in F$, we have $e_{F\cup\{\tuple s\}}(\tuple t)\in E$ since $e_{{F}}(\tuple t)\in E\cap J$ and $f(\beta e_{{F}}(\tuple t),\gamma e_{{F}}(\tuple t))\in J$.
This concludes the proof of our claim that $f_\alpha(N,N)=N$ for all $\alpha$.}

{Now, we use another local argument to obtain a new operation $g$ with the property that $g_\alpha$ is $\lor$ for all $\alpha$.
For this, it suffices to show that for every finite set $F\subseteq J\times I_\ell$ of pairs of injective tuples, there exists an injective binary polymorphism $g_F$ acting lexicographically on the order and such that for all $(\tuple s,\tuple t)\in F$, the orbit of $g_F(\tuple s,\tuple t)$ under $\Aut(\sH)$ is according to $\lor$.
We can set $g_\emptyset$ to be $f$.
Suppose that $g_F$ is already defined and $(\tuple s,\tuple t)\not\in F$.
Define $g_{F\cup\{(\tuple s,\tuple t)\}}(x,y):=f(g_F(x,y),f(x,\beta y))$, where $\beta$ is any automorphism of $\sH$ such that $\beta\tuple t\in J$.
Note that for any pair of injective tuples $\tuple s',\tuple t'$, $g_F(\tuple s',\tuple t')$ and $f(\tuple s',\beta\tuple y')$ are ordered like $\tuple s'$.
Moreover, if $\tuple s',\tuple t'\in N$ then by the previous paragraph we have $g_F(\tuple s',\tuple t'),f(\tuple s',\beta\tuple t')\in N$ and therefore $g_{F\cup\{(\tuple s,\tuple t)\}}(\tuple s',\tuple t')\in N$.
If $(\tuple s',\tuple t')\in F$ and one of them is in $E$, then $g_F(\tuple s',\tuple t')\in E$ and therefore also $g_{F\cup\{(\tuple s,\tuple t)\}}(\tuple s',\tuple t')\in E$.
Finally, if one of $\tuple s,\tuple t\in E$, then $f(\tuple s,\beta\tuple t)\in E$ and therefore $g_{F\cup\{(\tuple s,\tuple t)\}}(\tuple s,\tuple t)\in E$.}
\end{proof}

A combination of \Cref{lemma:binary_essential} and \Cref{lemma:essential_injection} immediately yields the following proposition.

\begin{restatable}{proposition}{canonicalinjection}\label{prop:canonical_injection}
Let $\sA$ be a first-order reduct of $\sH$ that is a model-complete core.
If\/ $\CA$ does not have a uniformly continuous clone homomorphism to $\Projs$, then $\CAH$ contains a binary injection $f$ { canonical with respect to $(\sH,<)$} such that one of the following holds:

\begin{itemize}
    \item $f$ acts like a semilattice operation on $\en$, or
    \item $f$ acts like a projection on $\en$.%
\end{itemize}
\end{restatable}

\begin{lemma}\label{lexicographic-projection}
	{Suppose that $\sA$  is a first-order reduct of $\sH$ such that $\CAH$  contains a binary injective operation $f$ that %
    acts like a projection on $\en$.
	Then for all $n\geq {1}$ and all $i\in\{1,\dots,{n}\}$, $\CAH$ contains an $n$-ary injective operation $p^n_i$ that is canonical with respect to $(\sH,<)$, that acts like the $i$th projection on $\en$, and that acts lexicographically on the order.}
\end{lemma}
\begin{proof}
{For $n=1$, we can set $p^1_1(x):=x$; let now $n\geq 2$.}
{Assume without loss of generality that $f$ acts like the first projection on $\en$. Since $(\sH,<)$ is a homogeneous Ramsey structure, $f$ interpolates a canonical function with respect to $(\sH,<)$, so we may assume it is itself canonical. 
Let $f^n_1\in\CA$ be the $n$-ary operation defined by 
\[f^n_1(x_1,\dots,x_n)=f(x_1,f(x_{2},\dots f(x_{n-2},f(x_{n-1},x_{n}))\dots))\; .\]
 This operation $f^n_1\in\CA$ is injective and acts like the first projection. It is also canonical with respect to $(\sH,<)$ since $f$ is. By permuting its variables, one obtains for every $i$ a function $f^n_i\in\CA$ which acts like the $i$th projection. By~\Cref{obs:lexicographic}, $f^n_i$ can be composed with a self-embedding of $\sH$ so that it additionally acts lexicographically on the order.}

\end{proof}

{The projection-like operations obtained in~\Cref{lexicographic-projection} are used to ``convert'' an operation $f$ that is inj-canonical %
with respect to $(\sH,<)$ into an operation $f'$ that is inj-canonical with respect to $\sH$.
This is then used for our proof using smooth approximation (\Cref{lem:existssmoothapprox} and~\Cref{lemma:weaklycomm-semilattice}) to make up for the fact that not every polymorphism of $\sA$ locally interpolates an operation that is canonical with respect to $\sH$.
This technique is a new contribution to the theory of smooth approximation; in the other works in this line of research, this step is unnecessary as local interpolation of suitable canonical operations was in those settings possible (see, e.g., \cite[Lemma 34, Lemma 53]{SmoothApproximationsJournal}).}
\begin{lemma}\label{canonicity-projection}
    {Let $p_1,\dots,p_n$ be $n$-ary injective operations that are canonical with respect to $(\sH,<)$, that act lexicographically on the order, and where $p_i$ acts like the $i$th projection on $\en$ for all $i\in\{1,\dots,n\}$.
    Let $f\colon H^n\to H$ be an injective operation that is inj-canonical with respect to $(\sH,<)$ on a set $%
    {T}\subseteq H^n$ such that $(p_1(\tuple a),\dots,p_n(\tuple a))\in T$ for all $\tuple a\in T$.
    Then the operation $f\circ(p_1,\dots,p_n)$ is inj-canonical with respect to $\sH$ on $T$.}
\end{lemma}
\begin{proof}
    {Let $f'=f\circ(p_1,\dots,p_n)$. %
    Note that $f'$ is inj-canonical with respect to $(\rel H,<)$ on $T$, since $f$ and $p_1,\dots,p_n$ are and by the condition that $(p_1(\tuple a),\dots,p_n(\tuple a))\in T$ for all $\tuple a\in T$.
We show that it does not depend on the ordering for injective tuples.}

{Let $\tuple a^1,\dots,\tuple a^n\in {I_k}, %
\alpha_1,\dots,\alpha_n\in\Aut(\sH)$ be such that $(a^1_i,\dots,a^n_i)$ and $(\alpha_1a^1_i,\dots,\alpha_na^n_i)$ are in $T$ for all $i\in\{1,\dots,k\}$.
We see $(\tuple a^1,\dots,\tuple a^n)$ and $(\alpha_1\tuple a^1,\dots,\alpha_n\tuple a^n)$ as the columns of $k\times n$ matrices whose rows all belong to $T$.
By the fact that $(p_1(\tuple a),\dots,p_n(\tuple a))\in T$ for all $\tuple a\in T$, we get that the rows of the matrix $M$ with columns $(p_1(\tuple a^1,\dots,\tuple a^n),\dots,p_n(\tuple a^1,\dots,\tuple a^n))$ are all in $T$,
and similarly for the matrix $N$ with columns $(p_1(\alpha_1\tuple a^1,\dots,\alpha_n\tuple a^n),\dots,p_n(\alpha_1\tuple a^1,\dots,\alpha_n\tuple a^n))$.
Note that the columns of $M$ are all ordered (in the sense of the order on $(\sH,<)$) like $\tuple a^1$ since the operations $p_1,\dots,p_n$ act lexicographically on the order.
Similarly, the columns of $N$ are all ordered like $\alpha_1\tuple a^1$.}

	{Let $\sigma$ be a permutation of $\{1,\dots,k\}$.
	For an arbitrary $k$-tuple $\tuple a$, define $\tuple a^\sigma:=(a_{\sigma(1)},\dots,a_{\sigma(\ell)})$.
	Then since the injective orbits of $\sH$ are invariant under permutations of the coordinates, we obtain that $f'(\tuple a^1,\dots,\tuple a^n)$ and $f'((\tuple a^1)^\sigma,\dots,(\tuple a^n)^\sigma)$ are in the same orbit under $\Aut(\sH)$ for every permutation $\sigma$.
	Choose such a $\sigma$ for which $(\tuple a^1)^\sigma$ and $\alpha_1\tuple a^1$ are in the same orbit in $\Aut(\sH,<)$ (i.e., for which they are ordered similarly).}
	
	{Let now $M^\sigma$ be the matrix obtained from $M$ by permuting the rows according to $\sigma$.
	Each column of $M^\sigma$ is in the same orbit as the corresponding column of $N$ under $\Aut(\sH,<)$.
	Indeed, the column $i$ in $N$ is $p_i(\alpha_1\tuple a^1,\dots,\alpha_n\tuple a^n)$, which is in the orbit of $\tuple a^i$ under $\Aut(\sH)$ since $p_i$ acts like the $i$th projection on $\en$,
	and it is ordered like $\alpha_1\tuple a^1$ since $p_i$ acts lexicographically on the order.
	The $i$th column of $M^\sigma$ is $p_i((\tuple a^1)^\sigma,\dots,(\tuple a^n)^\sigma)$, which is in the orbit of $(\tuple a^i)^\sigma$ under $\Aut(\sH)$ and ordered like $(\tuple a^1)^\sigma$ for the same reasons.
	By the choice of $\sigma$, we get that these two columns are in the same orbit under $\Aut(\sH)$.}
	
	{Thus, the result of applying $f$ to $M^\sigma$ and $N$ rowwise, which are respectively the tuples $f'((\tuple a^1)^\sigma,\dots,(\tuple a^n)^\sigma)$ and $f'(\alpha_1\tuple a^1,\dots,\alpha_n\tuple a^n)$,
	are in the same orbit under $(\sH,<)$ since $f$ is canonical with respect to $(\sH,<)$ on $T$. In particular they are in the same orbit under $\Aut(\sH)$, and therefore $f'(\tuple a^1,\dots,\tuple a^n)$ and $f'(\alpha_1\tuple a^1,\dots,\alpha_n\tuple a^n)$ are in the same orbit under $\Aut(\sH)$, which we wanted to show.}
\end{proof}

\section{The NP-hard case}\label{sect:NP-hard}
In this section, we prove that if $\sA$ is a first-order reduct of $\sH$ which is a model-complete core, and such that $\CAH\actson\en$ is equationally trivial, then $\CA$ has a uniformly continuous clone homomorphism to the clone of projections $\Projs$.
We make a case distinction based on which case of \Cref{thm:1st_loop_lemma} applies.

\subsection{The case when there exists a very smooth approximation}\label{seubsect:case1}

In this section, we prove that if the first case of~\Cref{thm:1st_loop_lemma} applies, $\Csp(\sA)$ is NP-hard.

\begin{restatable}{lemma}{inclusionCAA}\label{lemma:CAA_eq-trivial}
Let $\sA$ be a first-order reduct of\/ $\sH$ that is a model-complete core.  
Suppose that $\CAH\actson\en$ is equationally trivial. Then so is $\CAA$. Moreover, $\CAH\subseteq\CAA$.
\end{restatable}
\begin{proof}
Suppose that $\CAA$ is equationally non-trivial. Then $\CA$ contains a binary injection by~\Cref{prop:canonical_injection}.
If this injection acts as a semilattice operation on $\en$, then it witnesses that $\CAH\actson\en$ is equationally non-trivial; we can therefore assume that this is not the case.

We claim that $\CAA$ has an operation $s$ which satisfies the pseudo-Siggers identity modulo $\Aut(\sA)$ on injective tuples{, i.e., for any $k\geq 1$ and for any injective tuples $\tuple x, \tuple y,\tuple z\in I_k$, the tuples $s(\tuple x,\tuple y,\tuple z,\tuple x,\tuple y,\tuple z)$ and $s(\tuple y,\tuple x,\tuple x,\tuple z,\tuple z,\tuple y)$ lie in the same orbit under $\Aut(\sA)$.} To see this, consider the actions of $\CAA$ on $I_n/{\Aut(\sA)}$ for all $n\geq 1$. Each of these actions has a Siggers operation $s_n$ {by~\cite{Siggers}}, and we can assume that the sequence $(s_n)_{n\geq 1}$ converges in $\CAA$ to a function $s$, which has the desired property.

Let $s'$ be a polymorphism of $\sA$ canonical with respect to $(\sH,<)$ that is locally interpolated by $s$ modulo $\Aut(\sH,<)$.
Then on all injective tuples $\tuple x,\tuple y,\tuple z$ of the same length, $s'$ satisfies the pseudo-Siggers %
{identity} on $\tuple x,\tuple y,\tuple z$ modulo $\Aut(\sA)$ {by the same argument as in the proof of Lemma 35 in~\cite{SmoothApproximations} which we provide for the convenience of the reader.} %

{Let us fix the tuples $\tuple x,\tuple y,\tuple z$; as $s'$ is locally interpolated by $s$ modulo $\Aut(\sH,<)$, there exists $\beta,\alpha_1,\dots,\alpha_6\in \Aut(\sH,<)$ such that
\begin{equation}\label{eq:locapp}
    s'(x_1,\dots,x_6)=\beta s(\alpha_1 x_1,\dots,\alpha_6 x_6)    
\end{equation}
for all $x_1,\dots,x_6$ from the set of all entries of $\tuple x,\tuple y,\tuple z$.} 
{Since $\tuple x,\tuple y$ and $\tuple z$ are injective and $s\in\CAA$, we can find $\gamma,\gamma'\in\Aut(\sA)$ so that
\begin{equation}\label{eq:can1}
    \gamma s(\tuple x,\tuple y,\tuple z,\tuple x,\tuple y,\tuple z) = s(\alpha_1\tuple x,\alpha_2\tuple y,\alpha_3\tuple z,\alpha_4\tuple x,\alpha_5\tuple y,\alpha_6\tuple z)
\end{equation}
\begin{equation}\label{eq:can2}
    \gamma' s(\tuple y,\tuple x,\tuple x,\tuple z,\tuple z,\tuple y) = s(\alpha_1\tuple y,\alpha_2\tuple x,\alpha_3\tuple x,\alpha_4\tuple z,\alpha_5\tuple z,\alpha_6\tuple y).
\end{equation}}
{Finally, as $s$ satisfies the pseudo-Siggers identity modulo $\Aut(\sA)$ on injective tuples, there exists $\delta\in\Aut(\sA)$ such that
\begin{equation}\label{eq:Sigg}
    s(\tuple x,\tuple y,\tuple z,\tuple x,\tuple y,\tuple z) = \delta s(\tuple y,\tuple x,\tuple x,\tuple z,\tuple z,\tuple y).
\end{equation}}
{We are now ready to prove that $s'$ satisfies the pseudo-Siggers identity on $\tuple x,\tuple y,\tuple z$. Indeed,
\begin{align*}
    s'(\tuple x,\tuple y,\tuple z,\tuple x,\tuple y,\tuple z) & \stackrel{(\ref{eq:locapp})}{=} \beta s(\alpha_1 \tuple x,\alpha_2 \tuple y,\alpha _3 \tuple z,\alpha_4 \tuple x,\alpha_5 \tuple y,\alpha_6 \tuple z)\\
    & \stackrel{(\ref{eq:can1})}{=} \beta\gamma s(\tuple x,\tuple y,\tuple z,\tuple x,\tuple y,\tuple z) \\
    & \stackrel{(\ref{eq:Sigg})}{=} \beta\gamma\delta s(\tuple y,\tuple x,\tuple x,\tuple z,\tuple z,\tuple y)\\
    & \stackrel{(\ref{eq:can2})}{=} \beta\gamma\delta(\gamma')^{-1} s(\alpha_1 \tuple y,\alpha_2 \tuple x,\alpha _3 \tuple x,\alpha_4 \tuple z,\alpha_5 \tuple z,\alpha_6 \tuple y)\\
    & \stackrel{(\ref{eq:locapp})}{=} \beta\gamma\delta(\gamma')^{-1}\beta^{-1} s'(\tuple y,\tuple x,\tuple x,\tuple z,\tuple z,\tuple y)\\
\end{align*}}
{and $\beta\gamma\delta(\gamma')^{-1}\beta^{-1} \in \Aut(\sA)$ as $\Aut(\sH,<)\subseteq \Aut(\sH) \subseteq \Aut(\sA)$.}
{Now, for all $1\leq i\leq 6$, let $p_i^6$ be a $6$-ary function that is canonical with respect to $(\sH,<)$, acting as the $i$th projection on $\en$, and acting lexicographically on the order, obtained by~\Cref{lexicographic-projection}.}

Set $s'':=s'(p_1^6(x_1,\ldots,x_6),\ldots,p_6^6(x_1,\ldots,x_6))$.
{By~\Cref{canonicity-projection} (applied in the case where $T$ is the full set), we get that $s''\in\CAH$}.
Moreover, $s''$ also satisfies the pseudo-Siggers equation modulo $\Aut(\sA)$ on injective tuples.
Say without loss of generality that $s''$ acts as the first projection on $\en$; the case where it flips $E$ and $N$ of the first coordinate is similar.
Let $\tuple a,\tuple b$ be injective increasing tuples in distinct orbits under $\Aut(\sA)$ (such tuples only exist if $\sA$ is not a first-order reduct of $(H;=)$, but we already know from~\Cref{lemma:binary_essential,lemma:essential_injection} that in those cases, $\Pol(\sA)$ consists of essentially unary operations only and thus the result is also true in this case).
Since $s''$ acts as the first projection on $\en$, we have that $\tuple a$ and $s''(\tuple a,\tuple b,\tuple a,\tuple b,\tuple b,\tuple b)$ are in the same orbit under $\Aut(\sH)$, and similarly $\tuple b$ and $s''(\tuple b,\tuple a,\tuple b,\tuple a,\tuple b,\tuple b)$ are in the same orbit under $\Aut(\sH)$.
However, we also know that $s''(\tuple a,\tuple b,\tuple a,\tuple b,\tuple b,\tuple b)$ and $s''(\tuple b,\tuple a,\tuple b,\tuple a,\tuple b,\tuple b)$ are in the same orbit under $\Aut(\sA)$ since $\tuple a,\tuple b$ are injective.
Thus, $\tuple a$ and $\tuple b$ are in the same orbit under $\Aut(\sA)$, a contradiction.

The proof that $\CAH\subseteq\CAA$ is almost identical as the proof of Lemma 32 in~\cite{SmoothApproximations} but we provide it for the convenience of the reader. Observe that the action $\CAH\actson\en$ is by essentially unary functions by~\cite{Post}. Moreover, the action $\CAH\actson\en$ determines the action of $\CAH$ on orbits of injective tuples under $\Aut(\sH)$ since $\sH$ is homogeneous in an $\ell$-ary language. Hence, for every $f\in\CAH$ of arity $n\geq 1$, there exists $1\leq i\leq n$ such that the orbit of $f(\tuple a_1,\ldots,\tuple a_n)$ under $\Aut(\sH)$ is either equal to the orbit of $\tuple a_i$ under $\Aut(\sH)$ for all injective tuples $\tuple a_1,\ldots,\tuple a_n$ of the same length, or the orbit of $f(\tuple a_1,\ldots,\tuple a_n)$ under $\Aut(\sH)$ is equal to the orbit under $\Aut(\sH)$ of a tuple obtained from $\tuple a_i$ by changing hyperedges to non-hyperedges and vice versa. In the first case, $f\in\CAA$ since the orbits under $\Aut(\sH)$ are refinements of the orbits under $\Aut(\sA)$. {In the second case, $\sA$ has an endomorphism $e$ changing hyperedges into non-hyperedges and vice versa, namely  $e(x):=f(x,\dots,x)$. Hence, $e\circ f\in\CAA$, and since $\sA$ is a model-complete core and thus $e$ locally invertible by automorphisms, also $f\in\CAA$.}
\end{proof}

The following is an adjusted version of the fundamental theorem of smooth approximations~{\cite[Theorem 13]{SmoothApproximations}}. The original version of this theorem does not apply to our case since for a first-order reduct $\sA$ of $\sH$, we do not have that $\CA$ locally interpolates $\CAA$ modulo $\Aut(\sH)$.
{We first extract the condition that allows for the proof of~\cite[Theorem 13]{SmoothApproximations} to work more generally, and after this show that this condition holds for the reducts of $\sH$ under consideration.}

\begin{theorem}\label{generalized-fundamental}
	{Let A be a set. Let $\fC,\fD$ be function clones on $A$.
	Let $(S,\sim)$ be a subfactor of $\fC$, and let $\eta$ be a $\fD$-invariant approximation of $\sim$.
	Let $\phi\colon\fD\to\fC$ be any arity-preserving map such that for all $u_1,\dots,u_m\in S$ and all $m$-ary $f\in\fD$, one has
	\[ f(u_1,\dots,u_m) \mathrel{(\eta\circ{\sim})} \phi(f)(u_1,\dots,u_m).\]
	Then the map $\fD\to \fC\actson S/{\sim}$ defined by sending $f\in\fD$ to the action of $\phi(f)$ on $S/{\sim}$ is a clone homomorphism.
	If $\sim$ has finite index, then this clone homomorphism is uniformly continuous.}
\end{theorem}
\begin{proof}
	{The map in the statement preserves arities, so it remains to show that it preserves compositions as well.
	Let $n,m\geq 1$, let $f\in\fD$ be $m$-ary, let $g_1,\ldots,g_m\in\fD$ be $n$-ary, and let $u_1,\ldots, u_n\in S$.
	Since $g_i(u_1,\dots,u_n) \mathrel{(\eta\circ{\sim})} \phi(g_i)(u_1,\dots,u_n)$ holds, there exists $v_i$ such that $g_i(u_1,\dots,u_n)\;\eta\; v_i$ and $v_i\sim \phi(g_i)(u_1,\dots,u_n)$.
	Note that $v_i\in S$ since $\sim$ is a relation on $S$. %
Then
\begin{align*}
f(g_1(u_1,\dots,u_n),\ldots,g_m(u_1,\dots,u_n))&\; \eta\; f(v_1,\ldots,v_m)\\
	&\mathrel{(\eta\circ{\sim})}  \phi(f)(v_1,\ldots,v_m) \\
	&\sim \phi(f)(\phi(g_1)(u_1,\dots,u_n),\ldots,\phi(g_m)(u_1,\dots,u_n)).
\end{align*}
where the first equivalence is due to the fact that $\eta$ is $\fD$-invariant, the second is by the assumption on $\phi$, and the last one is due to the fact that $\phi(g_i)(u_1,\dots,u_n)\sim v_i$ for all $i\in\{1,\dots,n\}$, that $\phi(f)\in\fC$, and that $\sim$ is $\fC$-invariant.
Overall, this gives us \[(f\circ(g_1,\dots,g_m))(u_1,\dots,u_n) \;(\eta\circ{\sim})\; (\phi(f)\circ(\phi(g_1),\dots,\phi(g_m)))(u_1,\dots,u_n).\]
Using the assumption on $\phi$ once again, we also get 
\[(f\circ(g_1,\dots,g_m))(u_1,\dots,u_n)\;(\eta\circ{\sim})\; \phi(f\circ(g_1,\dots,g_m))(u_1,\dots,u_n).\]
Combining the two and recalling that $\eta\circ\eta\subseteq \eta$, we obtain \[\phi(f\circ(g_1,\dots,g_m))(u_1,\dots,u_n) \; ({\sim}\circ \eta\circ{\sim})\; (\phi(f)\circ (\phi(g_1),\dots,\phi(g_m)))(u_1,\dots,u_n).\]
Finally, we have ${\sim}\circ\eta \circ{\sim}\subseteq{\sim}$  since $\eta$ is an approximation of $\sim$ and therefore
\[\phi(f\circ(g_1,\dots,g_m))(u_1,\dots,u_n) \sim (\phi(f)\circ (\phi(g_1),\dots,\phi(g_m)))(u_1,\dots,u_n).\]
It follows that $\phi(f\circ(g_1,\dots,g_m))$ and $\phi(f)\circ(\phi(g_1),\dots,\phi(g_m))$ induce the same function on $S/{\sim}$.
This concludes the proof that $f\mapsto\phi(f)\actson S/{\sim}$ is a clone homomorphism.}

{To show that it is uniformly continuous if $\sim$ has finite index, let $s_1,\dots,s_k\in S$ be a system of representatives for $\sim$.
Then for all $n\geq 1$, the finite set $F=\{s_1,\dots,s_k\}^n$ determines the image of any $f\in\fD$ under the clone homomorphism, which is therefore uniformly continuous.}
\end{proof}

\begin{restatable}{lemma}{fundamentaltheorem}\label{lem:existssmoothapprox}
Let $\sA$ be a first-order reduct of\/ $\sH$ that is a model-complete core. 
Assume that $\CAH\subseteq \CAA$, and that $\CAH$ does not contain a binary function that acts as a semilattice operation on $\en$. Let $k\geq 1$ be so that $I_k/{\Aut(\sA)}$ has at least two elements, and let $(S,\sim)$ be a subfactor of $\CAA\actson I_k$ with $\Aut(\sA)$-invariant equivalence classes.
Assume moreover that $\eta$ is a very smooth approximation of $\sim$ which is invariant under $\CA$. Then $\CA$ has a uniformly continuous 
clone homomorphism to $\CAA\actson S/{\sim}$.
\end{restatable}

\begin{proof}
{If $\CA$ has a uniformly continuous clone homomorphism to the clone of projections, then it has such a homomorphism to any clone; hence we may assume otherwise. Then~\Cref{prop:canonical_injection} and~\Cref{lexicographic-projection} imply that $\CA$ contains 
for all $n\geq 1$ and all $1\leq i\leq n$ an $n$-ary injection  $p_i^n$ which is canonical with respect to $(\sH,<)$, acts like the $i$th projection on $\en$, and  acts lexicographically on the order.} %
{We define a map $\phi$ from $\CA$ to $\CAA$ and show that it satisfies the condition of~\Cref{generalized-fundamental} {(formally, we apply the theorem to the actions of those clones on $I_k$ and the map that $\phi$ induces between those actions)}.}
For every $n\geq 1$ and every $n$-ary $f\in \CA$, we take any function $f'$ canonical on $(\sH,<)$ which is locally interpolated by $f$ modulo $\Aut(\sH,<)$, and then set {$\phi(f):=f'\circ (p_1^n,\ldots,p_n^n)$. This is an element of $\CAH$ by~\Cref{canonicity-projection} applied with $T=H^n$.}

{We now show that $f(\tuple u_1,\dots,\tuple u_n) \; {(\eta\circ{\sim})}\; \phi(f)(\tuple u_1,\dots,\tuple u_n)$ holds for all $\tuple u_1,\dots,\tuple u_n\in S$.
This is done similarly as in~\cite[Theorem 13]{SmoothApproximations}. 
Let $\tuple u_1,\dots,\tuple u_n\in S$.
Let $\alpha,\beta_1,\dots,\beta_n\in\Aut(\sH,<)$ be such that
\[f'(p_1^n(\tuple u_1,\dots,\tuple u_n),\dots,p_n^n(\tuple u_1,\dots,\tuple u_n))=\alpha f(\beta_1p_1^n(\tuple u_1,\dots,\tuple u_n),\dots,\beta_np_n^n(\tuple u_1,\dots,\tuple u_n)),\]
which exist since $f'$ is locally interpolated by $f$.
Since $p_i^n$ acts like the $i$th projection on $\en$ for all $i\in\{1,\dots,n\}$ and since $S\subseteq I_k$, we have that $\beta_ip_i^n(\tuple u_1,\dots,\tuple u_n)$ and $\tuple u_i$ are in the same orbit under $\Aut(\sA)$ (even under $\Aut(\sH)$) for all $i\in\{1,\dots,n\}$.
Since $\eta$ is very smooth with respect to $\Aut(\sA)$, we get $\beta_ip_i^n(\tuple u_1,\dots,\tuple u_n) \mathrel{\eta}\tuple u_i$ and since $\eta$ is $\Pol(\sA)$-invariant, we obtain
\begin{align*}
f(\tuple u_1,\dots,\tuple u_n) &\mathrel{\eta}  f(\beta_1p_1^n(\tuple u_1,\dots,\tuple u_n),\dots,\beta_np_n^n(\tuple u_1,\dots,\tuple u_n))\\
&\sim\alpha f(\beta_1p_1^n(\tuple u_1,\dots,\tuple u_n),\dots,\beta_np_n^n(\tuple u_1,\dots,\tuple u_n))\\
&=  f'(p_1^n(\tuple u_1,\dots, \tuple u_n),\dots,p_n^n(\tuple u_1,\dots,\tuple u_n))\\
&=\phi(f)(\tuple u_1,\dots, \tuple u_n),\end{align*}
where the second statement above follows from the fact that $\phi(f)(\tuple u_1,\dots,\tuple u_n)$ is in $S$, which means that $\alpha f(\beta_1p_1^n(\tuple u_1,\dots,\tuple u_n),\dots,\beta_np_n^n(\tuple u_1,\dots,\tuple u_n))$ is in $S$, and from the fact that the classes of $\sim$ are invariant under $\Aut(\sA)$.
By~\Cref{generalized-fundamental}, we obtain a uniformly continuous clone homomorphism from $\Pol(\sA)$ to $\CAA\actson S/{\sim}$.}
\end{proof}

\subsection{The case when there exists a weakly commutative function}\label{subsect:case2}

In this section, we prove that if $\sA$ is a first-order reduct of $\sH$ which has among its polymorphisms a weakly commutative function, then $\CAH$ contains a binary function that acts as a semilattice operation on $\en$. This allows us to derive a contradiction if the second case of~\Cref{thm:1st_loop_lemma} applies.
To prove this, we need the following consequence from~\cite{BB21} of the fact that $(\sH,<)$ is a Ramsey structure.

Let $\sC$ be a relational structure, and let $A,B\subseteq C$. We say that \emph{$A$ is independent from $B$ in $\sC$} if for every $n\geq 1$ and for all $\tuple a_1,\tuple a_2\in A^n$ such that $\tuple a_1,\tuple a_2$ satisfy the same first-order formulas over $\sC$, $\tuple a_1$ and $\tuple a_2$ satisfy the same first-order formulas over $\sC$ with parameters from $B$. We say that two substructures $\sA$ and $\sB$ of $\sC$ are \emph{independent} if $A$ is independent from $B$ in $\sC$ and $B$ is independent from $A$ in $\sC$. Finally, we say that a substructure $\sA$ of $\sC$ is \emph{elementary} if the identity mapping from $A$ to $C$ preserves the truth of all first-order formulas in the signature of $\sC$.

\begin{theorem}[Theorem 4.4 in~\cite{BB21}]\label{thm:Independent}
    Let $\sC$ be a countable homogeneous $\omega$-categorical Ramsey structure. Then $\sC$ contains two independent elementary substructures.
\end{theorem}

\begin{lemma}\label{lemma:independent_injections}
    {Let $\sA$ be a first-order reduct of $\sH$, %
    and let us assume that $\CAH$ contains a binary injection  which acts as a projection on $\en$. %
     Then there exist $p_1,p_2\in \CAH$ with the following properties.}

\begin{itemize}
    \item $p_1,p_2$ are both canonical with respect to $(\sH,<)$,
    \item $p_i$ acts like the $i$th projection on $\en$ for all $i\in [2]$,
    \item $p_i$ acts lexicographically on the order for all $i\in[2]$,
    \item for all $x,y\in H$, if $x<y$, then $p_1(x,y)<p_2(x,y)$, and if $y<x$, then $p_1(x,y)>p_2(x,y)$,
    \item the ranges of $p_1$ and $p_2$ are disjoint and independent as substructures of $\sH$.
\end{itemize}
\end{lemma}

\begin{proof}
{By~\Cref{lexicographic-projection}, $\CAH$ contains for all $i\in [2]$ binary injections $g_i$  which are canonical with respect to $(\sH,<)$, act like the $i$th projection on $\{E,N\}$, and act lexicographically on the order. 
Moreover, %
{\Cref{thm:Independent} yields that there exist injective endomorphisms $e_1,e_2$ of $(\sH,<)$ whose ranges lie in different independent elementary substructures of $(\sH,<)$. Setting $g'_i(x,y):=e_i\circ g_i(x,y)$ for $i\in[2]$, we} %
obtain binary injections %
with the same properties and whose ranges are disjoint and induce in $(\sH,<)$ substructures that are independent.}

Let us define a linear order $<^{\ast}$ on $U:=\textrm{im}(g'_1)\cup\textrm{im}(g'_2)$ as follows. We set $u<^{\ast}v$ if one of the following holds.

\begin{itemize}
    \item $u<v$ and $u,v\in \textrm{im}(g'_i)$ for some $i\in[2]$, or
    \item $u=g'_i(x_1,y_1), v=g'_j(x_2,y_2)$ for some $i\neq j\in[2]$, $x_1,x_2,y_1,y_2\in H$, and one of the following holds
    \begin{itemize}
        \item $i=1, j=2, x_2\leq y_2$, and $u\leq g'_1(x_2,y_2)$, or
        \item $i=1, j=2, x_1\leq y_1$, and $g'_2(x_1,y_1)\leq v$, or
        \item $i=2, j=1, x_2> y_2$, and $u\leq g'_2(x_2,y_2)$, or
        \item $i=2, j=1, x_1> y_1$, and $g'_1(x_1,y_1)\leq v$.
    \end{itemize}
\end{itemize}

It is easy to verify that $<^{\ast}$ is a linear order on $U$ { which we extend arbitrarily  to $H$. Since the order $<$ is free over $\sH$, the structure $(\sH,<^\ast)$ embeds into $(\sH,<)$ via an embedding $e$. Setting $p_i:=e\circ g_i'$ for all $i\in[2]$ then finishes the proof.}
\end{proof}

\begin{lemma}\label{lemma:weaklycomm-semilattice}
    Let $\sA$ be a first-order reduct of\/ $\sH$ which is a model-complete core.
    Suppose that $\CA$ does not have a uniformly continuous clone homomorphism to $\Projs$ and contains a binary function $f$ such that there exist $k\geq 1$, $S\subseteq I_k$, and an equivalence relation $\sim$ on $S$ with at least two $\Aut(\sA)$-invariant classes such that $f(\tuple a,\tuple b)\sim f(\tuple b,\tuple a)$ for all disjoint injective tuples $\tuple a,\tuple b\in H^k$ with $f(\tuple a,\tuple b),f(\tuple b,\tuple a)\in S$. Then $\CAH\actson\en$ contains a semilattice operation.
\end{lemma}

\begin{proof}
Since $\sA$ is a model-complete core, we have $f(\tuple a,\tuple a)\in S$ for all $\tuple a\in S$, and hence picking any $\tuple b\in S$ which is not $\sim$-equivalent to $a$, we see on the values $f(\tuple a,\tuple a),f(\tuple b,\tuple b),f(\tuple a,\tuple b)$, and $f(\tuple b,\tuple a)$ that $f$ must be essential. It follows from~\Cref{prop:canonical_injection}  %
 and from %
\Cref{lemma:independent_injections} that we may assume that $\CA$ contains functions $p_1,p_2$ as in~\cref{lemma:independent_injections}, or else the conclusion %
 follows immediately. Moreover, since its distinctive property is stable under diagonal interpolation modulo $\Aut(\sH)$, we can assume that $f$ is diagonally canonical with respect to $(\sH,<)$.

{By the diagonal canonicity of $f$, it follows}
that {$f$ is inj-canonical with respect to $(\sH,<)$ on $\nabla=\{(x,y)\in H^2\mid x<y\}$}.
Let $c_1{<}c_2\in H$ be arbitrary, let $U:=\{p_1(d,c_2)\mid d\in H, d<c_1\}$, and let $V:=\{p_2(c_1,d)\mid d\in H, c_2<d\}$.
Hence, for every $u\in U, v\in V$, it holds that $u<v$. Note that for every $m\geq 1$ and for every $\tuple a\in H^m$, there exist $\tuple b\in U^m, \tuple c\in V^m$ such that $\tuple a,\tuple b,\tuple c$ are in the same orbit under $\Aut(\sH,<)$. Note moreover that if $m\geq 1$ and $(\tuple a,\tuple b)$ is any pair of $m$-tuples such that $(a_i,b_j)\in U\times V$ for all $i,j\in[m]$, then $p_1(a_i,b_j)<p_2(a_i,b_j)$ for every $i,j\in[m]$. Setting $g(x,y):=f(p_1(x,y),p_2(x,y))$, we obtain a function {that is inj-canonical with respect to $\sH$ on $U\times V$ by~\Cref{canonicity-projection} (applied with the set $T=\nabla$)}. Hence, the restriction of $g$ to $U\times V$ naturally acts on $\en$, and we may assume it does so as an essentially unary function as otherwise we are done. A similar statement holds for its restriction to $V\times U$. If one of the two mentioned essentially unary functions on $\en$ is not a projection, then $\Aut(\sA)$ contains a function flipping $E$ and $N$. It follows that whenever $\tuple a,\tuple b\in S$ are so that $(a_i,b_j)\in U\times V$ for all $i,j\in\{1,\ldots,k\}$, then $g(\tuple a,\tuple b)$ and $g(\tuple b,\tuple a)$ are elements of $S$. If $\tuple a,\tuple b$ are moreover both increasing, then $g(\tuple a,\tuple b)$ and $f(\tuple a,\tuple b)$, as well as $g(\tuple b,\tuple a)$ and $f(\tuple b,\tuple a)$, belong to the same orbit under $\Aut(\sA)$. Since $f(\tuple a,\tuple b)\sim f(\tuple b,\tuple a)$, the two essentially unary functions mentioned above cannot depend on the same argument, as witnessed by choosing $\tuple a,\tuple b\in S$ from distinct $\sim$-classes. Hence, we may assume that the one from $U\times V$ depends on the first argument and the other one on the second.

More generally, since $g(x,y)=f(p_1(x,y),p_2(x,y))$, the function $g$ has the property that for all injective $\ell$-tuples $\tuple a,\tuple b$, where $\tuple a$ is increasing, the type of their image under this function in $\sH$ only depends on the relations of $\sH$ on each of $\tuple a$ and on $\tuple b$, respectively, and on the order relation on pairs $(a_i,b_j)$, where $i,j\in\{1,\ldots,\ell \}$; this type is precisely the same type obtained when applying the function to $\tuple a,\tuple b'$, where $\tuple b'$ is obtained from $\tuple b$ by changing its order to be increasing.

In the following, for any pair $(\tuple a,\tuple b)$ of injective increasing $\ell$-tuples, we shall consider the set $B$ of all pairs $(\tuple a',\tuple b')$ of injective increasing $\ell$-tuples such that the order relation on $(a_i,b_j)$ and that on $(a_i',b_j')$ agree for all $i,j\in\{1,\ldots,\ell \}$; we call this set an \emph{increasing diagonal order type}. We then have by the above that $g$ acts naturally on $\en$ within each such set $B$. 

Set $h(x,y):=g(e_1\circ g(x,y),e_2\circ g(x,y))$, where $e_1,e_2$ are self-embeddings of $(\sH,<)$ which ensure that for all injective increasing $\tuple a,\tuple b$, the diagonal order type of $(\tuple a,\tuple b)$ is equal to that of $(e_1\circ g(\tuple a,\tuple b),e_2\circ g(\tuple a,\tuple b))$. Then $h$ acts idempotently or as a constant function in its action on $\en$ within each increasing diagonal order type $B$ as above. In particular, $h$ then acts like the first projection on $\en$ when restricted to $U\times V$, and like the second when restricted to $V\times U$. 

We now rule out the possibility that $h$ acts like a constant function on $\en$ within some increasing diagonal order type of injective $\ell$-tuples. 
In the following, we assign to every increasing diagonal order type $B$ of pairs of increasing injective $\ell$-tuples a number $n_B$ as follows: if $((x_1,\ldots,x_\ell),(y_1,\ldots,y_\ell))\in B$, then $n_B$ is the minimal number in $\{0,\ldots,\ell\}$ such that for all $n_B<i\leq\ell$, it holds that $x_i< y_i$, and if $i>1$, then moreover $y_{i-1}<x_i$.

Suppose that within some increasing diagonal order type $B$, we have that $h$ acts as a constant function on $\en$. Pick such $B$ such that $n_B$ is minimal. Assume without loss of generality that the constant value of $h$ on $B$ is $E$. We claim that $\sA$ has an endomorphism onto a clique, which contradicts the conjunction of the assumptions that $\sA$ is a model-complete core and not a reduct of $(H,=)$. We prove this claim by showing that any finite injective tuple can be mapped to a clique by an endomorphism of $\sA$. Suppose that $m\geq 1$ and that $(a_1,\ldots,a_m)$ is an injective $m$-tuple of elements of $H$ which does not induce a clique, i.e., there is a subtuple of length $\ell$ which is not an element of $E$. Then $m\geq \ell$, and we may assume without loss of generality that $(a_1,\ldots,a_\ell)\not\in E$. By applying a self-embedding $e_1$ of $\sH$, we obtain an increasing tuple $(x_1,\ldots,x_m)$ such that $(x_1,\ldots,x_\ell)\in N$. Applying an appropriate self-embedding $e_2$ of $\sH$ to $(a_1,\ldots,a_m)$, we moreover obtain an increasing tuple $(y_1,\ldots,y_m)$ such that $((x_1,\ldots,x_\ell),(y_1,\ldots,y_\ell))\in B$ and such that $y_{i-1}<x_i<y_i$ for all $i>\ell$. Then the increasing diagonal order type $C$ for any pair $((x_{i_1},\ldots,x_{i_\ell}),(y_{i_1},\ldots,y_{i_\ell}))$ of subtuples, both increasing, has the property that $n_C\leq n_B$ (note that in order to compute $n_C$, the entries of both tuples receive the indices $1$ to $\ell$). If $n_C<n_B$, then $h$ acts idempotently on $\en$ within $C$, by the minimality of $n_B$; if on the other hand $n_C=n_B$, then $B=C$ and $h$ acts as a constant with value $E$. It follows that applying the endomorphism $h(e_1(x),e_2(x))$ to $(a_1,\ldots,a_m)$, one obtains a tuple which has strictly more subtuples in $E$ than $(a_1,\ldots,a_m)$. The claim follows.

We may thus henceforth assume that within each increasing diagonal order type $B$, we have that $h$ acts in an idempotent fashion on $\en$. Thus, within each such type,
$h$ acts as a semilattice operation or as a projection on $\en$. 

Let $0\leq j\leq \ell$, and consider the diagonal order type $T$ given by a pair $(\tuple c,\tuple d)$ of increasing injective $j$-tuples of elements of $H$; we call $j$ the \emph{length} of $T$. In the following, we shall say that $T$ is \emph{categorical} if for all pairs $(\tuple a,\tuple b)$ of increasing injective $\ell$-tuples which extend $(\tuple c,\tuple d)$ (we mean any extension, not just end-extension) 
the corresponding diagonal order type is one where $h$ behaves like the first projection on, or if a similar statement holds for the second projection, or for the semilattice behaviour (both semilattice behaviours are considered the same here). Note that for length $j=0$, every $T$ is non-categorical, by the behaviours on $U\times V$ and $V\times U$. Note also that for length $j=\ell$, every $T$ is trivially categorical. We claim that there exists $T$ of length $j=\ell-1$ which is not categorical. Suppose otherwise, and take any $T$ which is categorical and implies the behaviour as the second projection; this exists by the behaviour on $V\times U$. Let $(\tuple c,\tuple d)$ be a pair of injective $j$-tuples which are ordered such that they represent the diagonal order type $T$. Let $(\tuple a_1,\tuple b_1)$ be obtained from $(\tuple c,\tuple d)$ by extending both increasing tuples by a single element $c'$ and $d'$ at the end, respectively, in such a way that $c'<d'$. Let $(\tuple c_1,\tuple d_1)$ be tuple obtained from $(\tuple a_1,\tuple b_1)$ by taking away the first components. Then the order type represented by $(\tuple c_1,\tuple d_1)$ is categorical but not for the first projection. We continue in this fashion until we arrive at a pair $(a_\ell,b_\ell)$ in $U\times V$, a contradiction.

In the following, we assume that there exists $T$ of length $j=\ell-1$ which extends to diagonal order types where $h$ behaves like different projections; the other case (projection + semilattice) is handled similarly. We show by induction on $m\geq 1$ that for all tuples $\tuple a, \tuple b\in I^m$ such that for every $i\in\{1,\dots,m\}$, at most one of the tuples $(a_i, b_i)$ is in $N$,
there exists $u\in\CA$ such that $u(\tuple a, \tuple b)\in E^m$. A standard compactness argument then implies that $\CAH$ contains a function such that $\CAH\actson\en$ is a semilattice operation.

The base case $m=1$ is clearly achieved by applying an appropriate projection. For the induction step, let $\tuple a,\tuple b\in I^{m}$ for some $m\geq 2$. Since $\CA$ contains $p_1$, we may assume that the kernels of $\tuple a$ and $\tuple b$ are identical. We may then also assume that $a_i, a_j$ induce distinct sets whenever $1\leq i, j\leq m$ and $i\neq j$, for otherwise we are done by the induction hypothesis. By the induction hypothesis, we may assume that all components of $\tuple a$ are in $E$ except for the second, and all components of $\tuple b$ are in $E$ except for the first. It is sufficient to show that there exists an increasing diagonal order type on the pair $(\tuple a,\tuple b)$ (i.e., increasing diagonal order types for each of the pairs $(a_1,b_1),\ldots,(a_m,b_m)$ which are consistent with the kernels of $\tuple a$ and $\tuple b$) such that $h$ behaves like the first projection within the order type of $(a_1,b_1)$ and like the second within the order type of $(a_2,b_2)$. This, however, is obvious by our assumption.
\end{proof} 
\section{The power of Local Consistency for Hypergraph-SAT}\label{sect:main}\label{sect:bwidth}

In this section, we characterise those sets $\Psi$ of $\ell$-hypergraph formulas for which the problem $\ell$-Hypergraph-SAT($\Psi$) is solvable by local consistency methods, in case that $\Psi$ contains the formulas $E(x_1,\dots,x_\ell)$.
In the parlance of the previous sections, this is the same as a characterization of those first-order \emph{expansions} $\sA$ of $\sH$ such that $\CSP(\sA)$ has bounded width, where a first-order expansion of $\sH$ is a first-order reduct of $\sH$ having $E$ as one of its relations.

In the following, recall that $b_{\sH}$ is a constant such that a finite $\ell$-hypergraph embeds into $\sH$ if and only if all its induced sub-hypergraphs of size at most $b_{\sH}$ do.

\begin{theorem}\label{thm:hypergraphs_bwidth}
    Let $\sA$ be a first-order expansion of\/ $\sH$. Then precisely one of the following applies.

\begin{enumerate}
    \item The clone $\CA$ has no uniformly continuous minion homomorphism to the clone of affine maps over a finite module, and $\Csp(\sA)$ has relational width $(2\ell,\max(3\ell,b_{\sH}))$.
    \item The clone $\CA$ has a uniformly continuous minion homomorphism to the clone of affine maps over a finite module.
\end{enumerate}
\end{theorem}

We  prove~\Cref{thm:hypergraphs_bwidth} in a similar way as~\Cref{thm:hypergraph-dichotomy}. We prove that if $\sA$ is a first-order expansion of $\sH$ which is a model-complete core, then $\Csp(\sA)$ has bounded width if, and only if, $\CAH\actson\en$ is equationally non-affine.
If $\CAH\actson\en$ is equationally non-affine, then it follows from~\cite{MarotiMcKenzie, Maltsev-Cond} that for every $k\geq 3$, $\CAH$ contains an operation of arity $k$ that acts as a WNU operation on $\en$.
An easy modification of the proof of Theorem~2 in~\cite{SymmetriesEnough} for injectivisations of instances instead of instances themselves yields that $\Csp_{\injinstances}(\sA)$ has relational width $(2\ell,\max(3\ell, b_{\sH}))$.  
Moreover, $\CA$ contains a binary injection by \Cref{prop:canonical_injection}, and the result follows from~\Cref{corollary:CAH_eq-non-affine}.

Let us therefore suppose that $\CAH\actson\en$ is equationally affine. Moreover, we can assume that $\CAH\actson\en$ is equationally non-trivial as otherwise, $\CA$ has a uniformly continuous homomorphism to the clone of projections by the proof of~\Cref{thm:hypergraph-dichotomy}. We apply the second loop lemma of smooth approximations~\cite[Theorem 11]{SmoothApproximations}.

\begin{theorem}\label{thm:2nd_loop_lemma}
Let $k\geq 1$, and suppose that $\CAH\actson \en$ is equationally non-trivial. Let $(S,\sim)$ be a minimal subfactor of $\CAH\actson\en$ with $\Aut(\sH)$-invariant $\sim$-classes. Then one of the following holds:

\begin{itemize}
    \item $\sim$ is approximated by a $\CA$-invariant equivalence relation that is very smooth with respect to $\Aut(\sH)$;
    \item every $\CAH\actson\en$-invariant binary symmetric relation $R\subseteq I^2$ that contains a pair $(\tuple a,\tuple b)\in S^2$ such that $\tuple a,\tuple b$ are disjoint and such that $\tuple a\not\sim \tuple b$ contains a pseudo-loop modulo $\Aut(\sH)$, i.e., a pair $(\tuple c,\tuple c')$ where $\tuple c,\tuple c'$ belong to the same orbit under $\Aut(\sH)$.
\end{itemize}
\end{theorem}

In the formulation of the first item of~\Cref{thm:2nd_loop_lemma}, we are using~\cite[Lemma 8]{SmoothApproximations}. If the first case of the theorem applies, i.e., the equivalence relation $(S,\sim)$ on whose classes $\CAH$ acts by a function from a clone $\cM$ of affine maps over a finite module is approximated by a $\CA$-invariant equivalence relation that is very smooth with respect to $\Aut(\sH)$,
~\Cref{lemma:CAA_eq-trivial} (with $k=\ell$, $\CAA=\CAH$) implies that $\CA$ has a uniformly continuous clone homomorphism to $\CAH\actson (S,\sim)$, and hence to $\cM$.

If the second case of~\Cref{thm:2nd_loop_lemma} applies, we get a weakly commutative function by~\cite[Lemma 13]{SmoothApproximations}, and the same argument mentioned at the end of the first part of \Cref{sect:NP-hard} using \Cref{lemma:weaklycomm-semilattice} gives that $\CAH\actson\en$ contains a semilattice operation. In particular, $\CAH\actson\en$ is equationally non-affine, which is a contradiction.

\bibliographystyle{plain}
\bibliography{main}

\end{document}